\documentclass[preprint,aps,floats,nofootinbib]{revtex4}
\usepackage{graphicx}
\usepackage{amsmath}
\usepackage{amssymb}
\usepackage{slashed}
\makeatletter
\@addtoreset{equation}{section}
\makeatother
\usepackage{bm}
\newcommand{\tr}{{\mathop{\mbox{tr}}\nolimits}} % trace "tr"
\newcommand{\diag}{{\mathop{\mbox{diag}}\nolimits}} % diagonal matrix "diag"
\renewcommand{\Re}{\mathop{\mbox{Re}}} % Real part "Re"
\newcommand{\calD}{{\cal D}}

\newcommand{\calL}{{\cal L}}
\newcommand{\calM}{{\cal M}}

\newcommand{\calO}{{\cal O}}

\newcommand{\g}{{g}}
\newcommand{\rescaled}[1]{\breve{#1}}
\begin{document}
\date{February 16, 2016}
\title{Collider signals of $W'$ and $Z'$ bosons
\\ in the gauge-Higgs unification}
\author{Shuichiro Funatsu$^a$}
%\email{funatsu@het.phys.sci.osaka-u.ac.jp}
\author{Hisaki Hatanaka$^{b}$}
%\email{hatanaka@het.phys.sci.osaka-u.ac.jp}
\author{Yutaka Hosotani$^b$}
%\email{hosotani@phys.sci.osakau.ac.jp}
\author{Yuta Orikasa$^{c}$}
\affiliation{$^a$KEK Theory Center, Tsukuba, Ibaraki 305-0801, Japan}
\affiliation{$^b$Department of Physics, Osaka University, Toyonaka, Osaka 560-0043, Japan}
\affiliation{$^c$Institute of Experimental and Applied Physics, Czech Technical University, Prague 12800, Czech Republic}
%\email{orikasa@kias.re.kr}
%
\begin{abstract}
In the $SO(5)\times U(1)$ gauge-Higgs unification (GHU), Kaluza-Klein (KK) excited states of charged and neutral vector bosons, $W^{(1)}$, $W_R^{(1)}$, $Z^{(1)}$, $\gamma^{(1)}$ and 
$Z_R^{(1)}$, can be observed as $W'$ and $Z'$ signals in collider experiments. 
In this paper we evaluate the decay rates of the $W'$ and $Z'$, and $s$-channel cross sections mediated by $W'$ and $Z'$ bosons with final states involving the standard model (SM) fermion pair ($\ell \nu$, $\ell\bar{\ell}$, $q\bar{q}'$), $WH$, $ZH$, $WW$ and $WZ$.
$W'$ and $Z'$ resonances appear around $6.0$ TeV ($8.5$ TeV) for $\theta_H = 0.115$ (0.0737)
where $\theta_H$ is the Aharonov-Bohm phase in the fifth dimension in GHU.
For decay rates we find 
$\Gamma(W' \to WH) \simeq \Gamma(W' \to WZ)$ ($W' = W^{(1)}$, $W_R^{(1)}$), 
$\Gamma(W^{(1)} \to WH,WZ) \sim \Gamma(W_R^{(1)} \to WH,WZ)$,
$\Gamma(Z^{(1)} \to ZH) \simeq \sum_{Z'=Z^{(1)},\gamma^{(1)}} \Gamma(Z' \to WW)$, and
$\Gamma(Z_R^{(1)} \to ZH) \simeq \Gamma(Z_R^{(1)}\to WW)$.
$W'$ and $Z'$ signals of GHU can be best found at the LHC experiment 
in the processes $pp \to W' (Z') + X$  followed by
$W' \to t\bar{b}$, $WH$, and $Z' \to e^+ e^-$, $\mu^+\mu^-, ZH$ near the $W'$ and $Z'$ resonances.
For the lighter $Z'$ ($\theta_H = 0.115$) case, with forthcoming 30 fb$^{-1}$ data of the 13 TeV LHC experiment we expect about ten $\mu^+\mu^-$ events for the invariant mass range 3000 to 7000 GeV, though the number of the events becomes much smaller when $\theta_H = 0.0737$.
In the process with $WZ$ in the final state, it is confirmed that the leading contributions in the amplitude from the longitudinal polarizations of $W$ and $Z$
in the $s$-, $t$- and $u$-channels cancel with each other so that the unitarity is preserved, provided that all KK excited states in the intermediate states are taken into account. Deviation of the $WWZ$ coupling from the SM is very tiny.
Exotic partners $t_T^{(1)}$ and $b_Y^{(1)}$ of the top and bottom quarks with electric charge $+5/3$ and $-4/3$ have mass $M_{t_T^{(1)},b_Y^{(1)}} = 4.6$ TeV ($5.4$ TeV) for $\theta_H = 0.115$ ($0.0737$), becoming the lightest non-SM particles in GHU which can be singly produced in collider experiments.
\end{abstract}
\preprint{KEK-TH-1944, OU-HET/915}
\maketitle

\allowdisplaybreaks

%\tableofcontents

\section{Introduction}

At the LHC, $W'$ and $Z'$ are searched with various decay modes:
decay to lepton pairs ($\ell \bar{\ell}$, $\ell\nu$, $\nu\bar{\nu}$) \cite{Aaboud:2016cth,
CMS:2016abv,ATLAS:2014wra,Chatrchyan:2013lga,Khachatryan:2015pua,Aaboud:2016zkn}, 
a pair of top and bottom \cite{Aad:2014xra,Aad:2014xea,Chatrchyan:2012gqa},
$q\bar{q}$-dijet \cite{Aad:2015eha,ATLAS:2015nsi},
$WH(\to \ell\nu b\bar{b})$ and $ZH(\to \ell\bar{\ell}b\bar{b})$ \cite{Khachatryan:2015bma,Khachatryan:2016yji,Aaboud:2016lwx,Khachatryan:2016cfx}
and $WW$ and $WZ$ \cite{Aad:2015ipg,Chatrchyan:2012rva,Khachatryan:2014hpa,Khachatryan:2014xja,Aad:2015owa,ATLAS-CONF-2015-068,ATLAS-CONF-2015-073}. 
In the models with extra dimensions,
$W'$ and $Z'$ appear as Kaluza-Klein (KK) excited states of charged and neutral vector bosons.
The couplings of $W'$ and $Z'$ to fields in the standard model (SM) depend on the details of the models.
For example, in the minimal universal extra dimension (mUED) model \cite{Appelquist:2000nn},
the conservation of KK number forbids tree-level couplings of KK-excited states to the SM fields
so that production of $W'$ or $Z'$ in the collider experiment is highly suppressed.

Non-vanishing couplings of $W'$ and $Z'$ to the SM fields in models with extra-dimensions
originate from the violation of the KK number conservation
which reflects the translational invariance in the direction of the extra dimensional space. 
The KK number conservation is broken by,
for instance,
domain-wall like bulk mass terms,
brane localized mass terms to the fermion,
or warped extra dimensions.
It is well known that in the models in which SM fields live 
in the warped bulk space, couplings among fermions and KK-excited gauge bosons are non-vanishing and can be large.
In the minimal $SU(3)$ gauge-Higgs unification (GHU) model \cite{Kubo:2001zc} there are no $W'$ or $Z'$ couplings to the SM fields.
In a custodially-protected warped extra dimensional model \cite{Angelescu:2015yav}, $WW$ and $WH$ diboson signals have been studied.

The GHU provides a natural scenario for solving the gauge hierarchy
problem.  Many advances have been made in this direction recently \cite{%
Kubo:2001zc,Hosotani:1983xw, Davies1988, Hosotani1989a, Hatanaka:1998yp,%,
Hatanaka:1999sx,Scrucca:2003ra,Haba:2004qf, Agashe2005, Medina2007,%
Hosotani:2006qp,Sakamura:2006rf,Hosotani:2007qw,Hosotani:2007kn,%
Hosotani:2008tx,Hosotani:2008by,Haba:2009hw,Hosotani:2009qf, %
Serone:2009kf,%
Hosotani:2010hx,Hosotani:2011vr, Hasegawa2013,%
Funatsu:2013ni,Funatsu:2014fda,Funatsu:2014tka,Funatsu:2015xba,%
Burdman2003, HHKY2004, Lim2007,  Kojima:2011ad,%
Yamamoto2014, Kakizaki2014, Sakamura2014, Kitazawa2015,%
HY2015a, Yamatsu2016a,  Sakamura2016, Furui:2016owe,%
Adachi:2016zdi,Hasegawa:2016spd,%
Kojima:2016fvv, Hosotani2016b,%
deAnda:2014dba,%
Cossu:2013ora, Forcrand2015, Knechtli2016,%
HH-flat%
}. 
In the present paper, we evaluate couplings of KK excited gauge bosons to the SM gauge boson, Higgs boson, and fermions, and 
%evaluate cross sections not only with 
%In this paper, we 
study collider signals of $W'$ and $Z'$ in the $SO(5)\times U(1)$ GHU 
model which naturally incorporates the Higgs boson of mass $m_H = 125$ GeV
and gives almost the same phenomenology as the SM at low energies \cite{Funatsu:2013ni,Funatsu:2014fda,Funatsu:2014tka,Funatsu:2015xba}. 
In the previous work \cite{Funatsu:2014fda}, we reported that in hadron collider experiments large $Z'$ signals are expected due to the large couplings of right-handed fermions to the KK-excited gauge bosons \cite{Gherghetta:2000qt}.
In the present paper we evaluate cross sections not only with fermionic final states but also with
bosonic $WH$, $ZH$, $WW$ and $WZ$ final states.

In the warped space, in general, it is difficult to evaluate couplings among various fields
as they  should be calculated in their mass eigenstates.
One remarkable feature of the GHU is that 
the Higgs VEV can be eliminated by a large gauge-transformation 
and its effect is transmitted to the change in the boundary conditions 
so that one can obtain mass eigenstates easily.
In the previous work \cite{Funatsu:2015xba},
it is found that KK non-conserving couplings $HW^{(m)}W^{(n)}$ and $Z W^{(m)}W^{(n)}$ ($m\ne n$),
including $HWW^{(n)}$ and $ZWW^{(n)}$, are non-vanishing, and we used them in the calculation of the $H\to Z\gamma$ decay rate .

%In the present paper, we evaluate couplings of KK excited gauge bosons to
%SM gauge bosons, the Higgs boson and fermions in the $SO(5)\times U(1)$ GHU in warped space-time, and apply them to the collider phenomenology of $W'$ and $Z'$ bosons.

This paper is organized as follows.
In Sec.~\ref{sec:model}, the model is introduced .
In Sec.~\ref{sec:decaywidth} decay width and cross sections formulas are given.
% simplified model
In Sec.~\ref{sec:simplified} we introduce effective theories 
to qualitatively describe salient relations among various decay widths.
In Sec.~\ref{sec:numeric}, couplings and decay widths of $W'$ and $Z'$ are evaluated in GHU.
In Sec.~\ref{sec:cross_section} cross sections are evaluated. 
$W'$ and $Z'$ signals in $pp$ collision experiment at LHC are explored.
We also show how the unitarity in the process $d \bar u \to WZ$
 is ensured by including contributions from KK states of
vector bosons and fermions in the intermediate states.
% summary
Sec.~\ref{sec:summary} is devoted to summary.
% formula
In Appendix~\ref{sec:formula} $SO(5)$ generators and basis functions in the analysis are summarized.
In Appendices~\ref{sec:bosons} and \ref{sec:fermions} masses and wave functions for bosonic and fermionic KK states
are given, respectively.
In Appendix~\ref{sec:fermion-couplings} fermion couplings to vector bosons and the Higgs boson are summarized,
whereas cubic vector couplings and Higgs couplings are given in Appendix~\ref{sec:boson-couplings}.
In Appendices~\ref{sec:formula-decay} and \ref{sec:formula-scat},
formulae for decay widths and scattering cross sections are summarized, respectively.

%%%%%%%%%%%%%%%%%%%%%%%%%%%%%%%%%%%%%%%%%%%%%%%%%%%%%%%%%%%%%%%%%%%%%%%%%
\section{Model}\label{sec:model}
We consider five-dimensional (5D) gauge theory in the Randall-Sundrum space-time, whose metric is
\begin{eqnarray}
ds^2 &=& e^{-2\sigma(y)} \eta_{\mu\nu}dx^\mu dx^\nu + dy^2
= G_{MN} dx^M dx^N,
\end{eqnarray}
where $\eta_{\mu\nu} = \diag(-1,1,1,1)$. 
Here $\sigma(y) = k|y|$ for $-L \le y\le L$ and $\sigma(y+2L) = \sigma(y)$ is satisfied.
$k$ is the $AdS_5$ curvature.
$y=0$ and $y=L$ boundaries are referred to as the UV (ultraviolet) and IR (infrared) branes,
respectively.
The Kaluza-Klein (KK) mass scale is given by
\begin{eqnarray}
m_{KK} &\equiv& \frac{\pi k}{e^{kL} -1},
\end{eqnarray}
and when $kL \gtrsim 5$, it is followed by $m_{KK} \ll k$, $L^{-1}$.

This space-time has symmetric under the $Z_2$ reflection $y \to -y$,
and fundamental region of the extra dimension is given by $0 \le y \le L$.
In this region we introduce a new coordinate $z = e^{ky}$, with which the metric becomes
\begin{eqnarray}
ds^2 &=& \frac{1}{z^2} \left(\eta_{\mu\nu}dx^\mu dx^\nu  + \frac{dz^2}{k^2}\right).
\end{eqnarray} 
Note that $\partial_y = kz \partial_z$ and $V_y = kz V_z$ for a vector field $V_M$.

In the 5D bulk space there are $SO(5)$ and $U(1)_X$ gauge fields,
four $SO(5)$-vector fermions per generation
$\Psi_a^{\g}$ ($a=1,2,3,4$, $\g=1,2,3$), and $N_F$ $SO(5)$-spinor fermions $\Psi_{F_i}$ ($i=1,\cdots,N_F$). 
We note that each of $\Psi_{1}$ and $\Psi_{2}$ is an $SU(3)$-color triplet.

The bulk part of the action is given by
\begin{eqnarray}
S_{\rm bulk} &=& \int d^5 x \sqrt{-G} \biggl\{
 -\frac{1}{2}\tr G^{MR} G^{NS} F_{MN}^{(A)} F^{(A)}_{RS}
  - \frac{1}{4} G^{MR} G^{NS} F^{(B)}_{MN} F^{(B)}_{RS}
  \nonumber\\&& 
  + \frac{1}{2\xi_{(A)}} (f_{\rm gf}^{(A)})^2
  + \frac{1}{2\xi_{(B)}} (f_{\rm gf}^{(B)})^2
  + \calL_{GH}^{(A)}
  + \calL_{GH}^{(B)} 
\nonumber\\&&
  + \sum_{\g=1}^3 \sum_{a=1}^4 
  \bar{\Psi}^{\g}_{a} \calD(c^{\g}_{a}) \Psi^{\g}_{a}
  + \sum_{i=1}^{N_F} \bar{\Psi}_{F_i} \calD(c_{F_i}) \Psi_{F_i} \biggr\},
\\
  \calD(c_a) &\equiv& \Gamma^A e_{A}{}^{M}  
  \left(\partial_M + \frac{1}{8} \Omega_{MBC} [\Gamma^B, \Gamma^C]
    - i g_A A_M - i g_B Q_{X,a}  -  c_a k \epsilon(y) \right),
\end{eqnarray}
where $\epsilon(y) \equiv \sigma'/k$ is a sign function. 
$\Gamma^M$ denotes gamma matrices which is defined by $\{ \Gamma^M,\Gamma^N \} = 2\eta^{MN}$ ($\eta^{55} = +1$).
$e_A{}^M$ is an inverse fielbein, and
$\Omega_{MBC}$ is the spin connection.
$F_{MN}^{(A)} = \partial_M A_N - \partial_N A_M - i g_A [A_M, A_N]$ and 
$F_{MN}^{(B)} = \partial_M B_N - \partial_N B_M$.
$g_A$ and $g_B$ are 5D gauge couplings of $SO(5)$ and $U(1)_X$, respectively.
$g_w \equiv g_A/\sqrt{L}$ is the four-dimensional (4D) $SO(5)$ coupling.
$f_{\rm gf}^{(A)}$ and $f_{\rm gf}^{(B)}$ are gauge-fixing functions, and $\xi_{(A)}$ and $\xi_{(B)}$
are corresponding gauge parameters.
$\calL_{GH}^{(A)}$ and $\calL_{GH}^{(B)}$ denote ghost Lagrangians.

Bulk fermions are $SO(5)$-vectors. For the third generation, they are given by
\begin{eqnarray}
\Psi_1 &=& \left( \begin{pmatrix} T & t \\ B & b \end{pmatrix}, t' \right)
 = ((Q_1,q),t')
 = (\check{\Psi}_1^q, t'),
\nonumber
\\
\Psi_2 &=& \left( \begin{pmatrix} U & X \\ D & Y \end{pmatrix}, b' \right)
 = ((Q_2,Q_3),b')
 = (\check{\Psi}_2^q, b'),
\nonumber
\\
\Psi_3 &=& \left( \begin{pmatrix} \nu_\tau & L_{1X} \\ \tau & L_{1Y} \end{pmatrix} , \tau'\right)
 = ((\ell,L_1),\tau'),
 = (\check{\Psi}_3^\ell,\tau'),
\nonumber
\\
\Psi_4 &=& \left( \begin{pmatrix} L_{2X} & L_{3X} \\ L_{2Y} & L_{3Y} \end{pmatrix}, \nu'_\tau \right)
 = ((L_2,L_3),\nu'_\tau),
 = (\check{\Psi}_4^\ell, \nu'_\tau),
\end{eqnarray}
where $SO(4)$ vector is embedded in $(\bm{2},\bm{2})$-representation of $SU(2)_L\times SU(2)_R$  by
\begin{eqnarray}
\begin{pmatrix}
\psi_{11} & \psi_{12} \\ \psi_{21} & \psi_{22}
\end{pmatrix}
= \frac{1}{\sqrt{2}}(\psi_4 + i\vec{\sigma} \cdot \vec{\psi}) i\sigma_2
&=& 
\begin{pmatrix}
-i\psi_1 - \psi_2 & i\psi_3 + \psi_4 \\
i\psi_3 - \psi_4 & i\psi_1 - \psi_2 \end{pmatrix}.
\end{eqnarray}

The brane part of the action consists of scalar part $S_{\rm brane}^\Phi$ and brane-fermion part $S_{\rm brane}^{\chi}$.
The scalar part is given by
\begin{eqnarray}
S_{\rm brane}^{\Phi}
&=& \int d^5 x \sqrt{-G} \delta(y) [ - (D_\mu \Phi)^\dag (D^\mu\Phi) - \lambda_{\Phi} (|\Phi|^2 - w^2)^2],
\nonumber\\
D_\mu \Phi &=& \left( \partial_\mu  
- i \left\{ g_A \sum_{a_R=1}^3 A_\mu^{a_R} T^{a_R}
+  \frac{1}{2}g_B B_\mu\right\} \right)\Phi.
\end{eqnarray}

The fermion part of the brane action is
\begin{eqnarray}
S_{\rm brane}^{\chi}
&=& \int d^5 x \sqrt{-G} \delta(y) \{ \calL_{q} + \calL_{\ell} \},
\nonumber
\\
\calL_{q} &\equiv& 
\sum_{\g=1}^3 \sum_{\alpha=1}^3 (\hat{\chi}_{\alpha R}^{q,\g\dag} i \bar{\sigma}^\mu D_\mu \hat{\chi}_{\alpha R}^{q,\g})
- i \sum_{\g,\g'=1}^3 \bigl[
 \kappa_1^{q,\g\g'} \hat{\chi}_{1R}^{q,g\dag} \check{\Psi}_{1L}^{q,\g'} \tilde{\Phi}
+\tilde{\kappa}^{q,\g\g'} \hat{\chi}_{2R}^{q,g\dag} \check{\Psi}_{1L}^{q,\g'} \Phi
\nonumber\\&&
+\kappa_2^{q,\g\g'} \hat{\chi}_{2R}^{q,g\dag} \check{\Psi}_{2L}^{\g'} \tilde{\Phi}
+\kappa_3^{q,\g\g'} \hat{\chi}_{3R}^{q,g\dag} \check{\Psi}_{2L}^{\g'} \Phi
- \text{(H.c.)} 
\bigr],
\nonumber\\
\calL_{\ell} &\equiv& \sum_{\g=1}^3 \sum_{\alpha=1}^3 
(\hat{\chi}_{\alpha R}^{\ell,\g\dag} i \bar{\sigma}^\mu D_\mu \hat{\chi}_{\alpha R}^{\ell,\g})
- i \sum_{\g,\g'=1}^3 [
 \tilde{\kappa}^{\ell,\g\g'} \hat{\chi}_{3R}^{\ell,g\dag} \check{\Psi}_{3L}^{\ell,\g'} \tilde{\Phi}
+\kappa_1^{\ell,\g\g'} \hat{\chi}_{1R}^{\ell,g\dag} \check{\Psi}_{3L}^{\ell,\g'} \Phi
\nonumber\\&&
+\kappa_2^{\ell,\g\g'} \hat{\chi}_{2R}^{\ell,\g\dag} \check{\Psi}_{4L}^{\ell,\g'} \tilde{\Phi}
+\kappa_3^{\ell,\g\g'} \hat{\chi}_{3R}^{\ell,\g\dag} \check{\Psi}_{4L}^{\ell,\g'} \Phi
- \text{(H.c.)},
\\
D_\mu \hat{\chi} &=& \left(\partial_\mu - ig_A \sum_{a_L=1}^3 A_\mu^{a_L} T^{a_L}
- i Q_X g_B B_\mu \right) \hat{\chi},
\quad\tilde{\Phi}\equiv i\sigma_2 \Phi^*,
\end{eqnarray}
where 
\begin{eqnarray}
\hat{\chi}_{1R}^q &=& \begin{pmatrix} \hat{T}_R \\ \hat{B}_R \end{pmatrix}_{7/6},
\quad
\hat{\chi}_{2R}^q = \begin{pmatrix} \hat{U}_R \\ \hat{D}_R \end{pmatrix}_{1/6},
\quad
\hat{\chi}_{3R}^q = \begin{pmatrix} \hat{X}_R \\ \hat{Y}_R \end{pmatrix}_{-5/6},
\\
\hat{\chi}_{1R}^\ell &=& \begin{pmatrix} \hat{L}_{1XR} \\ \hat{L}_{1YR} \end{pmatrix}_{-3/2},
\quad
\hat{\chi}_{2R}^\ell = \begin{pmatrix} \hat{L}_{2XR} \\ \hat{L}_{2YR} \end{pmatrix}_{1/2},
\quad
\hat{\chi}_{3R}^\ell = \begin{pmatrix} \hat{L}_{3XR} \\ \hat{L}_{3YR} \end{pmatrix}_{-1/2},
\end{eqnarray}
are right-handed brane fermions. We note that each of 
$\hat{\chi}^q_{1R}$, $\hat{\chi}^q_{2R}$ and $\hat{\chi}^q_{3R}$ is an $SU(3)$-color triplet.
We also have introduced $3\times3$ Yukawa coupling matrices $\kappa^q_{1,2,3}$, $\kappa^\ell_{1,2,3}$, $\tilde{\kappa}^{q}$ and $\tilde{\kappa}^{\ell}$.

\subsection{Orbifold symmetry breaking}

We impose $Z_2$ boundary conditions at boundaries $y=y_i$, $y_0 \equiv 0$, $y_1 \equiv L$.
\begin{eqnarray}
\begin{pmatrix}
A_\mu  \\ A_y 
\end{pmatrix}(x,y_i-y)
 &=& P_i \begin{pmatrix} A_\mu \\ - A_y
 \end{pmatrix} (x,y_i + y )P_i^{-1},
\\
\begin{pmatrix}
B_\mu  \\ B_y 
\end{pmatrix}(x,y_i- y)
 &=& \begin{pmatrix} B_\mu \\ - B_y 
 \end{pmatrix} (x,y_i + y),
\\
\Psi_a (x,y_i - y) &=& \gamma_5 P_i \Psi(x,y_i+y)_a, 
\end{eqnarray}
where 
\begin{eqnarray}
P_0^{\rm vec} = P_1^{\rm vec} = \diag(-1,-1,-1,-1,+1),
\quad
P_0^{\rm sp} = P_1^{\rm sp} &=& \begin{pmatrix} \bm{1} & \\ & -\bm{1} \end{pmatrix}
\end{eqnarray}
in the vector and spinor representations, respectively.
These boundary conditions break $SO(5)$ to $SO(4)\simeq SU(2)_L \times SU(2)_R$.
$A_\mu^{a_L,a_R}$ and $A_y^{\hat{a}}$ are even function against reflections at $y=y_i$ and can have their zero-modes. Zero modes of $A_\mu^{a_L,a_R}$ are the gauge fields of unbroken $SO(4)$ symmetry.

\subsection{Symmetry breaking by brane scalar}

Once $\Phi$ develops a VEV
\begin{eqnarray}
\langle \Phi \rangle =  \begin{pmatrix} 0 \\ w \end{pmatrix},
\end{eqnarray}
$SU(2)_R \times U(1)_X$ symmetry is broken to $U(1)_Y$
After $\Phi$ develops a VEV, the boundary conditions in the original gauge are given by
\begin{eqnarray}
\text{at $z=1$} &:& 
\partial_z A_\mu^{a_L} = \left(\partial_z - \frac{\kappa}{2k}\right) A_\mu^{1_R,2_R} 
= \left(\partial_z - \frac{\kappa'}{2k}\right) A_\mu^{3'_R} = \partial_z B_\mu^{Y'}= 0,
\nonumber\\ &&
A_\mu^{\hat{a}} = A_\mu^{\hat{4}} = 0,
\nonumber\\ &&
A_z^{a_L} = A_z^{a_R} + B_z = 0,
\quad
\partial_z \left(\frac{1}{z} A_z^{\hat{a}}\right) = \partial_z \left(\frac{1}{z} A_z^{\hat{4}}\right) = 0,
\nonumber\\
\text{at $z=z_L$} &:& \partial_z A_\mu^{a_L} = \partial_z A_\mu^{a_R} = \partial_z B_\mu^X =0,
\nonumber\\ &&
 A_\mu^{\hat{a}} = A_\mu^{\hat{4}} = 0,
\nonumber\\ &&
A_z^{a_L} = A_z^{a_R} = B_z = 0, 
\quad
\partial_z \left( \frac{1}{z} A_z^{\hat{a}}\right) = \partial_z \left( \frac{1}{z} A_z^{\hat{4}}\right) = 0.
\end{eqnarray}
Here we have defined
\begin{eqnarray}
\kappa \equiv \frac{g_A^2 w^2}{4} = \frac{g_w^2 L w^2}{4},
\quad
\kappa' \equiv \frac{(g_A^2 + g_B^2)w^2}{4},
\end{eqnarray}
and
\begin{eqnarray}
&&
\begin{pmatrix} A_M^{3'_R} \\ B^{Y'}_M \end{pmatrix}
\equiv \begin{pmatrix}
c_\phi & -s_\phi \\ s_\phi & c_\phi \end{pmatrix}
\begin{pmatrix} A_M^{3_R} \\ B_M \end{pmatrix},
\nonumber\\
&& 
c_\phi \equiv \cos\phi = \frac{g_A}{\sqrt{g_A^2 + g_B^2}},
\quad
s_\phi \equiv \sin\phi = \frac{g_B}{\sqrt{g_A^2 + g_B^2}}.
\end{eqnarray}
where a mixing angle $\phi$ is defined by $\tan\phi \equiv g_B/g_A$.
KK modes of $A_\mu^{1,2_R}$ and $A_\mu'^{3_R}$ with $m_n \ll w$ obey effectively Dirichlet boundary conditions on the UV brane
: $A_\mu^{1_R,2_R} =  A_\mu^{3'_R}=0$ at $z=1$.

For fermions, non-vanishing $\langle \Phi \rangle$ also induces brane mass terms 
given by
\begin{eqnarray}
S_{\rm brane}^{\rm mass}
&=& \int d^5 x \sqrt{-G} \delta(y) \left\{
 \calL_{\rm quark}^{\rm mass} + \calL_{\rm lepton}^{\rm mass} \right\},
\\
\calL_{\rm quark}^{\rm mass} 
&=& 
\sum_{\g,\g'=1}^{3} \biggl[
-\sum_{\alpha=1}^3 
 i \mu_\alpha^{q,\g\g'}(\hat{\chi}_{\alpha R}^{q,g\dag} Q_{\alpha L}^{\g'} 
 - Q_{\alpha L}^{\g'\dag}\hat{\chi}_{\alpha R}^{q,\g})
-i \tilde{\mu}^{q,\g\g'}(\hat{\chi}_{2R}^{q,\g\dag} q_L^{\g'} - q_L^{\g'\dag} \hat{\chi}_{2R}^{q,\g})
\biggr],
\\
\calL_{\rm lepton}^{\rm mass} 
&=& \sum_{\g,\g'=1}^3 \biggl[
- \sum_{\alpha=1}^3 i \mu_{\alpha}^{\ell,\g\g'} 
(\hat{\chi}_{\alpha L}^{\ell,\g\dag} L_{\alpha L}^{\g'} - L_{\alpha L}^{\g'\dag} \hat{\chi}_{\alpha R}^{\ell,\g})
- i \tilde{\mu}^{\ell,\g\g'} (\hat{\chi}_{3R}^{\ell,\g\dag} \ell_L^{\g'} - \ell_L^{\g'\dag} \hat{\chi}_{3R}^{\ell,\g})
\biggr].
\end{eqnarray}
where
\begin{eqnarray}
\frac{\mu_\alpha^{q,\g\g'}}{\kappa_\alpha^{q,\g\g'}} 
= \frac{\tilde{\mu}^{q,\g\g'}}{\tilde{\kappa}^{q,\g\g'}} = 
\frac{\mu_\alpha^{\ell,\g\g'}}{\kappa_\alpha^{\ell,\g\g'}}
= \frac{\tilde{\mu}^{\ell,\g\g'}}{\tilde{\kappa}^{\ell,\g\g'}} = w.
\end{eqnarray}
For $w \gg m_{KK}$, and  $\mu_\alpha,\,\tilde{\mu} \gg \sqrt{m_{KK}}$, 
exotic fermions couple to brane fermions to become very heavy,
so that only quark and leptons remain at low energy.

Since the gauge field $A_y$ plays the role of the Higgs boson, the Yukawa couplings of quarks and leptons
are diagonal in the flavor space, and flavor mixing can be induced by non-diagonal brane mass terms.
For simplicity we assume that all brane mass terms are flavor-diagonal:
\begin{eqnarray}
&&
\mu_{\alpha}^{q,\g\g'} = \delta^{\g\g'} \mu_\alpha^q,
\quad
\mu_{\alpha}^{\ell,\g\g'} =  \delta^{\g\g'} \mu_{\alpha}^\ell,
\quad
\alpha = 1,2,3,
\nonumber
\\
&&
\tilde{\mu}^{q,\g\g'} = \delta^{\g\g'} \tilde{\mu}^q,
\quad
\tilde{\mu}^{\ell,\g\g'} = \delta^{\g\g'} \tilde{\mu}^\ell.
\end{eqnarray}

\subsection{Electroweak symmetry breaking}

$A_z^{\hat{a}}$ ($a=1,2,3$ and $4$) have their zero modes:
\begin{eqnarray}
A_z^{\hat{a}}(x,z) &=& \phi^a(x) \sqrt{\frac{2}{k(z_L^2-1)}}\cdot z + \cdots,
\end{eqnarray}
(where ``$\cdots$'' includes higher-KK modes)
and can develop a VEV.
We assume that $A_z$ develops a VEV in the direction of $T^{\hat{4}}$ and we
parameterize it by $\langle \phi^a \rangle = v_W \delta^{4a}$.
Then we define the Wilson-line phase parameter $\theta_H$ by
\begin{eqnarray}
\exp\left[\frac{i}{2}\theta_H (2\sqrt{2} T^{\hat{4}})\right]
&=& \exp\left[ig_A \int_0^{z_L} \langle A_z \rangle dz\right],
\end{eqnarray}
so that we obtain
\begin{eqnarray}
\theta_H &=& \frac{1}{2} g_A v_W \sqrt{\frac{z_L^2-1}{k}}
\sim \frac{g_2 v_W}{2} \frac{\pi \sqrt{kL}}{m_{KK}},
\end{eqnarray}
where $g_w \equiv g_A/\sqrt{L}$ is the 4D $SO(4)\simeq SU(2)_L \times SU(2)_R$ gauge coupling constant.
We also have a formula of $W$-boson mass.
\begin{eqnarray}
m_W &\simeq& \frac{m_{KK}}{\pi \sqrt{kL}}|\sin\theta_H|,
\end{eqnarray}
and for $\theta_H \ll 1$, $m_W = \frac{1}{2}g_w v_W$ is obtained.
This may be compared with the SM formula $m_W = \frac{1}{2} g_w v_H$, $v_H = 246$ GeV.

To solve the equations of motion, we move to the twisted gauge in which $\langle \tilde{A}_z \rangle = 0$.
This is achieved by
\begin{eqnarray}
\tilde{A}_M &=& \Omega A_M \Omega^{-1} + \frac{i}{g_A} \Omega (\partial_M \Omega^{-1}),
\quad
\tilde{\Psi} = \Omega \Psi,
\\
\Omega &=& \exp\left[i\theta(z)\sqrt{2}T^{\hat{4}}\right],
\quad
\theta(z) = \theta_H \frac{z_L^2-z^2}{z_L^2-1}.
\end{eqnarray}
Using $\Omega$ and making use of $SO(5)$ algebra,
we find gauge transformation as
\begin{eqnarray}
A_M^{a_L} &=& \frac{1}{\sqrt{2}} \left\{
\tilde{A}_M^{a_+} + \tilde{A}_M^{a_-} \cos\theta(z) - \tilde{A}_M^{\hat{a}} \sin\theta(z)
\right\},
\nonumber \\
A_M^{a_R} &=& \frac{1}{\sqrt{2}} \left\{
\tilde{A}_M^{a_+} - \tilde{A}_M^{a_-} \cos\theta(z) + \tilde{A}_M^{\hat{a}} \sin\theta(z)
\right\},
\nonumber\\
A_M^{\hat{a}} &=& \tilde{A}_M^{a_-} \sin\theta(z) + \tilde{A}_M^{\hat{a}}\cos\theta(z),
\quad a = 1,2,3,
\nonumber
\\
A_\mu^{\hat{4}} &=& \tilde{A}_\mu^{\hat{4}},
\quad
A_z^{\hat{4}} = \tilde{A}^{\hat{4}}_z - \frac{\sqrt{2}}{g_A}\theta'(z),
\end{eqnarray}
where 
$\tilde{A}_M^{a_\pm} \equiv (\tilde{A}_M^{a_L} \pm \tilde{A}_M^{a_R})/\sqrt{2}$.

\subsection{Kaluza-Klein towers}

\subsubsection{Gauge bosons}
$SO(5)\times U(1)_X$ gauge fields are decomposed into Kaluza-Klein towers given by
\begin{eqnarray}
A_\mu(x^\mu,z) + B_\mu (x^\mu,z) T^B
&=& 
  \hat{W}_\mu + \hat{W}_\mu^\dag
+ \hat{W}_{R\mu} + \hat{W}^\dag_{R\mu}
+ \hat{Z}_\mu
+ \hat{Z}_{R\mu}
+ \hat{A}^\gamma_\mu + \hat{A}^{\hat{4}}_\mu,
\label{eq:gauge-tower}
\end{eqnarray}
where $T^B$ is a $U(1)_X$ generator.
$\hat{W}$, $\hat{Z}$ and $\hat{A}^\gamma$ are KK towers for $W$, $Z$ bosons and photons in the SM, respectively. We note that each of $\hat{W}$ and $\hat{Z}$ towers contains two KK towers so that 
there are eleven KK towers in \eqref{eq:gauge-tower}.

Each tower has an expansion of the form
\begin{eqnarray}
\hat{A}^C_\mu &=& \sum_{n} A_\mu^{(n)}(x^\mu) 
\left\{ h_{A^{(n)}}^{L}(z) T^{-_L} + h_{A^{(n)}}^R(z) T^{-_R}
+ \hat{h}_{A^{(n)}}(z) T^{\hat{-}} \right\},
\nonumber
\\
\hat{A}^N_\mu &=& \sum_{n} A_\mu^{(n)}(x^\mu) 
\biggl\{ h_{A^{(n)}}^{L}(z) T^{3_L} + h_{A^{(n)}}^R(z) T^{3_R}
+ \hat{h}_{A^{(n)}}(z) T^{\hat{3}} + h_{A^{(n)}}^B(z) T^B \biggr\},
\nonumber
\\
\hat{A}_\mu^{\hat{4}} &=& \sum_{n} A_\mu^{(n)\hat{4}}(x) h_{A^{(n)\hat{4}}}(z) T^{\hat{4}},
\end{eqnarray}
for $\hat{A}^C = \hat{W}$, $\hat{W}_R$ and  $\hat{A}^N = \hat{Z}$, $\hat{Z}_R$ and $\hat{A}^\gamma$. 
$T^\pm = (T^1 \pm i T^2)/\sqrt{2}$.
Explicit forms of KK towers of gauge fields 
are summarized in Appendix~\ref{sec:bosons} and also found in \cite{Funatsu:2014fda}.

\subsubsection{$A_z$ and $B_z$}

$A_z$ and $B_z$ are expanded, in the twisted gauge, as
\begin{eqnarray}
\tilde{A}_z(x,z) &=& \sum_{a=1}^3 \hat{G}^a + \sum_{a=1}^3 \hat{D}^a + \hat{H},
\nonumber\\
B_z(x,z) &=& \hat{B} = \sum_{n=1}^\infty B^{(n)}(x) u_{{B}^{(n)}}(z) T^B.
\end{eqnarray}
$\hat{D}$ and $\hat{G}$ towers are expanded as 
\begin{eqnarray}
\hat{S}^- &=& \sum_{n} S^{-(n)}(x) \left\{ 
  u_{S^{(n)}}^{L}(z) T^{-_L}
+ u_{S^{(n)}}^{R}(z) T^{-_R}
+ \hat{u}_{S^{(n)}}(z) T^{\hat{-}} \right\}, 
\nonumber
\\
\hat{S}^3 &=& \sum_{n} S^{3(n)}(x) \left\{ 
  u_{S^{(n)}}^{L}(z) T^{3_L} 
+ u_{S^{(n)}}^{R}(z) T^{3_R} 
+ \hat{u}_{S^{(n)}} T^{\hat{3}} \right\}, 
\quad S = D,\,G,
\end{eqnarray}
whereas $\hat{H}$ is expanded as
\begin{eqnarray}
\hat{H} &=& \sum_{n=0}^\infty H^{(n)}(x) u_{H^{(n)}}(z) T^{\hat{4}}.
\end{eqnarray}
$H^{(0)}$ corresponds to the SM Higgs boson.
$\hat{D}$ contain two KK towers so that $\tilde{A}_z$ contain ten KK towers. 
We note that KK modes other than $H^{(0)}$ are Nambu-Goldstone bosons and eaten by KK excited states of corresponding vector bosons.

\subsubsection{Fermions}

Bulk fermions are also decomposed into KK towers as follows.
For example, 
quark bulk fermions $\Psi_1$ and $\Psi_2$ in the third generation
are decomposed into
\begin{eqnarray}
(\Psi_1 + \Psi_2) &=& \hat{t}_{T(5/3)}
+\hat{t}_{(2/3)}+
\hat{t}_{B(2/3)}+
\hat{t}_{U(2/3)}
\nonumber\\&& 
+
\hat{b}_{(-1/3)}+
\hat{b}_{D(-1/3)}+
\hat{b}_{X(-1/3)}+
\hat{b}_{Y(-4/3)},
\end{eqnarray}
where numbers in subscripts denote the electric charge of fields.
$\hat{t}$ and $\hat{b}$ are towers for top and bottom quarks, respectively, whereas others are towers for non-SM exotic partners of quarks.
We note that each of $\hat{t}$ and $\hat{b}$ contains two KK towers. 
In all $\Psi_1$ and $\Psi_2$ contain ten KK towers of fermions.
In the same way KK towers of quarks of the first and second generations are expressed as 
$\hat{u}_T+\hat{u} + \cdots$, $\hat{d} + \hat{d}_D +\cdots$, and so on.

Lepton bulk fermions $\Psi_3$ and $\Psi_4$ (in the third generation) are decomposed as
\begin{eqnarray}
(\Psi_3 + \Psi_4) &=&
\hat{\nu}_{\tau(0)} + \hat{\tau}_{(-1)}
+ 
\hat{\tau}_{1X(-1)} +
\hat{\tau}_{1Y(-2)}
\nonumber\\&&
 +
\hat{\nu}_{\tau 2X(+1)} + 
\hat{\nu}_{\tau 2Y(0)} + 
\hat{\nu}_{\tau 3X(0)} +
\hat{\tau}_{3Y(-1)},
\end{eqnarray}
where $\hat{\tau}$ and $\hat{\nu_{\tau}}$ are towers for tau and tau-neutrino, respectively, and others are towers for non-SM exotic lepton partners.

Hereafter we work with rescaled bulk fermions $\rescaled{\Psi} \equiv z^2\Psi$. 
One can find mass spectra of KK towers corresponding to 
$\hat{t}_T$, $\hat{b}_Y$, $\hat{t}$ and $\hat{b}$ in references
 \cite{Hosotani:2008tx,Hosotani:2009qf}.
For the fermions with $Q_{EM}=+2/3$, integrating brane fermions
and utilizing orbifold boundary conditions,
boundary conditions at $z=1_+$ in the original gauge are given by
\begin{eqnarray}
\frac{\mu_2}{2k} [\mu_2 \rescaled{U}_L + \tilde{\mu} \rescaled{t}_L] - D_+^{(2)} \rescaled{U}_L
&=& 0,
\nonumber\\
\frac{\mu_1^2}{2k} \rescaled{B}_L - D_+^{(1)} \rescaled{B}_L &=& 0,
\nonumber\\
\frac{\tilde{\mu}}{2k} [\mu_2 \rescaled{U}_L + \tilde{\mu} \rescaled{t}_L] -  D_+^{(1)} \rescaled{t}_L &=& 0,
\nonumber\\
\rescaled{t}'_L &=& 0,
\end{eqnarray}
where $D_\pm^{(a)} \equiv \pm(d/dz) + (c_a/z)$.
Bulk fermions in the twisted gauge, $\tilde{U}$, $\tilde{t}$, $\tilde{B}$ and $\tilde{t}'$ are obtained by gauge transformation
\begin{eqnarray}
\begin{pmatrix} (\tilde{t} - \tilde{B})/\sqrt{2} \\ \tilde{t}' \end{pmatrix}
&=& \begin{pmatrix} \cos\theta(z) & \sin\theta(z) \\ -\sin\theta(z) & \cos\theta(z) \end{pmatrix}
\begin{pmatrix} (t-B)/\sqrt{2} \\ t' \end{pmatrix},
\nonumber\\
\tilde{U} &=& U, 
\nonumber\\ 
\tilde{t} + \tilde{B} &=& t + B.
\end{eqnarray}
where 
$\theta(z) = \theta_H (z_L^2-z^2)/(z_L^2-1)$.
We express KK expansion of bulk fermions in the twisted gauge as
\begin{eqnarray}
\begin{pmatrix} \rescaled{\tilde{U}} \\ \rescaled{\tilde{t}} \\ \rescaled{\tilde{B}} \\ \rescaled{\tilde{t}}' 
\end{pmatrix}
= \sqrt{k}\sum_{n} \psi_L^{(n)}(x) \begin{pmatrix}
a_U C_L^{(2)} \\ a_t C_L^{(1)} \\ a_B C_L^{(1)} \\ a_{t'} S_L^{(1)}
\end{pmatrix}(z,\lambda_n)
+\sqrt{k}\sum_{n} \psi_R^{(n)}(x) \begin{pmatrix}
a_U S_R^{(2)} \\ a_t S_R^{(1)} \\ a_B S_R^{(1)} \\ a_{t'} C_R^{(1)}
\end{pmatrix}(z,\lambda_n),
\label{eq:fermion-KK}
\end{eqnarray}
where $C_L^{(a)}(z,\lambda_n) = C_L(z;\lambda_n,c_a)$ and so on.
 $\psi_L^{(n)}$ and $\psi_R^{(n)}$ are 4D left- and right-handed
fermions with the mass $m_n = k\lambda_n$, respectively.
When we assume that $\mu_1^2,\, \mu_2^2,\, \tilde{\mu}^2,\, \tilde{\mu}\mu_2 \gg k\lambda_n$,
then the boundary conditions at $z=1$ are simplified as follows
\begin{eqnarray}
\rescaled{B_L} = 0,
\quad
\mu_2 \rescaled{U}_L + \tilde{\mu} \rescaled{t}_L = 0,
\quad
\tilde{\mu} D_+^{(2)} \rescaled{U}_L - \mu_2 D_+^{(1)} \rescaled{t}_L = 0,
\quad
\rescaled{t}'_L = 0.
\end{eqnarray}
Substituting left-handed KK modes in \eqref{eq:fermion-KK} into the above conditions,
we can obtain KK masses and corresponding eigenstates. 
KK fermions with $Q_{EM} = -1/3$ are also obtained in a similar way.

Masses and couplings of fermions are summarized in Appendices~\ref{sec:fermions}
and \ref{sec:fermion-couplings}, respectively.
It is seen in Tables. \ref{tbl:fermion-masses5} and \ref{tbl:fermion-masses4},
that $t_T^{(1)}$ and $b_Y^{(1)}$, which are exotic partners of the top and bottom quarks with electric charge $+5/3$ and $-4/3$, respectively, have mass $M_{t_T^{(1)},b_Y^{(1)}} = 4.6$ TeV ($5.4$ TeV) for $\theta_H = 0.115$ ($0.0737$).
$t_T^{(1)}$ and $b_Y^{(1)}$ are the lightest non-SM states in GHU which can be singly produced in the colliders.

%%%%%%%%%%%%%%%%%%%%%%%%%%%%%%%%%%%%%%%%%%%%%%%%%%%%%%%%%%%%%%%%%%%%%%%
\section{Decay width and cross sections}\label{sec:decaywidth}

The relevant parts of the Lagrangian for the present study consist of 
cubic interactions among vector and Higgs bosons 
\begin{eqnarray}
\calL_{\rm eff}^{\rm boson}
&=& \calL_{W'WZ} + \calL_{W'WH} + \calL_{Z'WW} + \calL_{Z'ZH},
\nonumber
\\
\calL_{W'WZ} &=& \sum_{W' = W^{(1)},W_R^{(1)}} i g_{W'WZ} (\eta^{\alpha\gamma}\eta^{\beta\delta} - 
\eta^{\alpha\delta}\eta^{\beta\gamma}) (W^{\prime-}_\gamma Z_\delta \partial_\alpha W_\beta^{+}
+W^{-}_\gamma Z_\delta \partial_\alpha W_\beta^{\prime+}
\nonumber\\&&
+Z_\gamma W_\delta^{\prime+} \partial_\alpha W_\beta^{-}
+Z_\gamma W_\delta^{+} \partial_\alpha W_\beta^{\prime-}
%\nonumber\\&&
+W_\gamma^{\prime+}W_\delta^{-} \partial_\alpha  Z_\beta
+W_\gamma^{+}W_\delta^{\prime-} \partial_\alpha  Z_\beta),
\nonumber
\\
\calL_{W'WH} &=& g_{W^{(1)}WH} [H W^{+}_\mu W^{-(1)\mu} + HW^{-}_\mu W^{+(1)\mu}],
\nonumber
\\
\calL_{Z'WW}  &=& \sum_{Z' = Z^{(1)},\gamma^{(1)},Z_R^{(1)}} 
ig_{Z'WW}
(\eta^{\alpha\gamma}\eta^{\beta\delta} - 
\eta^{\alpha\delta}\eta^{\beta\gamma}) (W^{-}_\gamma Z'_\delta \partial_\alpha W_\beta^{+}
+W^{-}_\gamma Z'_\delta \partial_\alpha W_\beta^{+}
\nonumber\\&&
+Z'_\gamma W_\delta^{+} \partial_\alpha W_\beta^{-}
+Z'_\gamma W_\delta^{+} \partial_\alpha W_\beta^{-}
+W_\gamma^{+}W_\delta^{-} \partial_\alpha  Z'_\beta
+W_\gamma^{+}W_\delta^{-} \partial_\alpha  Z'_\beta),
\nonumber
\\
\calL_{Z'ZH} &=& \sum_{Z' = Z^{(1)},\gamma^{(1)},Z_R^{(1)}} g_{Z'ZH} [H Z_\mu Z'^\mu],
\end{eqnarray}
and the $W'$ and $Z'$ couplings to fermions 
\begin{eqnarray}
\calL_{\rm eff}^{\rm fermion} &=& \calL_{W'f\bar{f}} + \calL_{Z'f\bar{f}},
\nonumber
\\
\calL_{W'f\bar{f}} &=& \sum_{W' = W^{(1)},W_R^{(1)}}
\biggl\{
\sum_{(\ell,\nu_\ell)} 
\left[g_{W'\ell\nu}^L W'^-_\mu \bar{\ell}\gamma^\mu \frac{1-\gamma_5}{2} \nu_{\ell} + (\text{H.c.})\right]
\nonumber\\&&
+ \sum_{\rm color} \sum_{(U,D)}
\left[g_{W'UD}^L W'^-_\mu \bar{D}\gamma^\mu \frac{1-\gamma_5}{2} U + (\text{H.c.})\right]
+ (L \to R, \gamma_5 \to -\gamma_5)
\biggr\},
\nonumber
\\
\calL_{Z'f\bar{f}} &=& \sum_{Z' = Z^{(1)},\gamma^{(1)}, Z_R^{(1)}}
\biggl\{
\sum_{l=\ell,\nu_\ell} 
\left[g_{Z'l}^L Z'^-_\mu \bar{l}\gamma^\mu \frac{1-\gamma_5}{2} \nu_{l} + (\text{H.c.})\right]
\nonumber\\&&
+ \sum_{\rm color} \sum_{q=u,d,s,c,b,t}
\left[g_{Z'q}^L Z'^-_\mu \bar{q}\gamma^\mu \frac{1-\gamma_5}{2} q + (\text{H.c.}) \right],
+ (L \to R, \gamma_5 \to -\gamma_5)
\biggr\}.
\end{eqnarray}
The couplings $\calL_{\rm eff}^{\rm bosons}$ and $\calL_{\rm eff}^{\rm fermions}$ 
are summarized in Appendix~\ref{sec:boson-couplings} and Appendix~\ref{sec:fermion-couplings}, respectively.
 $Z'$ couplings to fermions are also given in Ref \cite{Funatsu:2014fda}.

Formulas for decay widths of $W'$ and $Z'$ are summarized in Appendix~\ref{sec:formula-decay}.
When $M_{W'},M_{Z'} \gg M_{W},M_{Z},M_{H}$,
partial decay widths of $W'=W^{(1)}$ and $Z'=Z^{(1)},\gamma^{(1)},Z_R^{(1)}$ are approximately given by
\begin{eqnarray}
\Gamma(W'\to WH) &\simeq& \frac{M_{W'}}{192\pi} \left(\frac{g_{HW'W}}{M_W}\right)^2,
\nonumber\\
\Gamma(Z' \to ZH) &\simeq& \frac{M_{Z'}}{192\pi} \left(\frac{g_{HZ'Z}}{M_Z} \right)^2,
\nonumber\\
\Gamma(W' \to WZ) &\simeq& \frac{M_{W'}}{192\pi} g_{W'WZ}^2 \frac{M_{W'}^4}{M_W^2 M_Z^2},
\nonumber\\
\Gamma(Z' \to W^+W^-) &\simeq& \frac{M_{Z'}}{192\pi} g_{Z'WW}^2 \frac{M_{Z'}^4}{M_W^4},
\end{eqnarray}
and
\begin{eqnarray}
\Gamma(W' \to f\bar{f}') &\simeq& N_c \frac{M_{W'}}{24\pi} ( |g_{W'ff'}^L|^2 + |g_{W'ff'}^R|^2),
\nonumber\\
\Gamma(Z' \to f\bar{f}) &\simeq& N_c \frac{M_{Z'}}{24\pi} ( |g_{Z'f}^L|^2 + |g_{Z'f}^R|^2),
\end{eqnarray}
where $N_c$ is the number of color of fermions $f,f'$.
For later use, we define ratios of partial decay widths  as follows.
\begin{eqnarray}
\frac{\Gamma(W' \to WH)}{\Gamma(W' \to WZ)} 
&\simeq& \frac{
\frac{g_{HW'W}^2}{M_W^2}}{
g_{W'WZ}^2\frac{M_{W'}^4}{M_W^2M_Z^2}}
= \frac{g_{HW'W}^2}{g_{W'WZ}^2} \frac{M_Z^2}{M_{W'}^4}
\equiv \eta_{W'},\label{eq:decay-ratio-W}
\\
\frac{\Gamma(Z'\to ZH)}{\Gamma(Z'\to WW)}
&\simeq& \frac{ \frac{g_{HZ'Z}^2 }{M_Z^2}  }{ g_{Z'WW}^2 \frac{M_{Z'}^4}{M_W^4} }
= \frac{g_{HZ'Z}^2}{g_{Z'WW}^2} \frac{M_W^4}{M_{Z'}^4 M_Z^2}
\equiv \eta_{Z'}.
\label{eq:decay-ratio-Z}
\end{eqnarray}

Formulas of $s$-channel cross sections mediated by $W'$ or $Z'$ are summarized in
Appendix~\ref{sec:formula-scat}.
When $\sqrt{s} \gg M_{W},M_Z,M_H$, 
cross sections for the processes $f\bar{f}' \to \{ W,\,W^{(1)}\} \to WH$
and $f\bar{f} \to \{ Z,\, Z^{(1)},\, Z_R^{(1)} \} \to ZH$ are approximately given by
\begin{eqnarray}
\lefteqn{
\sigma(f\bar{f}' \to \{ W,\, W^{(1)}\} \to WH)
} \nonumber\\
&\simeq& \frac{1}{N_c^i}\frac{1}{384\pi}\frac{s}{M_W^2}
\biggl\{
\sum_{V=W,W^{(1)}} \frac{g_{HVW}^2 [ |g_{Vff'}^L|^2 + |g_{Vff'}^R|^2]}{(s - M_V^2)^2 + M_V^2 \Gamma_V^2}
\nonumber\\&&
+ 2 \Re \left[
\frac{
g_{H W W} g_{H W^{(1)} W} [ (g_{Wff'}^L) (g_{W^{(1)}ff'}^{L})^* 
+ (g_{W ff'}^R) (g_{W^{(1)} ff'}^{R})^*]}{
{}[(s-M_{W}^2) + i M_{W}\Gamma_{W}]
{}[(s-M_{W^{(1)}}^2) - i M_{W^{(1)}}\Gamma_{W^{(1)}}]}
\right]
\biggl\},
\\
\lefteqn{
\sigma(f\bar{f} \to \{ Z,\, Z^{(1)},\, Z_R^{(1)} \} \to ZH)
}\nonumber\\
&\simeq& \frac{1}{N_c^i} \frac{1}{384\pi} \frac{s}{M_Z^2} \biggl\{
\sum_{V=Z,Z^{(1)},Z_R^{(1)}} 
\frac{g_{HVZ}^2 [ |g_{Vf}^L|^2 + |g_{Vf}^R|^2]}{
(s - M_V^2)^2 + M_V^2 \Gamma_V^2}
\nonumber\\&&
+ \sum_{\substack{V_1,V_2 = Z,Z^{(1)},Z_R^{(1)} \\ V_1\ne V_2}}
\Re\biggl[
\frac{
g_{HV_1 Z} g_{HV_2 Z} [ (g_{V_1 f}^L) (g_{V_2 f}^{L})^* + (g_{V_1 f}^R) (g_{V_2 f}^{R})^*]
}{
{}[(s-M_{V_1}^2) + i M_{V_1}\Gamma_{V_1}]
{}[(s-M_{V_2}^2) - i M_{V_2}\Gamma_{V_2}]}
\biggr]
\biggr\}.
\end{eqnarray}

For processes $f\bar{f} \to WW$ and $f\bar{f}' \to WZ$,
careful treatments are necessary.
Each amplitude contains not only $s$- channel diagrams but also $t$- and $u$- channel diagrams.
Therefore the total amplitude $\calM$ will be given by
\begin{eqnarray}
\calM &=& \calM^{SM} + \calM^{NP},
\nonumber\\
\calM^{SM} &=& \calM_{s}^{SM} + \calM_{t}^{SM} + \calM_{u}^{SM},
\quad
\calM^{NP} =  \calM_{s}^{NP} + \calM_{t}^{NP} + \calM_{u}^{NP},
\end{eqnarray}
where $\calM_{s,t,u}^{SM}$ and $\calM_{s,t,u}^{NP}$ are $s$-, $t$- and $u$-channel amplitudes for the SM fields and new physics parts, respectively.
The square of the total amplitude is given by
\begin{eqnarray}
|\calM|^2 &=& |\calM^{SM}|^2 + |\calM^{NP}|^2 + (\text{interference}),
\end{eqnarray}
where the interference terms contain products of $\calM^{SM(*)}$ and $\calM^{NP(*)}$.
When the energy of the initial state $\sqrt{s}$ is much larger than the EW scale, each of $\calM_{s,t,u}^{SM}$ grows due to the longitudinal part of the vector bosons in the final state. 
In the SM  it is known that the growing contributions from longitudinal parts cancels with each other precisely, and that the unitarity of the amplitude $\calM^{SM}$ is protected. 
In our model, couplings among SM fields are very close to those of the SM values,
so that ${\cal M}^{SM}$ is well-behaved.
In the vicinity of  $M_{W'}$ and $M_{Z'}$ productions, $|\calM^{NP}_s|^2$ dominates over
the interference terms.
The relevant formulas for cross sections for the SM part  $|\calM^{SM}|^2$ are found in \cite{Brown:1978mq,Brown:1979ux}. 
For the NP part near the resonance $\sqrt{s} \sim M_{W'},\, M_{Z'} \gg M_W, M_Z,M_H$, 
one can neglect small $M_{t,u}^{NP}$.
% so that $|\calM|^2$ is approximated by $|\calM^{SM}|^2 + |\calM^{NP}_s|^2$.
In the precesses $f\bar{f} \to WW$ and $f\bar{f}' \to WZ$,
we approximate $|\calM|^2$ by $|\calM^{SM}|^2 + |\calM^{NP}_s|^2$, 
though the interference term need to be included for more rigorous treatment.
Cross sections of $s$-channel processes $f\bar{f}' \to \{ W^{(n)} \} \to WZ$ and $f\bar{f} \to \{ \gamma^{(n)},\, Z^{(n)},\, Z_R^{(n)} \} \to W^+W^-$ are given in Appendix~\ref{sec:formula-scat}.
Hence the NP part of the cross sections are approximately given by
\begin{eqnarray}
\lefteqn{
\sigma(f\bar{f}' \to \{  W^{(n)} \} \to WZ)}
\nonumber\\
&\simeq& \frac{1}{N_c^i} \frac{1}{384\pi} \frac{s^3}{M_W^2 M_Z^2} 
\biggl\{ 
\sum_{\substack{
V=W^{(n)} \\
n\ge1
}} \frac{[|g_{Vff'}^L|^2 + |g_{Vff'}^R|^2] g_{VWZ}^2}{(s-M_V^2)^2 + M_V^2 \Gamma_V^2}
\nonumber\\&&
+ \sum_{\substack{
(V_1,V_2) = (W^{(n)},W^{(m)})
\\
1\le n<m
}}
2\Re \biggl[ 
\frac{g_{V_1 WZ}g_{V_2 WZ}
{} [(g_{V_1 ff'}^L)(g_{V_2 ff'}^L)^* + (g_{V_1 ff'}^R)(g_{V_2 ff'}^R)]
}{
{} [(s-M_{V_1}^2) + i M_{V_1} \Gamma_{V_1} ][(s-M_{V_2}^2) - i M_{V_2}\Gamma_{V_2}]}
\biggr]
\biggr\},
\nonumber\\
\\
\lefteqn{
\sigma(f\bar{f} \to \{ \gamma^{(n)},\, Z^{(n)},\, Z_R^{(n)} \} \to W^+W^-)
}\nonumber\\
&\simeq& \frac{1}{N_c^i} \frac{1}{384\pi} \frac{s^3}{M_W^4} \biggl\{
\sum_{\substack{
V=Z^{(n)},\gamma^{(n)},Z_R^{(n)}\\
n \ge1
}}
 \frac{g_{VWW}^2 \left[|g_{Vf}^L|^2 + |g_{Vf}^R|^2 \right] }{(s-M_V^2)^2 + M_V^2 \Gamma_V^2}
\nonumber\\&&
+ \sum_{\substack{(V_1,V_2) = \{ Z^{(n)},\gamma^{(n)},Z_R^{(n)}\}
\\
n\ge1
\\
M_{V_1} < M_{V_2}}} 2\Re \biggl[ 
\frac{g_{V_1 WW} g_{V_2 WW} [(g_{V_1 f}^L) (g_{V_2 f}^{L})^* + (g_{V_1 f}^R) (g_{V_2 f}^{R})^*]
}{
{}[(s-M_{V_1}^2) + i M_{V_1} \Gamma_{V_1}]
{}[(s-M_{V_2}^2) - i M_{V_2} \Gamma_{V_2}]}
\biggr]
\biggr\}.
\nonumber\\
\end{eqnarray}
We note that at around the resonance points $s\simeq M_{W^{(1)}},M_{Z^{(1)}},
M_{Z_R^{(1)}}$, we see that the ratios of the
cross sections behave
\begin{eqnarray}
\left.
\frac{\sigma(f\bar{f}' \to W^{(1)} \to WH)}{\sigma(f\bar{f}' \to W^{(1)} \to WZ)}
\right|_{s=M_{W^{(1)}}}
&\simeq& 
\frac{ \displaystyle
\frac{M_{W^{(1)}}^2}{M_W^2} g_{HW^{(1)}W}^2}{ \displaystyle
\frac{M_{W^{(1)}}^6}{M_W^2 M_Z^2} g_{W^{(1)} WZ}^2
} 
= 
\frac{g_{HW^{(1)}W}^2}{g_{W^{(1)} WZ}^2}
\frac{M_Z^2}{M_{W^{(1)}}^4} = \eta_{W^{(1)}},
\label{eq:ratio-section-W}
\end{eqnarray}
and
\begin{eqnarray}
\left.
\frac{\sigma(f\bar{f}' \to Z' \to ZH)}{\sigma(f\bar{f}' \to Z' \to WW)}
\right|_{s=M_{Z'}}
&\simeq& 
\frac{ \displaystyle
\frac{M_{Z'}^2}{M_Z^2} g_{HZ'Z}^2}{ \displaystyle
\frac{M_{Z'}^6}{M_W^4} g_{Z'WW}^2} 
= 
\frac{g_{HZ^{(1)}Z}^2}{g_{Z^{(1)} WW}^2}
\frac{M_W^4}{M_Z^2 M_{Z'}^4} = \eta_{Z'},
\label{eq:ratio-section-Z}
\end{eqnarray}
for $Z' = Z^{(1)},\, Z_R^{(1)}$.
It is observed that ratios \eqref{eq:decay-ratio-W} and \eqref{eq:decay-ratio-Z}
coincide with  \eqref{eq:ratio-section-W} and \eqref{eq:ratio-section-Z}, respectively.

For SM fermions $f^{(\prime)},\, F^{(\prime)}$,
the cross section of the process $f\bar{f}' \to \{ W,\, W^{(1)} \} \to F\bar{F}'$ 
is given by
\begin{eqnarray}
\lefteqn{\sigma(f\bar{f}' \to \{ W,\, W^{(1)} \} \to F\bar{F}')}\nonumber\\
&\simeq& \frac{N_c^f}{N_c^i}\frac{s}{48\pi}\biggl\{
\sum_{V=W,W^{(1)}} 
\frac{ \left[|g_{Vff'}^L|^2 + |g_{Vff'}^R|^2 \right] \left[|g_{VFF'}^L|^2 + |g_{VFF'}^R|^2 \right]}{
(s-M_V^2)^2 + M_{V}^2 \Gamma_{V}^2 }
\nonumber\\&&
+ 2\Re \left[\frac{ 
%\left[ (g_{Wff'}^L) (g_{W^{(1)}ff'}^L)^* +  (g_{Wff'}^R) (g_{W^{(1)}ff'}^R)^* \right] 
%\left[ (g_{WFF'}^L) (g_{W^{(1)}FF'}^L)^* + (g_{WFF'}^R) (g_{W^{(1)}FF'}^R)^* \right]
K(f,f') K(F,F')
 }{ 
 \left[(s-M_W^2)+i M_W \Gamma_W \right] 
 \left[(s-M_{W^{(1)}}^2) -i M_{W^{(1)}} \Gamma_{W^{(1)}} \right] }
 \right]
\biggr\},
\nonumber\\
K(f,f') &\equiv& (g_{Wff'}^L)(g_{W^{(1)}ff'}^L)^* + (g_{Wff'}^R)(g_{W^{(1)}ff'}^R)^*, 
\end{eqnarray}
where $N_c^f$ is the number of color of final-state fermions. A similar formula for processes 
$f\bar{f} \to Z,Z' \to f'\bar{f}'$ are found in \cite{Funatsu:2014fda}.

%%%%%%%%%%%%%%%%%%%%%%%%%%%%%%%%%%%%%%%%%%%%%%%%%%%%%%%%%%%%%%%%%%%%%%%%%%%
\section{Effective theories}\label{sec:simplified}

Before jumping to calculate the couplings numerically, 
we formulate effective theories which yield qualitative understandings of various couplings and decay widths.

\subsubsection{4D $SO(4)\times U(1)$ model}

Let us consider a 4D $SO(4)\times U(1)_X$ model with two scalars $\Phi_R$ and $\Phi_H$.
Gauge couplings of $SO(4) \simeq SU(2)_L \times SU(2)_R$ and $U(1)_X$ are denoted by $g_w$ and $g_b$, respectively.
$\Phi_R$ is $(\bm{0},\bm{2})_{-1/2}$ in $SU(2)_L \times SU(2)_R \times U(1)_X$
and $\Phi_H$ is a $SO(4)$-vector $(\bm{2},\bar{\bm{2}})$ corresponding to the Higgs boson.
When $\Phi_R$ develops a VEV $\langle \Phi_R \rangle = (0,\mu)/\sqrt{2}$,
the $SU(2)_R \times U(1)_X$ symmetry is broken to $U(1)_Y$.
We take $\mu = {\cal O}(m_{KK})$.
On the other hand non vanishing $\Phi_H$ breaks $SU(2)_L \times SU(2)_R$ to $SU(2)_V$ whose generators are $(T^{a_L} + T^{a_R})/\sqrt{2}$.
With both $\langle \Phi_R \rangle$ and $\langle \Phi_H \rangle$ non-vanishing,
$SO(4)\times U(1)_X$ symmetry is broken to $U(1)_{\rm em}$.
The mass matrices of gauge bosons
in $(A_\mu^{a_L},A_\mu^{a_R})$ ($a=1,2$)
and $(A_\mu^{3_L}, A_\mu^{3_R}, B_\mu)$ are given by
\begin{eqnarray}
&&
{\cal M}_C = \begin{pmatrix} M_{LL}^2 & - M_{LL}^2 \\ -M_{LL}^2 & M_{LL}^2 + M_{RR}^2 \end{pmatrix},
\quad
{\cal M}_N = \begin{pmatrix} M_{LL}^2 & - M_{LL}^2 & 0 \\
- M_{LL}^2 & M_{LL}^2 + M_{RR}^2 & - M_{RB}^2 \\
0 & - M_{RB}^2 & M_{BB}^2  \end{pmatrix}, 
\nonumber
\\
&&M_{LL}^2 = \frac{g_w^2 v^2}{4},
\quad
M_{RR}^2 = \frac{g_w^2 \mu^2}{4},
\quad
M_{RB}^2 = \frac{g_w g_b \mu^2}{4},
\quad
M_{BB}^2 = \frac{g_b^2 \mu^2}{4},
\end{eqnarray}
respectively. 
${\cal M}_C$ has two eigenvalues corresponding to the mass-squared of $W$ and $W_R$ bosons, respectively, which are given by
\begin{eqnarray}
M_W^2 &=& \bar{M}_W^2 (1 + \calO(v^2/\mu^2)),
\quad  \bar{M}_W \equiv \frac{g_w v}{2},
\nonumber
\\
M_{W_R}^2 &=& \frac{g_w^2 \mu^2}{4} (1 + \calO(v^2/\mu^2)).
\end{eqnarray}
${\cal M}_N$ has three eigenvalues, which correspond to mass-squared of the photon, $Z$-boson, and $Z_R$-boson, respectively. Here
\begin{eqnarray}
M_{\gamma} &=& 0,
\nonumber
\\
M_{Z}^2 &=& \frac{g_w^2v^2}{4} \cdot \frac{g_w^2 + 2g_b^2}{g_w^2 + g_b^2}
 (1 + \calO(v^2/\mu^2))
\nonumber\\
&=& \bar{M}_Z^2 (1 + \calO(v^2/\mu^2)),
\quad 
\bar{M}_Z \equiv \frac{g_w v}{2 \cos\theta_W},
\nonumber
\\
M_{Z_R}^2 &=& \frac{(g_w^2 + g_b^2)\mu^2}{4} (1 + \calO(v^2/\mu^2))
\nonumber\\
&=& \frac{g_w^2 \mu^2}{4} \frac{\cos^2\theta_W}{\cos^2\theta_W- \sin^2\theta_W}
(1 + \calO(v^2/\mu^2)).
\end{eqnarray}

Diagonalizing these mass matrices we obtain mass eigenstates $W,W'$ and $Z,Z',\gamma$.
Mixing matrices are given by
\begin{eqnarray}
\begin{pmatrix} A_\mu^{a_L} \\ A_\mu^{a_R} \end{pmatrix}
&=& 
\begin{pmatrix}
\cos\theta_C & \sin\theta_C \\
-\sin\theta_C & \cos\theta_C \end{pmatrix}
\begin{pmatrix} W^a_\mu \\ W_{R\mu}^a \end{pmatrix},
\quad a=1,2,
\nonumber\\
\begin{pmatrix} A_\mu^{3_L} \\ A_\mu^{3_R} \\ B_\mu \end{pmatrix}
&=& 
\begin{pmatrix}
\vec{v}_\gamma,\,
\vec{v}_- \cos\theta_N + \vec{v}_+ \sin\theta_N,\,
\vec{v}_+ \cos\theta_N - \vec{v}_- \sin\theta_N
\end{pmatrix}
\begin{pmatrix} \gamma_\mu \\ Z_\mu \\ Z_{R\mu} \end{pmatrix},
\nonumber
\\
\vec{v}_\gamma &=& \begin{pmatrix} \sin\theta_W \\ \sin\theta_W 
\\ \sqrt{\cos2\theta_W} \end{pmatrix},
\quad
\vec{v}_- = \begin{pmatrix}
\cos\theta_W \\
\displaystyle -\frac{\sin^2\theta_W}{\cos\theta_W} 
\\
-\tan\theta_W \sqrt{\cos2\theta_W}
\end{pmatrix},
\quad
\vec{v}_+ = \begin{pmatrix}
0 \\ 
\displaystyle \frac{\sqrt{\cos2\theta_W}}{\cos\theta_W}  \\ -\tan\theta_W
\end{pmatrix},
\nonumber\\
\end{eqnarray}
where mixing angles are determined so as to mixing matrices properly diagonalize 
mass matrices: 
\begin{eqnarray}
\tan(2\theta_C) &=& \frac{2v^2}{\mu^2},
\nonumber
\\
\tan(2\theta_N) &=& \frac{2g_w^3v^2\sqrt{g_w^2+2g_b^2}}{w^2(g_w^2+g_b^2)^2- 2g^2g_b^2v^2}.
\end{eqnarray}
The weak mixing angle is given by
\begin{eqnarray}
\sin\theta_W &=& \frac{g_b}{\sqrt{g_w^2 + 2g_b^2}},
\end{eqnarray}
so that $\gamma_\mu$ couples to $W^\pm$ bosons with $e = g_w \sin\theta_W$.

Now we can calculate couplings among these mass eigenstates. 
For vector-boson trilinear $V_1V_2V_3$ couplings and $V_1V_2H$ couplings,
we obtain
\begin{eqnarray}
g_{WW\gamma} &=& g_w \sin\theta_W = e,
\nonumber
\\
g_{WWZ} &=& g_w \cos\theta_W + \calO(v^4/\mu^4),
\nonumber
\\
g_{Z_RWW} &=& -g_w \frac{(\cos^2\theta_W - \sin^2\theta_W)^{3/2}}{\cos^3\theta_W}
\frac{v^2}{\mu^2},
\nonumber
\\
g_{ZW_RW} &=& -g_w \frac{1}{\cos\theta_W} \frac{v^2}{\mu^2},
\end{eqnarray}
and
\begin{eqnarray}
g_{HWW} &=& g_w \bar{M}_W(1 + \calO(v^2/\mu^2)),
\nonumber
\\
g_{HZZ} &=& \frac{g_w}{\cos\theta_W} \bar{M}_Z (1 + \calO(v^2/\mu^2)),
\nonumber
\\
g_{HW_RW} &=& - g_w \bar{M}_W(1 + \calO(v^2/\mu^2)),
\nonumber
\\
g_{HZ_R Z} &=& - g_w \bar{M}_Z \frac{\sqrt{\cos^2\theta_W-\sin^2\theta_W}}{\cos\theta_W}(1 + \calO(v^2/\mu^2)),
\end{eqnarray}
where $\bar{M}_W = g_wv/2$ and $\bar{M}_Z = \bar{M}_W/\cos\theta_W$ are used.

Hence we see that
%\begin{enumerate}
%\item
(a) $g_{ZWW}$ is very close to its SM-value $g_w \cos\theta_W$ and the deviation from the SM-value will be suppressed by a factor $\calO(v^4/m_{KK}^4)$,
%\item
(b) Deviations of $g_{HZZ}$ and $g_{HWW}$ couplings from their SM values
are both suppressed by $v^2/m_{KK}^2$,
%\item
(c) $g_{Z'WW}$ and $g_{ZW_RW}$ are suppressed by a factor $\calO(v^2/\mu^2)$.
\begin{eqnarray}
g_{W_R WZ},\, g_{Z_R WW} \sim g_{WWZ} \cdot\frac{v^2}{\mu^2},
\end{eqnarray}
%\item
(d) In contrast, values of $HWW_R$ and $HZZ_R$ approximately
equal to $HWW$ and $HZZ$ couplings in magnitudes but opposite in signs, respectively.
\begin{eqnarray}
g_{HW_RW} \sim - g_{HWW},
\quad
g_{HZ'Z} \sim - g_{HZZ}.
\end{eqnarray}
%\item
(e)
For the decay widths of $W_R$ and $Z_R$ one finds
\begin{eqnarray}
\frac{\Gamma(W_R\to ZW)}{\Gamma(W_R\to WH)} &\simeq& \left(\frac{g_{ZW_RW}}{g_{HW_RW}}
\frac{M_{W_R}^2}{\bar{M}_Z} \right)^2 = 1 + \calO(v^2/\mu^2),
\label{eq:decay-relation-WR}
\\
\frac{\Gamma(Z_R\to WW)}{\Gamma(Z_R\to ZH)} 
&\simeq& \left(\frac{g_{Z_RWW}}{g_{HZ'Z}} 
\frac{M_{Z_R}^2\bar{M}_Z}{\bar{M}_W^2}\right)^2 = 1 + \calO(v^2/\mu^2).
\label{eq:decay-relation-ZR}
\end{eqnarray}
We will find in the following section that the relation \eqref{eq:decay-relation-WR} will be satisfied,
and that \eqref{eq:decay-relation-ZR} needs to be generalized to incorporate  KK-$\gamma$ and KK-$Z$ bosons.
%\end{enumerate}

\subsubsection{5D $SO(5)\times U(1)_X$ GHU in the flat-space limit}

We also explore the flat-space limit of the warped $SO(5)\times U(1)_X$ GHU by taking $kL \to 0$ limit while keeping $L=\pi R$ finite \cite{HH-flat}. 
A similar model is seen in \cite{Serone:2009kf}.
In this limit $m_{KK} = 1/R$.
The $W$-boson mass and the AB phase $\theta_H$ are related with each other by
$\sqrt{2}\sin(m_W \pi R) = \sin\theta_H$.
The couplings among vector bosons and Higgs are summarized as follows.
Vector boson trilinear couplings are given by
\begin{eqnarray}
g_{\gamma^{(0)} W^{(n)} W^{(m)}} &=& \delta_{mn} e,
\quad
g_{\gamma^{(1)} W^{(n)} W^{(m)}} = \delta_{mn} 2\sqrt{2}e, 
\nonumber
\\
g_{Z^{(0)}W^{(0)}W^{(0)}} &=& g_w\cos\theta_W + \calO(m_W^4 R^4),
\nonumber
\\
g_{Z^{(1)} W^{(0)} W^{(0)}},\, g_{Z^{(0)} W^{(1)} W^{(0)}} &=& \calO(m_W^3 R^3),
\nonumber
\\
g_{Z_R^{(1)} W^{(0)} W^{(0)}} &=& \frac{8\sqrt{2}\sqrt{\cos2\theta_W}}{\pi \cos\theta_W} g_w m_W^2 R^2 + \calO(m_W^4 R^4),
\nonumber
\\
g_{Z^{(0)} W_R^{(1)} W^{(0)}} &=& \frac{8\sqrt{2}}{\pi} g_w m_W m_Z R^2 + \calO(m_W^4R^4).
\end{eqnarray}
Higgs vector-boson couplings are given by
\begin{eqnarray}
g_{H W^{(m)} W^{(n)}} &=& \delta_{mn} (-1)^n g_w m_{W^{(n)}},
\nonumber
\\
g_{H Z^{(m)} Z^{(n)}} &=&\delta_{mn} (-1)^n \frac{g_w}{\cos\theta_W} m_{Z^{(n)}},
\nonumber
\\
g_{H Z_R^{(1)} Z^{(0)}} &=& -\frac{2\sqrt{2}\sqrt{\cos2\theta_W}}{\pi \cos\theta_W } g_w m_Z[ 1 + \calO(m_W^2R^2) ],
\nonumber
\\
g_{H W_R^{(1)} W^{(0)}} &=&  -\frac{2\sqrt{2}g_w}{\pi} m_{W} [1  + \calO(m_W^2R^2)],
\nonumber
\\
g_{H \gamma^{(n)} Z^{(m)}},\, g_{H \gamma^{(n)} Z_R^{(m)}} &=& 0.
\end{eqnarray}
Here we note that $m_{W_R^{(1)}} = m_{Z_R^{(1)}} = 1/(2R)$.

It seems that $W$ and $W_R$, $Z$ and $Z_R$ are mixed in an ordinary manner as seen in the 4D model, whereas
$W$ and $W^{(n)}$, $Z$ and $Z^{(n)}$ are very weakly mixed.
This will be understood by mass-mixings in the original gauge.
Gauge fields for charged bosons are 
$A_\mu^{(n)\pm_L}, A_\mu^{(m)\pm_R}, A_\mu^{(m)\hat{\pm}}$, ($a=1,2$, $n=0,1,2,\cdots$ and $m=1,2,\cdots$).
For up to first KK excited states,
the mass matrix in the $(A_\mu^{(0)\pm_L},\,  A_\mu^{(1)\pm_R},\,
A_\mu^{(1)\pm_L},\, A_\mu^{(1)\hat{\pm}})$ basis
is given by
\begin{eqnarray}
\begin{pmatrix}
M_v^2 & - M_v^2 & 0 & 0 \\
-M_v^2 & M_v^2 + M_R^2 & - M_v'^2 & 0 \\
0 & -M_v'^2 & M_v^2 + M_L^2 & 0 \\
0 & 0 & 0 & 2 M_v^2 + M_X^2 \end{pmatrix},
\end{eqnarray}
where
\begin{eqnarray}
M_v^2 &=& \frac{1}{4}g_w^2 v^2,
\quad
M_R^2 = \frac{1}{4R^2},
\quad
M_L^2 = M_X^2 = \frac{1}{R^2},
\nonumber\\
M_v'^2 &=& 
\frac{1}{4}g_w^2 v^2 \frac{1}{\pi R} \int_0^{2\pi R}
\cos\left(\frac{\pi R - y}{2R}\right)\cos\left(\frac{\pi R - y}{R}\right) dy.
\end{eqnarray}
$A_\mu^{(0)\pm_L}$ and $A_\mu^{(1)\pm_R}$ have a mixing term, whereas there is no mixing between
$A_\mu^{(0)\pm_L}$ and $A_\mu^{(1)\pm_L}$ in the mass matrix due to the KK-number conservation.
Therefore mixing between $A_\mu^{(0)\pm_L}$ and $A_\mu^{(1)\pm_L}$ is induced only through both 
$A_\mu^{(0)^\pm_L}$-$A_\mu^{(1)\pm_R}$ and $A_\mu^{(1)\pm_L}$-$A_\mu^{(1)\pm_R}$ mixing terms 
so that the mixing angle is suppressed.

\subsubsection{Comment on the warped $SO(5)\times U(1)_X$ GHU}

In the flat GHU case, $A_\mu^{(0)a_L}$-$A_\mu^{(1)a_L}$ mass terms vanish.
This is because the Higgs wave function is constant along the extra dimension.
The mass term, which is written as an overlap integral of wave functions  of $A_\mu^{(0)\pm_L}$, $A_\mu^{(1)\pm_L}$
and the Higgs boson,
vanishes by the orthonormality conditions of wave functions.
In the warped case, the Higgs wave function is not constant along the direction of the extra dimension. 
Therefore $A_\mu^{(0)a_L}$-$A_\mu^{(1)a_L}$ mass terms does not vanish.

%%%%%%%%%%%%%%%%%%%%%%%%%%%%%%%%%%%%%%%%%%%%%%%%%%%%%%%%%%%%%%%%%%%
\section{Couplings and decay widths}\label{sec:numeric}
Input parameters used in the numerical study is summarized in Table \ref{tbl:inputs}.
The $W$ boson mass at the tree level becomes $M_W^{\rm tree} = 79.9$ GeV.
In the following study, we have adopted model parameters $N_F=4$ and $z_L = 10^4$, $10^5$.
With $(N_F,\,z_L)$ given
bulk mass parameter of dark fermions, $c_F$, is determined such that the resultant Higgs mass becomes $M_H = 125$ GeV. This procedure determine the value of $\theta_H$ and the bulk mass parameters of quarks and leptons.(See for the details \cite{Funatsu:2013ni,Funatsu:2014fda})
The resultant values of $\theta_H$ and $k$, $m_{KK}$ and first-KK gauge boson masses 
are tabulated in Table~\ref{tbl:masses}.
%
%and bulk mass parameters of quarks and leptons are tabulated in Table.~\ref{tbl:fermion-bulk}.
We set fermion bulk mass parameters as $c_1^{g} = c_2^{g} \equiv \{c_u,c_c,c_t\}$ 
and $c_3^{\g} = c_4^{\g} \equiv \{c_e,c_\mu,c_\tau \}$. 
These parameters are tuned so that fermion masses coincide with values in Tables \ref{tbl:inputs}. 
These are listed in listed in Table~\ref{tbl:fermion-bulk},

\begin{table}[htbp]
\caption{Input parameters.
Masses of $Z$ boson, leptons and quarks in the unit of GeV.}\label{tbl:inputs}
\begin{tabular}{ccccccc}
\hline
%$M_W$ & 
$M_Z$
& $\sin^2\theta_W$
&
& $m_e(M_Z)$ 
& $m_\mu(M_Z)$
& $m_\tau(M_Z)$
\\
%79.9 &
 91.1876 & 0.23126 && $0.487 \times 10^{-3}$ & 0.103 & 1.75\\
\hline
$m_u(M_Z)$ & $m_d(M_Z)$ & $m_s(M_Z)$ & $m_c(M_Z)$
& $m_b(M_Z)$ & $m_t(M_Z)$  
\\
$1.27 \times 10^{-3}$ & $2.90 \times 10^{-3}$ & 0.055 & 0.619 & 2.89 & 171 \\
\hline
\end{tabular}
\end{table}
\begin{table}[htbp]
\caption{
Aharonov-Bohm phase $\theta_H$, the bulk mass parameter of dark fermions $c_F$, 
$AdS_5$ curvature $k$,
Kaluza-Klein scale $m_{KK}$ and masses of first KK gauge bosons are given in the unit of GeV with respect to $z_L$ for $N_F=4$ are summarized.
$Z^{(1)}$, $W^{(1)}$ and $\gamma^{(1)}$ are almost degenerate.
Their mass differences is $1-2$ GeV.}\label{tbl:masses}
\begin{tabular}{c|cc|cccccc}
$z_L$ 
& $\theta_H$ 
& $c_F$
& $k$
& $m_{KK}$ 
& $m_{Z^{(1)}}$ 
& $m_{W^{(1)}}$
& $m_{\gamma^{(1)}}$
& $m_{Z_R^{(1)}}=m_{W_R^{(1)}}$ \\
& [rad.] && [GeV] & [TeV] & [TeV] & [TeV] & [TeV] & [TeV] \\
\hline
$10^5$ & 0.115  & 0.3321 & $2.36\times10^8$ 
& 7.41 % 7405  
& 6.00 %6003 
& 6.00 % 6004 
& 6.01 % 6006 
& 5.67 %5668 
\\
$10^4$ & 0.0737 & 0.2561 & $3.29\times10^7$ 
& 10.3 %10348 
& 8.52 %8520 
& 8.52 %8520 
& 8.52 %8522 
& 7.92 %7922 
\\   
\end{tabular}
\end{table}
\begin{table}[htbp]
\caption{Fermion bulk mass parameters for quarks and leptons.}\label{tbl:fermion-bulk}
\begin{tabular}{c|c|c|c|c|c|c}
$z_L$ & $c_{u}$ & $c_{c}$ & $c_{t}$ & $c_{e}$ & $c_{\mu}$ & $c_{\tau}$ \\
\hline
$10^5$ & 1.55 & 1.05 & 0.227  & 1.72 & 1.22 & 0.950 \\
$10^4$ & 1.82 & 1.19 & 0.0366 & 2.04 & 1.41 & 1.07 \\
\end{tabular}
\end{table}

\subsection{Couplings}

The couplings among vector bosons $V$, Higgs $H$, and SM fermions $f_{\rm SM}$
are evaluated from overlap integrals where functions of $V$, $H$ and $f_{\rm SM}$ are inserted.
The detailed formulas are given in Appendices~\ref{sec:boson-couplings} and \ref{sec:fermion-couplings},
and Refs.~\cite{Funatsu:2014fda,Funatsu:2015xba}.

In Table~\ref{tbl:WfL-couplings}, 
the left-handed couplings of SM fermions to the $W$ boson and its KK excited states
are tabulated.
Similar results for $\theta_H=\pi/2$ are found in Ref.~\cite{Hosotani:2009qf}.
The right-hand couplings vanish within the accuracy of numerical calculation.
In this model couplings of the SM fermions to the $W_R$ boson vanish.

\begin{table}[htbp]
\caption{Masses and couplings of $W^{(n)}$ to left-handed SM fermions.}\label{tbl:WfL-couplings}
\begin{tabular}{c|c|c|c|c|c}
\hline\hline
\multicolumn{6}{c}{$N_F=4$, $z_L=10^5$ ($\theta_H = 0.115$) } \\
\hline
& $n=0$ & $1$ & $2$ & $3$ & $4$\\ 
\hline
$m_{W^{(n)}}$ [GeV] & 79.9 & 6004 & 9034 & 13378 & 16538 \\
\hline
$g^L_{W^{(n)}\ell\nu}/(g_w/\sqrt{2})$ &&&&\\
$(\ell,\nu)=(e,\nu_e)$       & 1.00019 & -0.3455 & -0.02507 & ~0.2510 & ~0.01937 \\
$\phantom{(\ell,\nu)=}(\mu,\nu_\mu)$   & 1.00019 & -0.3455 & -0.02507 & ~0.2510 & ~0.01937 \\
$\phantom{(\ell,\nu)=}(\tau,\nu_\tau)$ & 1.00019 & -0.3452 & -0.02505 & ~0.2507 & ~0.01934 \\
\hline
$g^L_{W^{(n)}UD}/(g_w/\sqrt{2})$  &&&&\\
$(U,D)=(u,d)$ 			& 1.00019 & -0.3455 & -0.02507 & ~0.2510 & ~0.01937 \\
$\phantom{(U,D)=}(c,s)$ 			& 1.00019 & -0.3454 & -0.02506 & ~0.2510 & ~0.01936 \\
$\phantom{(U,D)=}(t,b)$ 			& 0.9993 & ~1.2970 & ~0.06527 & -0.4342 & -0.03110 \\
\hline
\end{tabular}
\\
\vspace{2em}
\begin{tabular}{c|c|c|c|c|c}
\hline\hline
\multicolumn{6}{c}{$N_F=4$, $z_L=10^4$ ($\theta_H = 0.0737$)}\\
\hline
& $n=0$ & $1$ & $2$ & $3$ & $4$\\ 
\hline
$m_{W^{(n)}}$ [GeV] & 79.9 & 8520 & 12624 & 18852 & 23112 \\
\hline
$g^L_{W^{(n)}\ell\nu}/(g_w/\sqrt{2})$ &&&&\\
$(\ell,\nu)=(e,\nu_e)$       & 1.00009 & -0.3904 & -0.01861 & ~0.2901 & ~0.01461 \\
$\phantom{(\ell,\nu)=}(\mu,\nu_\mu)$   & 1.00009 & -0.3904 & -0.01861 & ~0.2901 & ~0.01461 \\
$\phantom{(\ell,\nu)=}(\tau,\nu_\tau)$ & 1.00009 & -0.3901 & -0.01858 & ~0.2896 & ~0.01457 \\
\hline
$g^L_{W^{(n)}UD}/(g_w/\sqrt{2})$  &&&&\\
$(U,D)=(u,d)$			& 1.00009 & -0.3904 & -0.01861 & ~0.2901 & ~0.01461 \\
$\phantom{(U,D)=}(c,s)$			& 1.00009 & -0.3904 & -0.01860 & ~0.2900 & ~0.01460 \\
$\phantom{(U,D)=}(t,b)$			& 0.9995 & ~1.7517 & ~0.04516 & -0.2925 & -0.01490 \\
\hline
\end{tabular}
\end{table}
It is seen that couplings to $W^{(0)}$ (the $W$ boson) are slightly larger than the SM value $g_w/\sqrt{2}$ for light quarks and leptons,
where those for top and bottom quarks is slightly smaller. 
Further the couplings of SM fermions except for $t$ and $b$ to $W^{(1)}$
are smaller than couplings to $W^{(0)}$ and their signs are opposite.
$W^{(1)}t\bar{b}$ coupling is larger than the SM value.

\begin{figure}[htbp]
\includegraphics[width=0.7\linewidth]{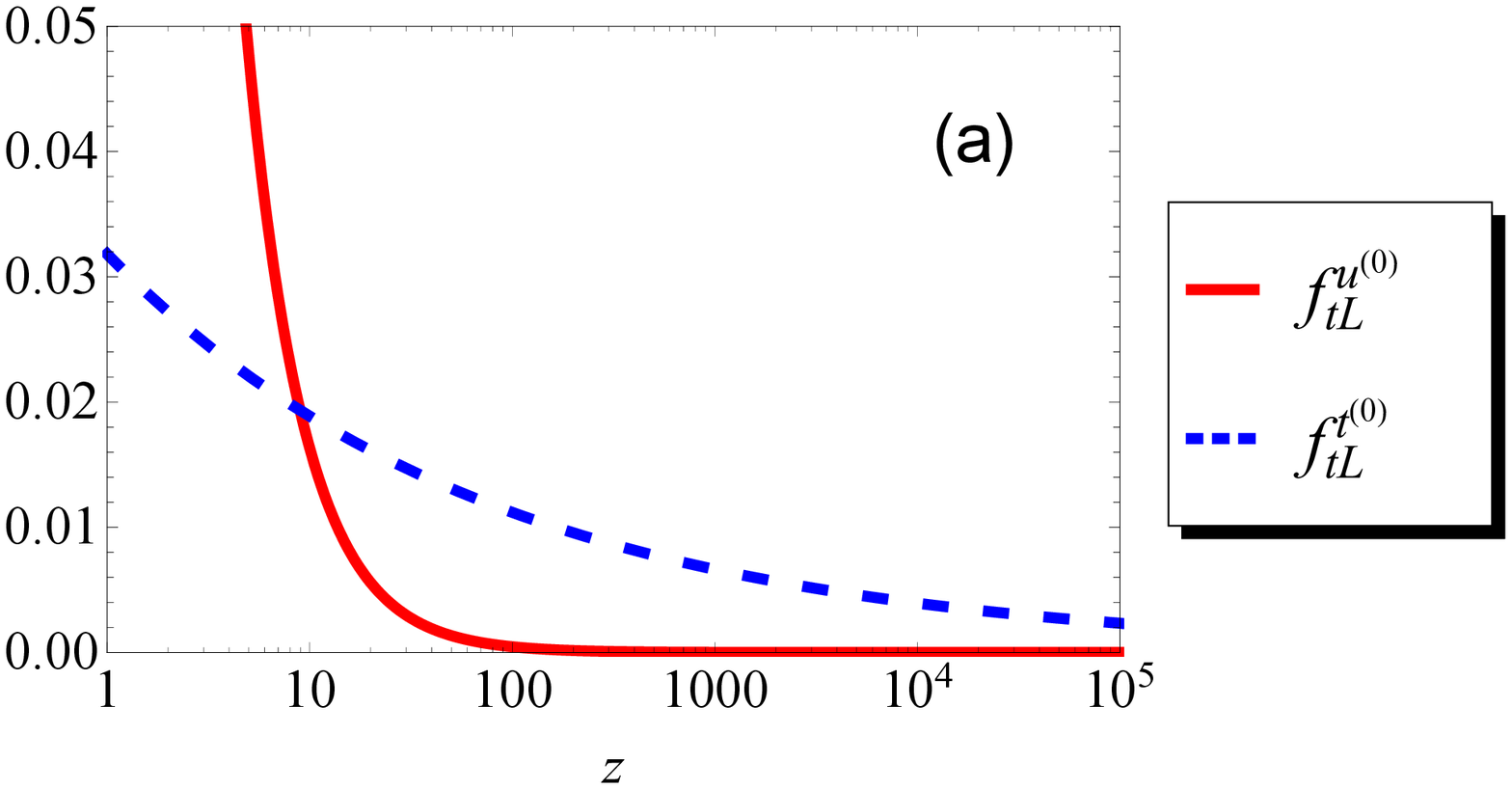}\\
\includegraphics[width=0.7\linewidth]{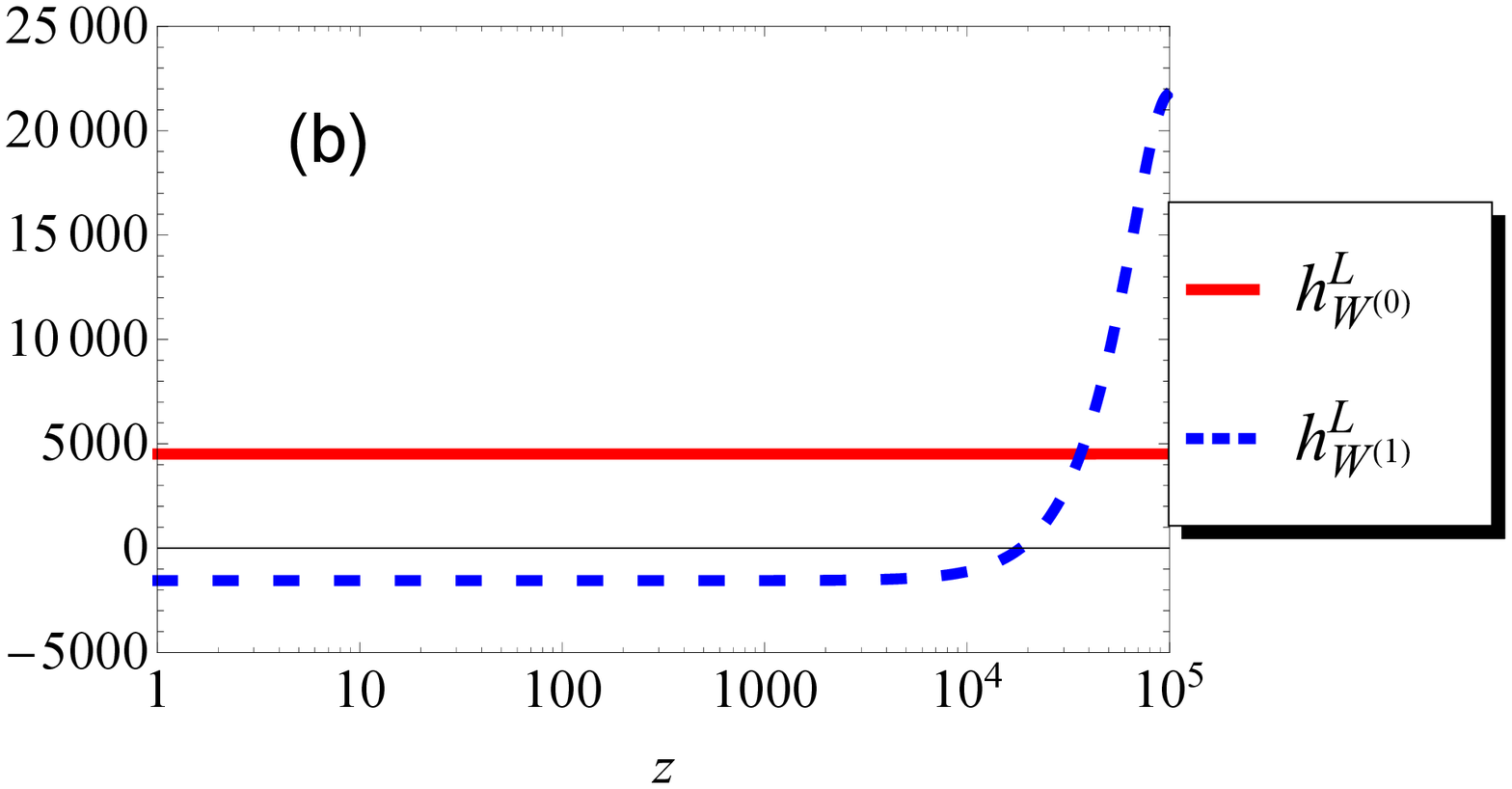}%
\caption{
Behavior of dominant component of the wave functions for fermions and $W^{(n)}$ bosons for $N_F=4$, $z_L=10^5$ ($\theta_H = 0.115$).
(a) Red-solid and blue-dashed lines are $f_{tL}^{t^{(0)}}$ for the top and $f_{tL}^{u^{(0)}}$ for the up quark, respectively.
(b) Red-solid and blue-dashed lines are $h^L_{W^{(0)}}$ for $W^{(0)}$ and $h^L_{W^{(1)}}$ for $W^{(1)}$ bosons, respectively.}\label{fig:Cplot}
\end{figure}

\begin{figure}[htbp]
\includegraphics[width=0.7\linewidth]{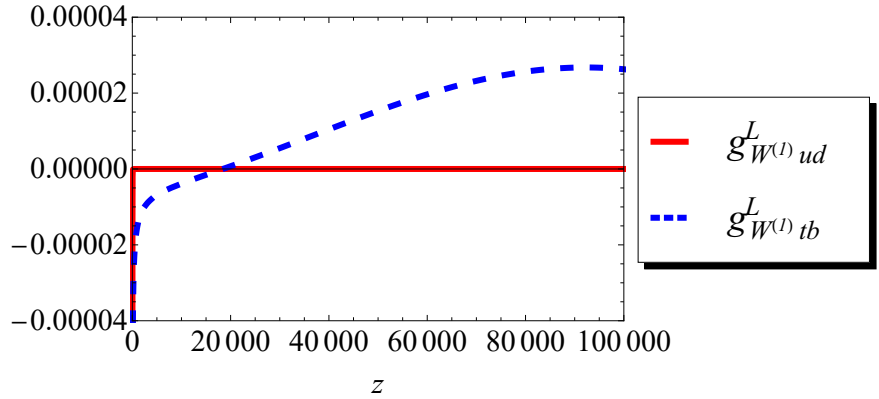}
\caption{
Integrand of the coupling $g_{W^{(1)}UD}^L$ in \eqref{eq:Wud-coupling}.
Red-solid and blue-dashed lines are for $(U,D)=(t,b)$ and $(u,d)$, respectively.}\label{fig:overlap}
\end{figure}

The difference between $W^{(1)}ud$ and $W^{(1)}tb$ couplings is understood as follows.
The couplings among left-handed up- and down-sector fermions $(U,D)=(u,d)$, $(t,b)$ and $W$ boson $g_{W^{(1)}UD}^L$
are given by the overlapping \eqref{eq:Wud-coupling}.
The integration in \eqref{eq:Wud-coupling} is dominated by the term $h_{W^{(n)}}^L f_{bL}^{D} f_{tL}^{U}$ ($(U,D)=(u^{(0)},d^{(0)}),\,(t^{(0)},b^{(0)})$).
In Fig.~\ref{fig:Cplot}, leading part of bulk wave functions for fermions and gauge bosons are plotted.
In Fig.~\ref{fig:Cplot}-(a), we see that $f_{tL}^{u^{(0)}}$ decreases much faster than $f_{tL}^{t^{(0)}}$.
In particular, near $z=z_L$, $f_{tL}^{t^{(0)}}$ has small but sizable value
whereas the value of $f_{tL}^{u^{(0)}}$ almost vanishes. 
In Fig.~\ref{fig:Cplot}-(b), $h^L_{W^{(0)}}$ is almost constant, 
whereas $h^L_{W^{(1)}}$ has negative values for small values of $z$ and has large positive value near $z=z_L$.
In Fig.~\ref{fig:overlap},
we plot overlapping of wave functions of up- and down-type fermions and $W^{(1)}$.
In the Figure, overlapping of wave functions of light quarks and $W^{(1)}$ has
large negative value only near the UV brane ($z=1$).
On the other hand, the overlapping of heavy quarks and $W^{(1)}$ takes negative value for small values of $z$ but becomes positive for larger $z$.
This difference in the integrand results in the differences in the signs and magnitudes of $g_{W^{(1)}tb}^L$, $g_{W^{(1)}ud}^L$ and $g_{W^{(0)}ud}$.


%\clearpage
In Table~\ref{tbl:higgs-couplings}, $HWW'$ ($W' = W^{(n)},W_R^{(n)}$)
and $HZZ'$ ($Z'= Z^{(n)},Z_R^{(n)}$) couplings are tabulated.
We note that $H\gamma^{(n)}Z^{(m)}$ and $H\gamma^{(n)}Z_R^{(m)}$ couplings vanish.
We find that
$g_{WWH}$ and $g_{ZZH}$ couplings are given by the SM values multiplied by $\cos\theta_H$.
$g_{W^{(1)}WH}$ and $g_{W_R^{(1)}WH}$ are a few times larger than $g_{WWH}$,
and similar relations holds among $g_{Z_R^{(1)}ZH}$, $g_{Z^{(1)}ZH}$ and $g_{ZZH}$.
All $g_{W'^{(n)}WH}$ and $g_{Z'(n)ZH}$ couplings become small as $n$ becomes larger.
\begin{table}[htbp]
\caption{Couplings of $W^{(n)}$, $Z^{(n)}$, $\gamma^{(n)}$ and $Z_R^{(n)}$
to $WH$, $ZH$ in the unit of GeV.}\label{tbl:higgs-couplings}
\begin{tabular}{c|c|c|c|c|c}
\multicolumn{6}{c}{$N_F=4$, $z_L=10^5$ ($\theta_H = 0.115$)} \\
\hline
& $n=0$ & $1$ & $2$ & $3$ & $4$\\ 
\hline
$g_{W^{(n)}WH}/(g_w\cos\theta_H)$
& 80.0 & 255 & 2.57 & 45.4 & 0.220 \\
$g_{Z^{(n)}ZH}/(g_w\cos\theta_H/\cos\theta_W)$ 
& 91.2 & 291 & 3.35 & 51.8 & 0.286 \\
$g_{W_R^{(n)}WH}/g_w $ 
& --- & 266 & 50.5 & 20.6 & 11.1 \\
$g_{Z_R^{(n)}ZH}/(g_w /\cos\theta_W)$ 
& --- & 223 & 42.3 & 17.2 & 9.27 \\
\hline
\end{tabular}\\
\vspace{5mm}
\begin{tabular}{c|c|c|c|c|c}
\multicolumn{6}{c}{$N_F=4$ $z_L=10^4$ ($\theta_H = 0.0737$)} \\
\hline
& $n=0$ & $1$ & $2$ & $3$ & $4$\\ 
\hline
$g_{W^{(n)}WH}/(g_w\cos\theta_H)$
& 80.0 & 225 & 1.89 & 39.2 & 0.169 \\
$g_{Z^{(n)}ZH}/(g_w\cos\theta_H/\cos\theta_W)$ 
& 91.2 & 257 & 2.46 & 44.8 & 0.220 \\
$g_{W_R^{(n)}WH}/g_w $ 
& --- & 238 & 45.1 & 18.4 & 9.89 \\
$g_{Z_R^{(n)}ZH}/(g_w /\cos\theta_W)$ 
& --- & 199 & 37.7 & 15.4 & 8.27 \\
\hline
\end{tabular}
\end{table}

%\clearpage
In Table~\ref{tbl:triple-gauge}, trilinear vector-boson couplings are tabulated.
The $g_{WWZ}$ coupling is very close to its SM value $g_w \cos\theta_W$.
The deviation is one part in $10^8$.
$g_{\gamma WW}$ is exactly $e$, which reflects the unbroken $U(1)_{\rm em}$ gauge symmetry. 
Couplings among the first KK and two SM vector bosons are suppressed by a factor of $\calO(10^{-4})$,  which is close to the square of the ratio of the weak boson mass to the 1st KK boson mass. 

\begin{table}[htbp]
\caption{
Trilinear vector-boson couplings of $W^{(n)}$, $Z^{(n)}$, $\gamma^{(n)}$ and $Z_R^{(n)}$ to $W^+W^-$, $WZ$}\label{tbl:triple-gauge}
\begin{tabular}{c|c|c|c|c|c}
\multicolumn{6}{c}{$N_F=4$, $z_L=10^5$ ($\theta_H = 0.115$)} \\
\hline
& $n=0$ & $1$ & $2$ & $3$ & $4$ \\
\hline
$g_{W^{(n)}WZ}/(g_w\cos\theta_W)$ 
& 0.9999998 & $-7.35 \times 10^{-4}$ & $-9.63 \times 10^{-7}$ & $-2.64 \times 10^{-5}$ & $-1.60 \times 10^{-7}$ \\ 
$g_{Z^{(n)}WW}/(g_w\cos\theta_W)$
& 0.9999998 & $-3.96 \times 10^{-4}$ &  $1.62 \times 10^{-8}$ & $-1.42 \times 10^{-5}$ & $-1.17 \times 10^{-7}$ \\
$g_{\gamma^{(n)}WW}/e$ 
& 1 & $-1.13\times 10^{-3}$ & $-4.04\times 10^{-5}$ & $-7.63\times 10^{-6}$ & $-2.07\times 10^{-6}$ \\
$g_{Z_R^{(n)}WW}/g_w$
& --- & $5.51\times 10^{-4}$ & $1.99\times 10^{-5}$ & $3.28\times 10^{-6}$ & $9.53\times 10^{-7}$ \\
$g_{W_R^{(n)}WZ}/g_w$
& --- & $7.51\times 10^{-4}$ & $2.71\times 10^{-5}$ & $4.47\times 10^{-6}$ & $1.30\times 10^{-6}$ \\
\hline
\end{tabular}
\\
\vspace{2em}
\begin{tabular}{c|c|c|c|c|c}
\multicolumn{6}{c}{$N_F=4$, $z_L=10^4$ ($\theta_H = 0.0737$)} \\
\hline
& $n=0$ & $1$ & $2$ & $3$ & $4$ \\
\hline
$g_{W^{(n)}WZ}/(g_w\cos\theta_W)$ 
& 0.99999995 & $-3.32 \times 10^{-4}$ & $-3.88 \times 10^{-7}$ & $-1.15 \times 10^{-5}$ & $-6.04 \times 10^{-8}$ \\ 
$g_{Z^{(n)}WW}/(g_w\cos\theta_W)$
& 0.99999995 & $-1.73 \times 10^{-4}$ & $-1.46 \times 10^{-8}$ & $-6.18 \times 10^{-6}$ & $-4.43 \times 10^{-8}$ \\
$g_{\gamma^{(n)}WW}/e$ 
& 1 & $-4.95 \times 10^{-4}$ & $-1.76\times 10^{-5}$ & $-3.46\times 10^{-6}$ & $-9.22\times 10^{-7}$ \\
$g_{Z_R^{(n)}WW}/g_w$
& --- & $2.53 \times 10^{-4}$ & $9.10\times 10^{-6}$ & $1.51\times 10^{-6}$ & $4.37\times 10^{-7}$ \\
$g_{W_R^{(n)}WZ}/g_w$
& --- & $3.45 \times 10^{-4}$ & $1.24\times 10^{-5}$ & $2.05\times 10^{-6}$ & $5.96\times 10^{-7}$ \\
\hline
\end{tabular}
\end{table}

%%%%%%%%%%%%%%%%%%%%%%%%%%%%%%%%%%%%%%%%%%%%%%%%%%%%%%%%%%%%%%%%%%%%%%%%%%
\subsection{Decay width}

In Tables~\ref{tbl:width-W} and \ref{tbl:width-WR}, decay widths of $W$ and $W_R$
boson are tabulated, respectively.
Since the $W^{(1)}$ boson couples equally to the light SM fermions except for $b$ and $t$ quarks, partial decay widths to light SM fermions are almost identical besides the QCD color factor.
The $W^{(1)}$ coupling to $t$ and $b$ quarks is larger than the couplings to other fermion pairs. The $W^{(1)}$ decay to $tb$ dominates over decay to other fermion pairs.
Partial decay widths of $W^{(1)}$ to $WZ$ and $WH$ are almost identical:
\begin{eqnarray}
\Gamma(W^{(1)}\to WH) &\simeq& \Gamma(W^{(1)}\to WZ),
\quad \therefore \eta_{W^{(1)}} \simeq 1,
\nonumber
\\
\Gamma(W_R^{(1)} \to WH) &\simeq& \Gamma(W_R^{(1)} \to WZ),
\quad \therefore \eta_{W_R^{(1)}} \simeq 1.
\label{eq:W-decay_relation}
\end{eqnarray}
We also find
\begin{eqnarray}
\Gamma(W^{(1)} \to WH) \sim \Gamma(W_R^{(1)} \to WH),
\quad
\Gamma(W^{(1)} \to WZ) \sim \Gamma(W_R^{(1)} \to WZ).
\label{eq:W-WR-decay_relation}
\end{eqnarray}
Since the $W_R^{(n)}$ boson does not couple to SM fermions,
and $W_R^{(1)}$ decay only to the SM bosons.

%%%%%%%%%%%%%%%%%%%%%%%%%%%%%%%%%%%%%%%%%%%%%%%%%%%%%%%%%%%%
In Tables~\ref{tbl:width-Z}, \ref{tbl:width-gamma} and \ref{tbl:width-ZR},
decay widths of $Z^{(1)}$, $\gamma^{(1)}$ and $Z_R^{(1)}$ are tabulated, respectively.
Compared with $W^{(1)}$ and $W_R^{(1)}$,
$\gamma^{(1)}$ and $Z_R^{(1)}$ have large total widths
\begin{eqnarray}
\Gamma_{\gamma^{(1)}}/M_{\gamma^{(1)}} &=& 
\begin{cases}
0.151 & \text{ for $N_F=4$, $z_L = 10^5$ ($\theta_H = 0.115$)}
\\
0.125 & \text{ for $N_F=4$, $z_L = 10^4$ ($\theta_H = 0.0737$)}
\end{cases}
\\
\Gamma_{Z_R^{(1)}}/M_{Z_R^{(1)}} &=& 
\begin{cases} 
0.129 & \text{ for $N_F=4$, $z_L = 10^5$ ($\theta_H = 0.115$)}
\\
0.133 & \text{ for $N_F=4$, $z_L = 10^4$ ($\theta_H = 0.0737$)}.
\end{cases}
\end{eqnarray}
From the tables, one finds that
\begin{eqnarray}
%\sum_{Z'=Z^{(1)},\gamma^{(1)}}
\Gamma(Z^{(1)}\to HZ) &\simeq& 
\sum_{Z'=Z^{(1)},\gamma^{(1)}} \Gamma(Z'\to W^+W^-),
\label{eq:Z-decay_relation}
\\
\Gamma(Z_R^{(1)}\to HZ) &\simeq& \Gamma(Z_R^{(1)}\to W^+W^-).
\label{eq:ZR-decay_relation}
\end{eqnarray}
$\gamma^{(1)}$ and $Z^{(1)}$ are almost degenerate.
The relation \eqref{eq:Z-decay_relation} follows from
the relation among Higgs-vector boson and trilinear vector boson couplings
\begin{eqnarray}
\frac{g_{\gamma^{(1)}ZH}^2 + g_{Z^{(1)} ZH}^2}{M_Z^2}
&\simeq& (g_{\gamma^{(1)}WW}^2 + g_{Z^{(1)} WW}^2 )\frac{M_{Z'}^4}{M_Z^4},
\end{eqnarray}
where $M_{\gamma^{(1)}}\simeq M_{Z^{(1)}} \equiv M_{Z'}$ and $ g_{\gamma^{(1)}ZH} = 0$.

%%%%%%%%%%%%%%%%%%%%%%%%%%%%%%
\begin{table}[htbp]
\caption{Partial and total decay width of $W^{-(1)}$
in the unit of GeV.}\label{tbl:width-W}
\begin{tabular}{c|ccccccccc}
\multicolumn{10}{c}{$N_F=4$, $z_L=10^5$ ($\theta_H = 0.115$)}\\
\hline
mode & $e^-\bar{\nu}_e$ & $\mu^-\bar{\nu}_\mu $ & $\tau^-\bar{\nu}_\tau$ & $d\bar{u}$ & $s\bar{c}$ & $b\bar{t}$
& $W^-Z$ & $W^-H$ & total \\
$\Gamma$ & 2.00 & 2.00 & 1.99 & 5.99 & 5.98 & 84.7 & 42.7 & 42.1  & 187\\
\hline
\end{tabular}
\vspace{5mm}
\\
\begin{tabular}{c|ccccccccc}
\multicolumn{10}{c}{$N_F=4$, $z_L=10^4$ ($\theta_H = 0.0737$)}\\
\hline
mode & $e^-\bar{\nu}_e$ & $\mu^-\bar{\nu}_\mu $ & $\tau^-\bar{\nu}_\tau$ & $d\bar{u}$ & $s\bar{c}$ & $b\bar{t}$  & $W^-Z$ & $W^-H$ & total
\\
$\Gamma$ & 3.63 & 3.63 & 3.62 & 10.88 & 10.88 & 219  & 46.9 & 47.2  & 346 \\
\hline
\end{tabular}
\end{table}

\begin{table}[htbp]
\caption{
Partial and total decay widths of $W_R^{-(1)}$ in the unit of GeV. 
}\label{tbl:width-WR}
\begin{tabular}{cccc}
\multicolumn{4}{c}{$N_F=4$, $z_L=10^5$ ($\theta_H =0.115$)}\\
\hline
SM fermions & $W^-Z$ & $W^-H$ & total \\
 0 & 43.4 & 43.4 & 86.8 \\
\hline
\end{tabular}
\\
\vspace{1em}
\begin{tabular}{cccc}
\multicolumn{4}{c}{$N_F=4$, $z_L=10^4$ ($\theta_H =0.0737$)}\\
\hline
SM fermions & $W^-Z$ & $W^-H$ & total \\
 0 & 48.7 & 48.8 & 97.4 \\
\hline
\end{tabular}
\end{table}

%%%%%%%%%%%%%%%%%%%%%%%%%%%%%%%%%%%%%
\begin{table}[htbp]
\caption{
Partial and total decay widths of $Z^{(1)}$ in the unit of GeV.
}\label{tbl:width-Z}
\begin{tabular}{ccc|ccc|ccc|ccc|cc|c}
\multicolumn{15}{c}{$N_F=4$, $z_L=10^5$ ($\theta_H=0.115$)}\\
\hline
$e^+e^-$ & $\mu^+\mu^-$ & $\tau^+\tau^-$ 
& $\nu_e \bar{\nu}_e$ & $\nu_\mu \bar{\nu}_\mu$ & $\nu_\tau\bar{\nu}_\tau$ 
& $u\bar{u}$ & $c\bar{c}$ & $t\bar{t}$
& $d\bar{d}$ & $s\bar{s}$ & $b\bar{b}$ 
& $W^+ W^-$ & $ZH$ & total 
\\
 40.4 & 35.7 & 32.1 & 1.30 & 1.30 & 1.30 & 53.3 & 46.1 & 48.5 & 15.7 & 13.8 & 45.7 &  16.1 & 54.8 & 406 \\
\hline
\end{tabular}\\
\vspace{1em}
$N_F=4$, $z_L=10^4$ ($\theta_H =0.0737$)\\
\begin{tabular}{ccc|ccc|ccc|ccc|cc|c}
\hline
$e^+e^-$ & $\mu^+\mu^-$ & $\tau^+\tau^-$ 
& $\nu_e \bar{\nu}_e$ & $\nu_\mu \bar{\nu}_\mu$ & $\nu_\tau\bar{\nu}_\tau$
& $u\bar{u}$ & $c\bar{c}$ & $t\bar{t}$ 
& $d\bar{d}$ & $s\bar{s}$ & $b\bar{b}$
& $W^+ W^-$ & $ZH$ & total 
\\
 48.1 & 42.5 & 38.0 & 2.36 & 2.36 & 2.36 & 64.4 & 55.5 & 84.4 & 20.3 & 18.1 & 106.8 & 17.8 & 61.1 & 564  \\
\hline
\end{tabular}
\end{table}
\begin{table}[htbp]
\caption{
Partial and total decay width of $\gamma^{(1)}$ in the unit of GeV.
}\label{tbl:width-gamma}
\begin{tabular}{ccc|ccc|ccc|ccc|cc|c}
\multicolumn{15}{c}{$N_F=4$, $z_L=10^5$ ($\theta_H=0.115$)}\\
\hline
$e^+e^-$ & $\mu^+\mu^-$ & $\tau^+\tau^-$ 
& $\nu_e \bar{\nu}_e$ & $\nu_\mu \bar{\nu}_\mu$ & $\nu_\tau\bar{\nu}_\tau$
& $u\bar{u}$ & $c\bar{c}$ & $t\bar{t}$ 
& $d\bar{d}$ & $s\bar{s}$ & $b\bar{b}$ 
& $W^+ W^-$ & $ZH$ & total
\\
 133.0 & 117.4 & 105.7 & 0 & 0 & 0 & 171.0 & 147.2 & 93.0 & 42.8 & 36.8 & 23.3 & 39.3 & 0 & 909 \\
\hline
\end{tabular}
\vspace{5mm}
\begin{tabular}{ccc|ccc|ccc|ccc|cc|c}
\multicolumn{15}{c}{$N_F=4$, $z_L=10^4$ ($\theta_H = 0.0737$)}\\
\hline
$e^+e^-$ & $\mu^+\mu^-$ & $\tau^+\tau^-$ 
& $\nu_e \bar{\nu}_e$ & $\nu_\mu \bar{\nu}_\mu$ & $\nu_\tau\bar{\nu}_\tau$
& $u\bar{u}$ & $c\bar{c}$ & $t\bar{t}$ 
& $d\bar{d}$ & $s\bar{s}$ & $b\bar{b}$
& $W^+ W^-$ & $ZH$ & total
\\
 158.9 & 140.2 & 125.2 &  0 & 0 & 0 & 204.6 & 175.0 & 101.0 & 51.1 & 43.8 & 25.3 & 43.6 & 0 & 1068 \\
\hline
\end{tabular}
\end{table}
\begin{table}[htbp]
\caption{
Partial and total decay widths of $Z_R^{(1)}$ in the unit of GeV.
}\label{tbl:width-ZR}
\begin{tabular}{cccccc}
\multicolumn{6}{c}{$N_F=4$, $z_L=10^5$ ($\theta_H = 0.115$)}\\
\hline
$e^+e^-$ & $\mu^+\mu^-$ & $\tau^+\tau^-$ 
& $\nu_e \bar{\nu}_e$ & $\nu_\mu \bar{\nu}_\mu$ & $\nu_\tau\bar{\nu}_\tau$ 
\\
 71.3  & 63.8 & 58.0 
& ${\cal O}(10^{-16})$ % $4.67\times 10^{-16}$ 
& ${\cal O}(10^{-11})$ % $3.97\times 10^{-11}$ 
& ${\cal O}(10^{-7})$ % $9.63\times 10^{-7}$ 
\\
\hline
 $u\bar{u}$ & $c\bar{c}$ & $t\bar{t}$ 
&$d\bar{d}$ & $s\bar{s}$ & $b\bar{b}$ 
\\  
 92.0 & 80.4 & 146.2 & 23.0 & 20.1 & 111.9 
\\
\hline
$W^+ W^-$ & $ZH$ &&&& total \\
 30.5 & 30.7 &&&& 729 \\
\hline
\end{tabular}\\
\vspace{5mm}
\begin{tabular}{cccccc}
\multicolumn{6}{c}{$N_F=4$, $z_L=10^4$ ($\theta_H = 0.0737$)}\\
\hline
$e^+e^-$ & $\mu^+\mu^-$ & $\tau^+\tau^-$ 
& $\nu_e \bar{\nu}_e$ & $\nu_\mu \bar{\nu}_\mu$ & $\nu_\tau\bar{\nu}_\tau$
\\
 83.9  & 75.2 & 68.1 
& ${\cal O}(10^{-13})$ % $9.20\times 10^{-13}$ 
& ${\cal O}(10^{-10})$ % $3.96\times 10^{-10}$ 
& ${\cal O}(10^{-6})$ % $2.52\times 10^{-6}$ 
\\
\hline
$u\bar{u}$ & $c\bar{c}$ & $t\bar{t}$ 
& $d\bar{d}$ & $s\bar{s}$ & $b\bar{b}$
\\  
 108.5 & 94.6 & 268.4 & 27.1 & 23.7 & 240.3 
\\
\hline
$W^+ W^-$ & $ZH$ &&&& total \\
 34.1 & 34.3 &&&& 1058 \\
\hline
\end{tabular}
\end{table}

%%%%%%%%%%%%%%%%%%%%%%%%%%%%%%%%%%%%%%%%%%%%%%%%%%%%%%%%%%%%%%%%%%%%%%%
\section{Cross section}\label{sec:cross_section}

In this section we evaluate cross sections in $pp$-collisions for various final states.
In the numerical evaluation we use CTEQ5 parton distribution functions \cite{Lai:1999wy}.

\subsection{$W' \to tb$, $\mu\nu$ and $Z' \to \ell^+\ell^-$}
\begin{figure}[htbp]
\includegraphics{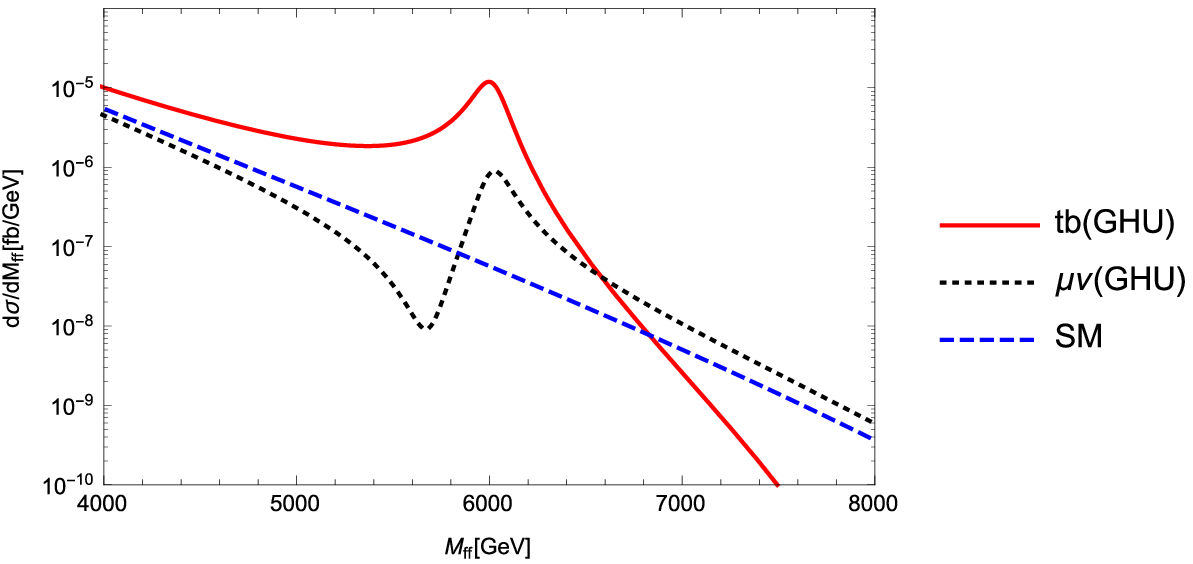}
\caption{$pp(d\bar{u}) \to \{ W^{-},\, W^{(1)-} \} \to b\bar{t},\mu^-\bar{\nu}_\mu$ differential cross sections $d\sigma/d M_{\rm ff}$ at $\sqrt{s_{pp}} = 14$ TeV for $N_F=4$, $z_L=10^5$ ($\theta_H = 0.115$).
$M_{\rm ff}$ is the invariant mass of $(b,\bar{t})$ or $(\mu,\bar{\nu})$. 
Red-solid, black-dotted lines are for the $b\bar{t}$ and $\mu^{-}\bar{\nu}$ final states in GHU.
Blue-dashed line is the cross section in the SM.
Cross sections for the processes $pp \to W^{(1)-} \to f\bar{f}'$, $(f,f') = (e^-,\nu_e), (\tau^-, \nu_\tau)$ are almost identical with that of $\mu^-\bar{\nu}$, 
whereas cross section for $(f,f')= (d,u),(s,c)$ final states is three times as large as that of $\mu^-\bar{\nu}$ due to the color factor.}\label{fig:Wp-fermions}
\end{figure}

In Figure~\ref{fig:Wp-fermions}, the differential cross sections of processes 
$pp \to \{ W^-,\, W^{(1)-} \}\to b\bar{t},\mu^-\bar{\nu}$ are plotted.
For light fermion doublet paris $\ell\bar{\nu},d\bar{u},s\bar{c}$ in the final state,
due to the flipped-signs of the couplings to $W^{(1)}$, a clear deficit of cross section 
just below the resonance $M_{\mu\nu} \sim M_{W^{(1)}}$ is observed.
For processes with $b\bar{t}$ final state, a deficit of cross section is observed above the resonance
$M_{tb} > M_{W^{(1)}}$,
since the $W^{(1)}\bar{t}b$ coupling has opposite sign relative to the $W^{(1)}\bar{u}d$ coupling.

\begin{figure}[htbp]
\includegraphics[width=0.8\linewidth]{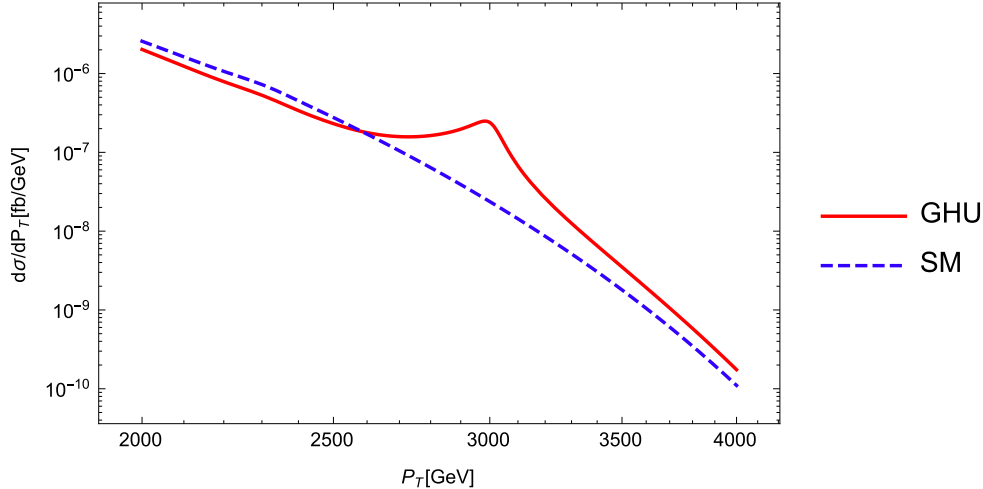}
\caption{
$d\sigma(pp\to e^-\bar{\nu}_e)/dp_T$ as a function of $p_T$ at $\sqrt{s_{pp}}=14\,\text{TeV}$
for $N_F=4$, $z_L=10^5$ ($\theta_H = 0.115$).
Red-solid and blue-dashed lines are for the GHU model and the SM, respectively.
}\label{fig:pT}
\end{figure}

When the final state contains a neutrino, 
the transverse momentum distribution $d\sigma/dp_T$ with respect to the transverse momentum of charged lepton, $p_T$, gives information on the mass of $W'$.
The transverse-momentum distribution at parton-level is given in Appendix~\ref{sec:formula-scat}. 
The $p_T$ distribution in $pp$ collision is given by
\begin{eqnarray}
\frac{d\sigma(pp\to e^-\bar{\nu}+X)}{dp_T}(p_T)
&=&
 \int_0^{1} d\tau \biggl\{
 \frac{dp_T(d\bar{u}\to e^-\bar{\nu})}{dp_T} (s_{pp}\tau,p_T) \cdot
 \frac{dL_{d\bar{u}}}{d\tau}(\tau) \biggl\},
\end{eqnarray}
where the parton luminosity $dL_{d\bar{u}}/d\tau$ is given
in terms of parton distribution functions $f_{q}(x_1,Q)$ by
\begin{eqnarray}
\frac{dL_{d\bar{u}}}{d\tau}(\tau) &=& 
\int_0^1 dx_1 \int_0^1 dx_2 [
f_d(x_1,Q) f_{\bar{u}}(x_2,Q) + f_d(x_2,Q) f_{\bar{u}}(x_1,Q)] \delta(\tau-x_1x_2),
\nonumber\\
Q &=& s_{pp}\tau.
\end{eqnarray}
In Figure~\ref{fig:pT} the $p_T$-distribution $d\sigma(pp\to e^-\bar{\nu})/d p_T$ is plotted.
In the figure, Jacobian peak at $p_T = M_{M^{(1)}}/2 \simeq 3\,\text{TeV}$ is observed.

\begin{figure}[htbp]
\includegraphics{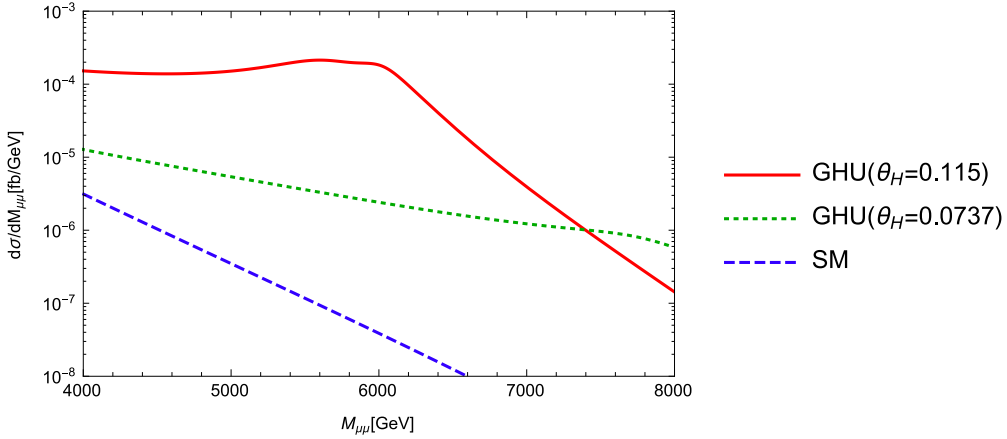}
\caption{Differential cross section $d\sigma/dM_{\mu\mu}$ of the process $pp(u\bar{u},d\bar{d}) \to \{ \gamma,\, Z,\, Z^{(1)},\, \gamma^{(1)},\, Z_R^{(1)} \} \to \mu^+\mu^- $ at $\sqrt{s} = 14$ TeV. 
$M_{\mu\mu}$ is the invariant mass of $\mu^+ \mu^-$.
Solid [red], dotted [green] and dashed [blue] lines are the $Z'$ resonance in the GHU for $N_F=4$, $z_L = 10^5$ ($\theta_H = 0.115$), 
$N_F=4$, $z_L=10^4$ ($\theta_H = 0.0737$) and the SM, respectively.
The cross section for the $e^+e^-$ final state is identical to the $\mu^+\mu^-$ final state.
}\label{fig:Zp-fermions}
\end{figure}

%%%%%%%%%%%%%%%%%%%%%%%%%%%%%%%%%
%
% Zprime to ff
For the process $pp(u\bar{u},d\bar{d}) \to Z' \to \ell^+ \ell^-$,
we  show the plot of the differential cross section $d\sigma/d M_{\mu\mu}$ in Figure~\ref{fig:Zp-fermions}.
In  the plot the updated decay widths of $Z'$ ($Z' = \gamma^{(1)}$, $Z^{(1)}$ and $Z_R^{(1)}$) has been used,
which takes bosonic final states ($W^+ W^-$ and $ZH$) into account,
and is  $\calO(10\%)$ bigger than that used in the previous paper \cite{Funatsu:2014fda}.
We stress that the rate of $Z'$ production is rather large,
and it is promising to see the $Z'$ events at the current LHC Run 2.
Since in this model $Z'$ bosons have large widths, at the early stage of
LHC experiment, sporadic events of high-energy $\mu^+\mu^-$ final states will be seen.
For $\theta_H = 0.115$ ($M_{Z^{(1)},\gamma^{(1)}}\sim6.0\,\text{TeV}$, $M_{Z_R^{(1)}}\sim 5.7\,\text{TeV}$), with the 30 fb$^{-1}$ and $\sqrt{s_{pp}} = 13\,\text{TeV}$ data, expected numbers of events in GHU $N_{GHU}$ and SM signal $N_{SM}$, and significance $S$ are
$N_{GHU}/N_{SM}(S) = 8.4/3.6\,(2.2)$, $3.9/0.26\,(3.7)$, $2.6/0.02\,(4.4)$, $2.9/0.004\,(6.0)$,
$0.65/0.0002\,(3.0)$ and $0.01/1\times10^{-5}\,(0.34)$ for bins (GeV) $[2000,3000]$, $[3000,4000]$, $[4000,5000]$, $[5000,6000]$, $[6000,7000]$ and $[7000,8000]$, respectively.
In this case, an excess of high-energy ($M_{\mu\mu} \gtrsim 3000\,\text{GeV}$) events is expected.
For smaller $\theta_H$ (heavier $M_{Z'}$), the signals becomes smaller and more data is required for confirming/rejecting the model.
For $\theta_H = 0.0737$ ($M_{Z^{(1)},\gamma^{(1)}} \sim 8.5\,\text{TeV}$, $M_{Z_R^{(1)}}\sim 7.9\,\text{TeV}$), with the 1000 fb$^{-1}$ and $\sqrt{s_{pp}}=14\,\text{TeV}$ data, $N_{GHU}/N_{SM}(S) = 140/155\,(1.21)$, $23/12\,(2.7)$, $8.5/1.3\,\,(4.2)$, $3.7/0.14\,(4.12)$, $0.7/0.006\,(2.3)$ and $0.3/0.0004\,(1.7)$
 for bins (GeV) $[2000,3000]$, $[3000,4000]$, $[4000,5000]$,
 $[5000,6000]$, $[6000,7000]$ and $[7000,8000]$, respectively.

%%%%%%%%%%%%%%%%%%%%%%%%%%%%%%%%%%%%%%%%%%%%%%%%%%%%%%%%%%%%%%%%%%%%%%%%%%%%%
\subsection{$W' \to WH$ and $Z' \to ZH$}
\begin{figure}[htbp]
\includegraphics{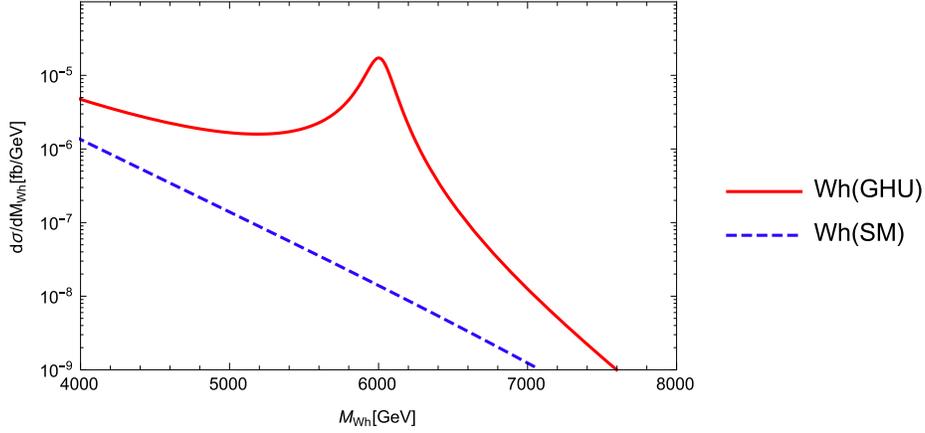}
\caption{Differential cross section $d\sigma/M_{Wh}$ of the process $pp(d\bar{u}) \to \{ W^{-},\, W^{(1)-} \} \to W^- H$  at $\sqrt{s_{pp}} = 14$ TeV for $N_F=4$, $z_L=10^5$ ($\theta_H = 0.115$).
$M_{Wh}$ is the invariant mass of $WH$.
Red-solid and blue-dashed lines show cross sections in GHU and in the SM, respectively.
}\label{fig:Wp-WH}
\end{figure}

\begin{figure}[htbp]
\includegraphics{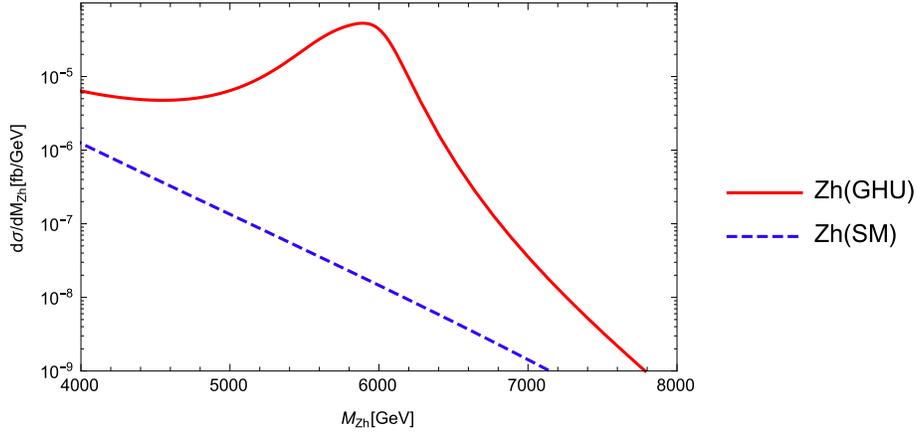}
\caption{Differential cross section $d\sigma/dM_{Zh}$ of the process $pp(u\bar{u},d\bar{d}) \to \{ Z,\, Z^{(1)},\, Z_R^{(1)} \} \to ZH $ cross section at $\sqrt{s_{pp}} = 14$ TeV for $N_F=4$, $z_L=10^5$ ($\theta_H = 0.115$).
$M_{Zh}$ is the invariant mass of $ZH$.
Red-solid and blue-dashed lines show cross sections in GHU and in the SM, respectively.
}\label{fig:Zp-ZH}
\end{figure}

In Figures~\ref{fig:Wp-WH} and \ref{fig:Zp-ZH}, differential cross sections of processes
$pp \to \{ W^-,\, W^{(1)-} \} \to W^-H$ and
$pp \to \{ Z,Z^{(1)},Z_R^{(1)} \} \to ZH$ are plotted, respectively.
Compared with the $WH$ mode, cross section for $ZH$ mode is bigger
and the width is wider.
It is due to the fact that $Z^{(1)}$ and $Z_R^{(1)}$ have large couplings to the right-handed quarks,
and their widths are large.

% WH
% ZH

%%%%%%%%%%%%%%%%%%%%%%%%%%%%%%%%%%%%%%%%%%%%%%%%%%%%%%%%%%%%%%%%%%%%%%%%%%%%%%
\subsection{$W' \to WZ$ and $Z' \to WW$}
\begin{figure}[htbp]
\includegraphics{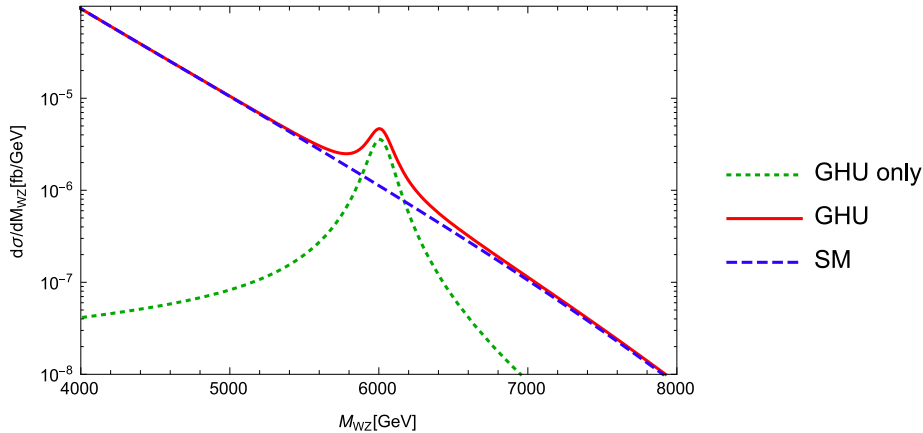}
\caption{Differential cross section $d\sigma/dM_{WZ}$ of the process 
$pp(d\bar{u}) \to \{ W^{-}, W^{(1)-} \} \to W^-Z$ at $\sqrt{s_{pp}} = 14$ TeV for $N_F=4$, $z_L=10^5$ ($\theta_H =0.115$).
$M_{\rm WZ}$ is the invariant mass of $WZ$.
Green-dotted line shows the $s$-channel $W'$ signals in GHU model.
Blue-dashed line shows the SM prediction including $s$-, $t$- and $u$-channels \cite{Brown:1979ux}.
Red-solid line is the sum of SM and GHU signals.}\label{fig:Wp-WZ}
\end{figure}

\begin{figure}[htbp]
\includegraphics{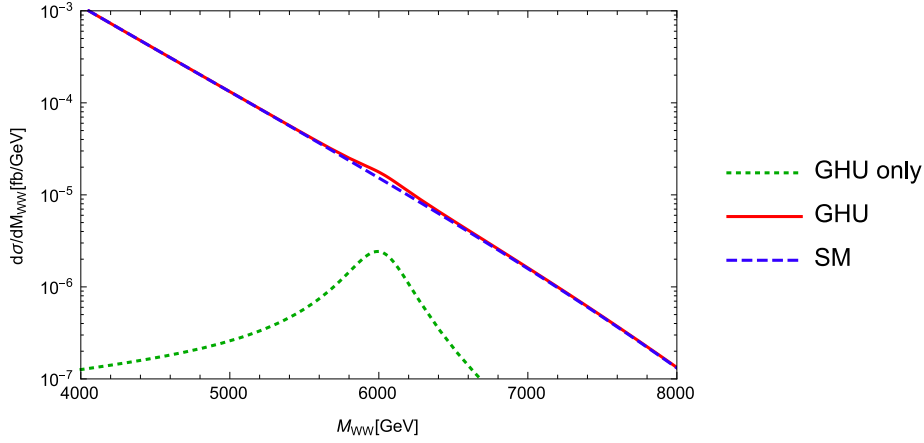}
\caption{%
Differential cross section $d\sigma/dM_{WW}$
of the process $pp(u\bar{u},d\bar{d}) \to \{ \gamma,\, Z,\, Z^{(1)},\,\gamma^{(1)},\, Z_R^{(1)} \}\to W^+W^-$ at $\sqrt{s_{pp}} = 14$ TeV for $N_F=4$, $z_L=10^5$ ($\theta_H = 0.115$).
$M_{WW}$ is the invariant mass of $W^+W^-$.
Green-dotted line shows the s-channel $W'$ signals in GHU model.
Blue-dashed line shows the SM prediction including $s$-, $t$- and $u$-channels \cite{Brown:1978mq}.
Red-solid line is the sum of SM and GHU signals.
}\label{fig:Zp-WW}
\end{figure}

% WZ and WW
In Figures~\ref{fig:Wp-WZ} and \ref{fig:Zp-WW}, differential cross sections of the processes 
$pp \to \{ \gamma,\, Z,\, Z^{(1)},\, \gamma^{(1)},\, Z_R^{(1)} \} \to W^+W^-$ and 
$pp \to \{ W^-,\, W^{(1)-} \} \to W^- Z$ are plotted, respectively.
For the $WZ$ final states, the signal of the resonance of $W'$ is a few times larger 
than that of the SM.
For the $WW$ final states, the contribution from $Z'$ resonances is much smaller than the SM cross section so that the signal is hard to see.

We comment that there are no $s$-channel processes with $ZZ$ final states mediated by vector bosons.
The process mediated by KK gravitons \cite{Davoudiasl:1999jd,Davoudiasl:2000wi}
can be ignored, as the couplings of KK gravitons to the SM fields are suppressed by $k/M_{Pl} \ll 1$,
where $M_{Pl}$ is the Planck mass. 

% WZ
% WW 

%%%%%%%%%%%%%%%%%%%%%%%%%%%%%%%%%%%%%%%%%%%%%%%%%%%%%%%%%%%%%%%%
\subsection{Unitarity in $f\bar{f}' \to WZ$}

\begin{figure}[htbp]
\centerline{%
\includegraphics[width=5cm]{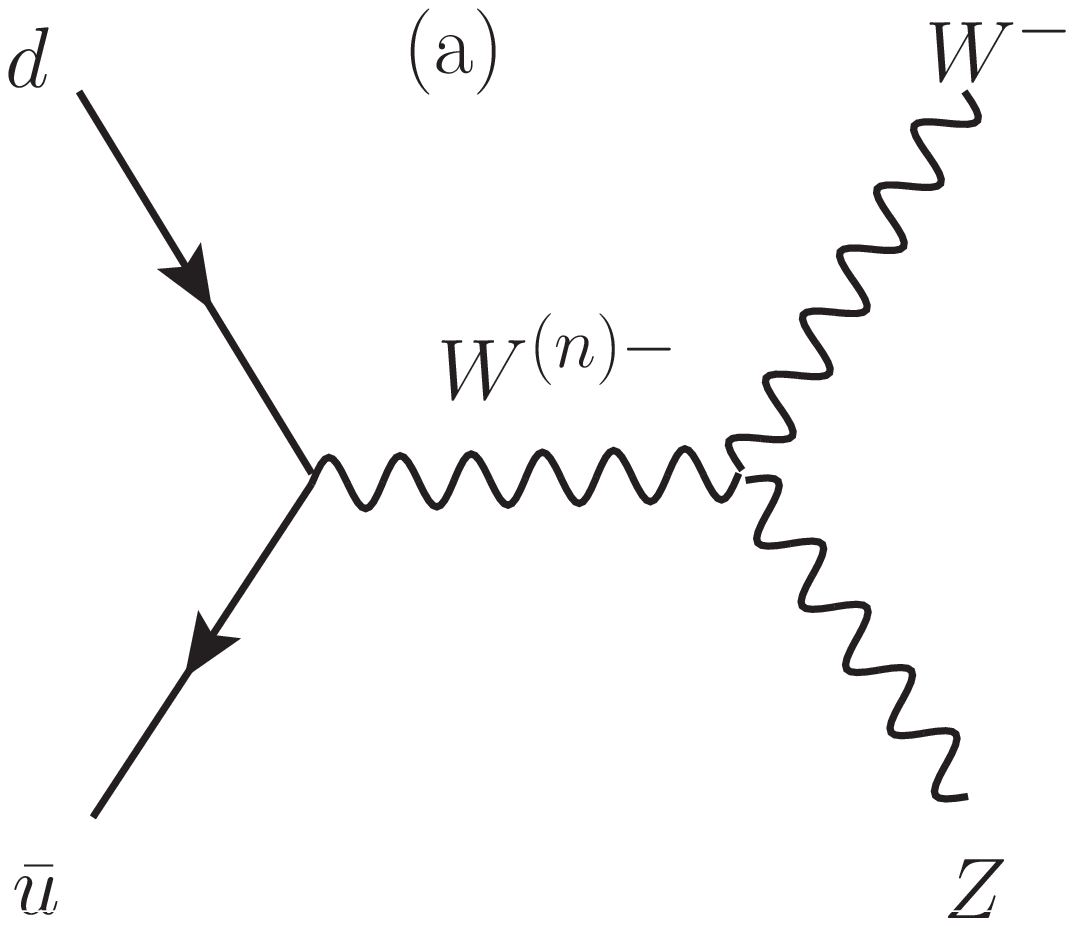}%
\includegraphics[width=4cm]{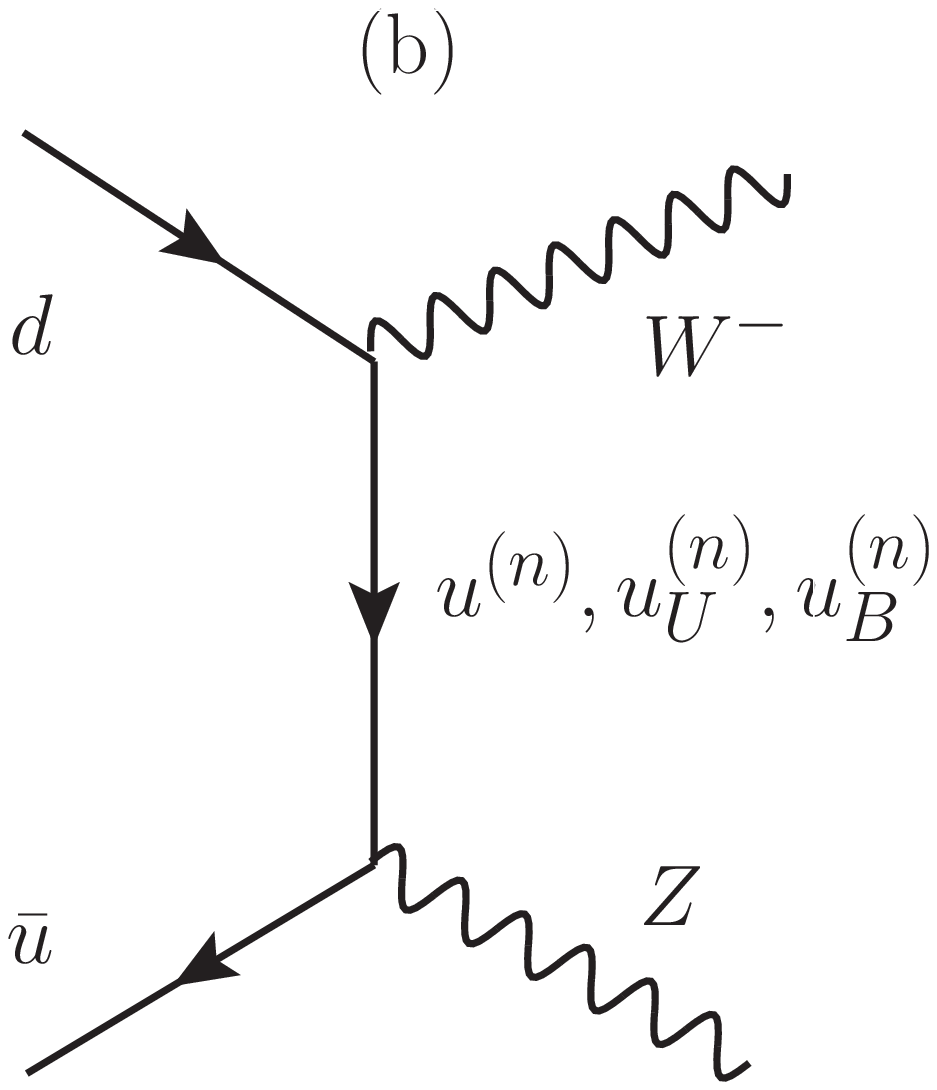}%
\includegraphics[width=5cm]{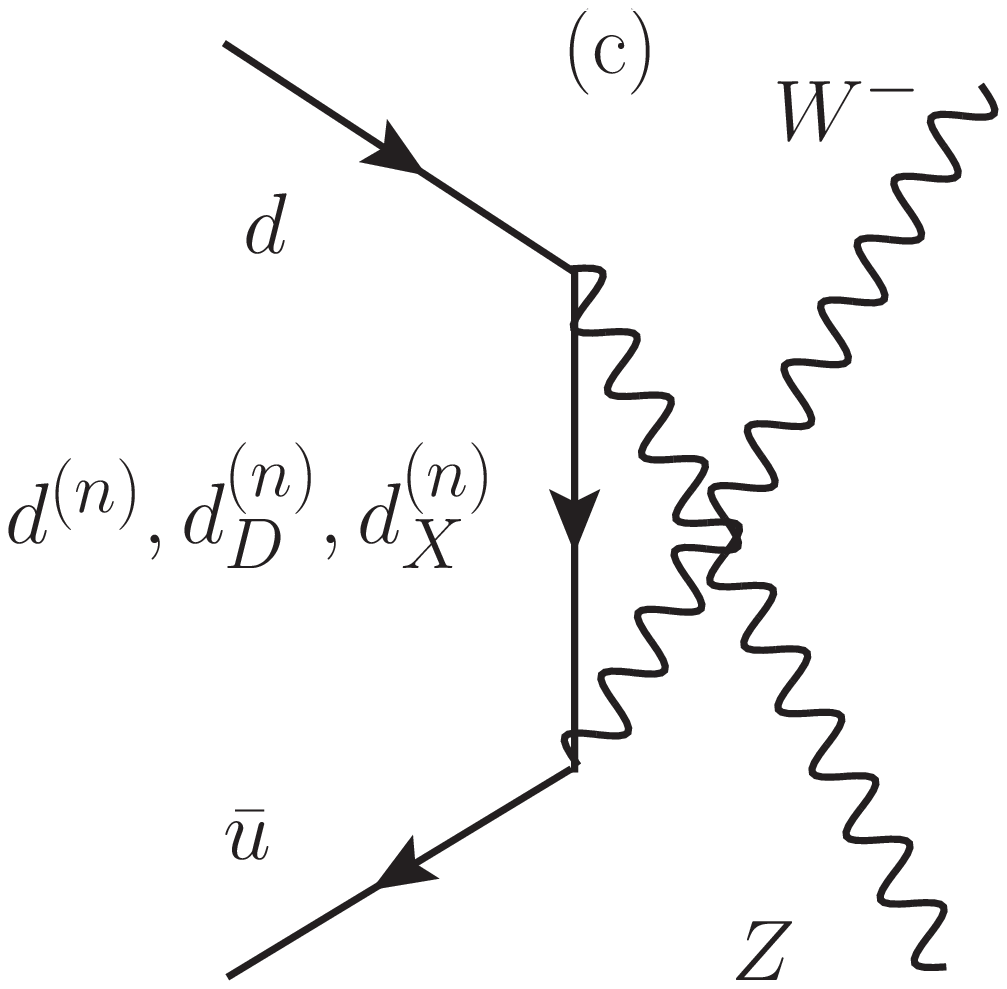}}
\caption{
Diagrams of $d\bar{u} \to WZ$ at a energy scale above $M_{KK}$.
(a), (b) and (c) represent $s$-, $t$- and $u$-channels, respectively.
$W^{(n)}$ is the $n$-th KK state of $W$,
whereas ${\cal D}^{(n)}$ and ${\cal U}^{(n)}$ are $n$-th KK states of $d$ and $u$,
and partners $d_{D,X}^{(n)}$, $u_{U,B}^{(n)}$, respectively.}\label{fig:stu}
\end{figure}

It is important to see how the unitarity is ensured 
when vector bosons are involved in the final states.
The unitarity in the $W$ boson scattering, $W W \to W W$  in the gauge-Higgs unification has been studied in \cite{Haba:2009hw} by using position-space propagators.
In the present paper we are considering $f\bar{f} \to V' \to WW$, $WZ$.
In these processes, $t$- and $u$-channel amplitudes
must be included to cancel the growing part of the the $s$-channel amplitude at $\sqrt{s} \gg m_{V'}$
 (See Fig.~\ref{fig:stu}).

Let us consider the process $d(p_1)\bar{u}(p_2) \to W^-(k_1)Z(k_2)$.
When the initial states are given by $d_R \overline{u_L}$,
there are no $s$-channel contribution as $W^{(n)}$ do not couple to the right-handed quarks. 
For the final state bosons with longitudinal polarization,
$\varepsilon^\mu(k_1) \simeq k_1^\mu/M_W$ and $\varepsilon^\nu(k_2) \simeq k_2^\nu/M_Z$,
$t$- and $u$-channel amplitudes at very high-energy $\sqrt{s} \gg m_{KK}$ are expressed as
\begin{eqnarray}
\calM_t &\sim& \frac{1}{M_W M_Z}
\sum_{{\cal U}} g_{Wd{\cal U}}^R g_{Zu{\cal U}}^R
\bar{v}(p_2) \slashed{k}_2 P_R \frac{(\slashed{p}_1-\slashed{k}_1)}{(p_1-k_1)^2} \slashed{k}_1 P_R u(p_1)
\nonumber\\
&=& 
\frac{1}{2M_W M_Z}
\sum_{{\cal U}} g_{Wd{\cal U}}^R g_{Zu{\cal U}}^R
\bar{v}(p_2) (\slashed{k}_2-\slashed{k}_1) P_R u(p_1),
\\
\calM_u &\sim& \frac{1}{M_W M_Z}
\sum_{{\cal D}}
g_{Zd{\cal D}}^R g_{Wu{\cal D}}^R 
\bar{v}(p_2)  \slashed{k}_1 P_R\frac{(\slashed{p}_1-\slashed{k}_2)}{
(p_1-k_2)^2} \slashed{k}_2 P_R u(p_1)
\nonumber\\
&=& \frac{1}{2M_WM_Z}
\sum_{{\cal D}} g_{Zd{\cal D}}^R g_{Wu{\cal D}}^R
\bar{v}(p_2) (\slashed{k}_1-\slashed{k}_2) P_R u(p_1),
\end{eqnarray}
where $P_{L/R}\equiv(1\mp\gamma_5)/2$, and ${\cal U}$ and ${\cal D}$ denote
KK-excited states with $Q_{EM}=+2/3$ and $-1/3$, respectively. 
Here we have retained contributions only from the first KK states of fermions,
as the $Wu{\cal D}^{(n)}$, $Wd{\cal U}^{(n)}$, $Zu{\cal U}^{(n)}$ and  $Zd{\cal D}^{(n)}$ couplings ($n \ge 2$), etc.,  are all negligibly small.

In order for the growing parts of $\calM_t$ and $\calM_u$ to cancel with each other,
the relation
\begin{eqnarray} 
\sum_{{\cal U}}
g_{W{\cal U}d}^{R} g_{Z u{\cal U}}^{R} 
&\simeq& 
\sum_{{\cal D}}
g_{Z{\cal D}d}^{R} g_{W u {\cal D}}^{R}
\label{eq:TU-relation}
\end{eqnarray}
should be satisfied.
With the values in Tables~\ref{tbl:WuD-couplings}-\ref{tbl:ZdD-couplings},
one finds that in \eqref{eq:TU-relation}
\begin{eqnarray}
(\text{L.H.S}) &\simeq& g_{Wdu^{(1)}}^R g_{Zuu^{(1)}}^R + g_{Wdu_B^{(1)}}^R g_{Zuu_B^{(1)}}^R
\nonumber\\
&=& [(-2.95)\cdot(-0.65) + (1.29)\cdot(-1.48)]\times 10^{-4} g_w^2/\sqrt{2},
\\
(\text{R.H.S}) &\simeq& g_{{dd^{(1)}}}^R g_{Wud^{(1)}}^R + g_{Zdd_X^{(1)}}^R g_{Wud_X^{(1)}}^R
\nonumber\\
&=& [(1.48)\cdot(-1.3) + (0.65)\cdot(2.95)] \times 10^{-4} g_w^2/\sqrt{2}.
\end{eqnarray}
We observe that the relation \eqref{eq:TU-relation} is well satisfied.
We note that 
\begin{eqnarray}
g_{Wdu^{(1)}}^R = - g_{Wud_X^{(1)}}^R,
\quad
g_{Zuu^{(1)}}^R = - g_{Zd{d_X}^{(1)}}^R,
\nonumber\\
g_{Wdu_B^{(1)}}^R = -g_{Wud^{(1)}}^R,
\quad
g_{Zdd^{(1)}}^R = - g_{Zu{u_B}^{(1)}}^R.
%\nonumber\\
%g_{Wdu_B^{(1)}} = g_{Wud_X^{(1)}}=0,
\end{eqnarray}

When the initial states are given by $d_L\overline{u_R}$,
there are contributions from $s$-channel amplitudes.
The condition for the cancellations is given by (c.f. Chapter 21 of \cite{Peskin:1995ev})
\begin{eqnarray}
\sum_{n=0}^\infty g_{W^{(n)}ud}^{L} \, g_{W^{(n)}WZ}
&\simeq& 
  \sum_{{\cal U}'} g_{Wd {\cal U'}}^{L} g_{Z u {\cal U}'}^{L}
- \sum_{{\cal D}'} g_{Zd {\cal D}'}^{L} g_{W u {\cal D}'}^{L},
\label{eq:STU-relation}
\end{eqnarray}
where ${\cal U}'$ and ${\cal D}'$ represent all SM and non-SM fermions
in the first generation with $Q_{EM} = 2/3$ and $-1/3$, respectively. 
From Tables~\ref{tbl:WuD-couplings}, \ref{tbl:WdU-couplings}, \ref{tbl:ZuU-couplings} and \ref{tbl:ZdD-couplings} in Appendix~\ref{sec:fermion-couplings},
one finds that $Wu {\cal D}^{(n)}$, $W d {\cal U}^{(n)}$, $Zu{\cal U}^{(n)}$ and $Zd{\cal D}^{(n)}$ couplings ($n \ge 1$) are all small.
Hence the right-hand-side of \eqref{eq:STU-relation} will be approximately given by
\begin{eqnarray}
(\text{R.H.S.}) &\simeq& g_{Wdu} (g_{Zuu} - g_{Zdd}) = 0.877163 \cdot g_w^2/\sqrt{2},
\end{eqnarray} 
The left-hand side is approximately given, with use of Tables.~\ref{tbl:WfL-couplings} and \ref{tbl:triple-gauge}, by
\begin{eqnarray}
(\text{L.H.S.}) &\simeq& \sum_{n=0}^4 g_{W^{(n)}ud} g_{W^{(n)}WZ} =
0.877162 \cdot g_w^2/\sqrt{2}.
\end{eqnarray}
It is recognized that \eqref{eq:STU-relation} is also quantitatively well-satisfied.

In an analogous way one can confirm the unitarity of the amplitude of $f\bar{f} \to W^+ W^-$. 
In this case KK bosons of $\gamma$, $Z$ and $Z_R$ are involved in the $s$-channel amplitudes.

%%%%%%%%%%%%%%%%%%%%%%%%%%%%%%%%%%%%%%
\section{Summary}\label{sec:summary}

In this paper we have studied the collider signals of $W'$ and $Z'$ 
in the $SO(5) \times U(1)_X$ gauge-Higgs unification.

First we evaluated the couplings of $W'$ and $Z'$ to the SM fields.
We found that the $W^{(1)}$ couplings to light fermions and to top-bottom 
are different in signs, which is explained from the different behavior of wave functions of fields along the extra dimension.

Next we evaluated the decay rates of neutral and charged KK vector bosons.
The total decay widths of $Z'$ are large.
 $\Gamma_{Z'}/M_{Z'} = 15\%$, $6.6\%$ and $13\%$ for $Z' = \gamma^{(1)}$, $Z^{(1)}$ and $Z_R^{(1)}$,
respectively. On the other hand, $W^{(1)}$ has a narrow total width: $\Gamma_{W^{(1)}}/M_{W^{(1)}} \simeq 3\%$.
Several interesting relations among decay modes \eqref{eq:W-decay_relation}, \eqref{eq:W-WR-decay_relation}, \eqref{eq:Z-decay_relation} and \eqref{eq:ZR-decay_relation} are found.
In the warped space $W^{(1)}$ and $W_R^{(1)}$ can decay to $WH$ and $WZ$.
Decay width of $W'$ to $WH$ and $WZ$ are all nearly equal with each other.
For $Z'$ it is found that $\Gamma(Z^{(1)} \to ZH) \simeq \Gamma(Z^{(1)} \to WW) + \Gamma(\gamma^{(1)}\to WW)$
and $\Gamma(Z_R \to ZH) \simeq \Gamma(Z_R \to WW)$.
These properties of $W_R$ and $Z_R$ are qualitatively understood in terms of the 4D $SO(4) \times U(1)$ model introduced in Section 4.

Further we have numerically evaluated the $s$-channel cross sections of $W'$ and $Z'$ in the LHC.
We studied not only processes with fermionic final states
but also bosonic $WH$, $ZH$, $WW$ and $WZ$ final states.
$W'$ and $Z'$ signals of GHU can be found at the LHC experiment in the processes $pp \to W'(Z') + X$, $W' \to tb,WH$, and $Z' \to e^+e^-,\mu^+ \mu^-,ZH$ near the $W'$ and $Z'$ resonances.
For $\theta_H=0.115$ 
($M_{Z^{(1)},\gamma^{(1)}}\simeq 6.0\,\text{TeV}$ and $M_{Z_R^{(1)}}\simeq 5.7\,\text{TeV}$), with the data of 30 fb$^{-1}$, $\sqrt{s_{pp}}=13\,\text{TeV}$ at LHC, 
an excess of the events of $\mu^+\mu^-$ with invariant mass is expected.
(e.g. expected signal[background] is $3.9$ $[0.29]$ events for the bin (GeV) $[3000,4000]$).

In the process with $WZ$ in the final state, it is found that in the amplitude the leading contributions from the longitudinal polarizations of $W$ and $Z$ in the
$s$-, $t$- and $u$-channels cancel with each other so that the unitarity is preserved, provided that both KK vector bosons and KK fermions in the intermediate states are taken into account. 
We have confirmed numerically that this cancellation of the leading terms in the amplitude with
6 digits of precision by taking into account contributions of up to the $4$-th level of KK excited states.
 
We also found that the non-SM 1st KK excited state of fermions can be much lighter than other KK states. 
Especially the 1st KK excited top and bottom partners ($t_{U,B,T}^{(1)}$ and $b_{D,X,Y}^{(1)}$) 
are the lightest non-SM particles and can be singly produced in colliders.
It is seen in Tables~\ref{tbl:fermion-masses5} and \ref{tbl:fermion-masses4}
that $t_{U,B,T}^{(1)}$ and $b_{D,X,Y}^{(1)}$, which are exotic partners of the top and bottom quarks respectively, have mass $M_{t_{U,B,T}^{(1)},b_{D,X,Y}^{(1)}} = 4.6$ TeV ($5.4$ TeV) for $\theta_H = 0.115$ ($0.0737$).
$t_T^{(1)}$ and $b_Y^{(1)}$ have electric charges $+5/3$ and $-4/3$ and can be observed in
the processes $t + W^+ \to t_T \to t + W^+$ and $b + W^- \to b_Y \to b + W^-$
in colliders \cite{Contino:2008hi,AguilarSaavedra:2009es,Mrazek:2009yu}.

The gauge-Higgs unification scenario is promising.
It gives many predictions to be tested at LHC and future colliders.
The 4D Higgs boson appears as the gauge boson in the extra dimension.
The gauge hierarchy problem is solved.
The AB phase $\theta_H$ is the important parameter in GHU.
Many of the physical quantities are determined by $\theta_H$.
The universal relations among $\theta_H$ and $m_{KK}$,
Higgs cubic and quartic couplings have been found.
Corrections to the decay rates for $H\to\gamma\gamma$, $Z\gamma$ due to infinitely many KK states
turn out finite and small.
$Z'$ and $W'$ are predicted around $6$ - $8$ TeV.
Discovery of $Z'$ and $W'$ is most awaited.

\begin{acknowledgments}
This work was supported in part by
the Japan Society for the Promotion of Science, Grants-in-Aid for Scientific Research No 15K05052 (HH and YH).
\end{acknowledgments}

%%%%%%%%%%%%%%%%%%%%%%%%%%%%%%%%%%%%%
\appendix
\section{Basic formulas}\label{sec:formula}
\subsection{$SO(5)$ generators}
The $SO(5)$ generators in the spinor-representation are given by
\begin{eqnarray}
&&
T^{a_L} = \frac{1}{2} \begin{pmatrix} \sigma^a & \bm{0}\\ \bm{0} & \bm{0} \end{pmatrix},
\quad
T^{a_R} = \frac{1}{2} \begin{pmatrix} \bm{0} & \bm{0} \\ \bm{0} &  \sigma^a \end{pmatrix},
\nonumber\\
&& T^{\hat{a}} = \frac{1}{2\sqrt{2}} \begin{pmatrix} & i\sigma^{a} \\
-i \sigma^{a} & \end{pmatrix},
\quad
T^{\hat{4}} = \begin{pmatrix} & \bm{1} \\ \bm{1} & \end{pmatrix},
\end{eqnarray}
and $\tr (T^a T^b) = \frac{1}{2}\delta^{ab}$ is satisfied. Here $\sigma^a$ ($a=1,2,3$) are Pauli matrices. $T^{a_L}$ and $T^{a_R}$ are generators for $SU(2)_L$ and $SU(2)_R$ subgroups, respectively.

%%%%%%%%%%%%%%%%%%%%%%%%%%%%%%%
\subsection{Bulk wave functions}

\subsubsection{Gauge boson bulk functions}
Bulk functions of gauge bosons $C = C(z;\lambda)$ and $S = S(z;\lambda)$ are defined as solutions of
\begin{eqnarray}
\left(\frac{d^2}{dz^2} - \frac{1}{z}\frac{d}{dz} + \lambda^2 \right)
\begin{pmatrix}
C \\ S \end{pmatrix}
&=& 0,
\end{eqnarray}
with boundary conditions 
\begin{eqnarray}
C = z_L, \quad S = 0, \quad C' = 0, \quad S' = \lambda
\quad \text{at $z=z_L$.}
\end{eqnarray}
Here $C' \equiv (d/dz) C$ etc. 
The solutions are given by
\begin{eqnarray}
C(z;\lambda) &=& + \frac{\pi}{2} \lambda z z_L F_{1,0}(\lambda z, \lambda z_L),
\nonumber\\
C'(z;\lambda) &=& + \frac{\pi}{2} \lambda^2 z z_L F_{0,0}(\lambda z, \lambda z_L),
\nonumber\\
S(z;\lambda) &=& - \frac{\pi}{2} \lambda z F_{1,1}(\lambda z, \lambda z_L),
\nonumber\\
S'(z;\lambda) &=& - \frac{\pi}{2} \lambda^2 z F_{0,1}(\lambda z, \lambda z_L),
\end{eqnarray}
where $F_{\alpha,\beta}(u,v) \equiv J_\alpha(u) Y_\beta(v) - Y_\alpha(u)J_\beta(v)$ and $J_\alpha(x)$ and $Y_\alpha(x)$
are Bessel functions of the 1st and 2nd kind, respectively.
$C$, $S$ and $C'$, $S'$ satisfy
\begin{eqnarray}
CS' - SC' = \lambda.
\end{eqnarray}

%%%%%%%%%%%%%%%%%%%%%%%%
\subsubsection{Fermion bulk functions}
Fermion bulk functions $C_{L/R}(z;\lambda,c)$, $S_{L/R}(z;\lambda,c)$ are defined by
\begin{eqnarray}
C_L(z;\lambda,c) &=& +\frac{\pi}{2} \lambda \sqrt{z z_L} F_{c+\frac{1}{2},c-\frac{1}{2}}(\lambda z,\,\lambda z_L),
\nonumber\\
S_L(z;\lambda,c) &=& -\frac{\pi}{2} \lambda \sqrt{z z_L} F_{c+\frac{1}{2},c+\frac{1}{2}}(\lambda z,\,\lambda z_L),
\nonumber\\
C_R(z;\lambda,c) &=& +\frac{\pi}{2} \lambda \sqrt{z z_L} F_{c-\frac{1}{2},c+\frac{1}{2}}(\lambda z,\,\lambda z_L),
\nonumber\\
S_R(z;\lambda,c) &=& +\frac{\pi}{2} \lambda \sqrt{z z_L} F_{c-\frac{1}{2},c-\frac{1}{2}}(\lambda z,\,\lambda z_L).
\end{eqnarray}
These satisfy
\begin{eqnarray}
&&
D_+ \begin{pmatrix} C_L \\ S_L \end{pmatrix} = \lambda \begin{pmatrix} S_R \\ C_R \end{pmatrix},
\quad
D_- \begin{pmatrix} C_R \\ S_R \end{pmatrix} = \lambda \begin{pmatrix} S_L \\ C_L \end{pmatrix},
\nonumber\\
&& D_\pm(c) \equiv \pm\frac{d}{dz}+ \frac{c}{z},
\nonumber\\
&& C_L C_R - S_L S_R = 1,
%\nonumber\\
%\begin{eqnarray}
%&& \{ D_+(c) D_-(c) - \lambda^2 \} \begin{pmatrix} C_R \\ S_R \end{pmatrix} = 0,
%\quad
% \{ D_-(c) D_+(c) - \lambda^2 \} \begin{pmatrix} C_L \\ S_L \end{pmatrix} = 0,
%\end{eqnarray}
\end{eqnarray}
and
\begin{eqnarray}
 C_R = C_L = 1,
\quad
S_R = S_L = 0,
%\nonumber\\
%&& 
%D_-(c) C_R = D_+(c) C_L = 0,
%\quad
%D_-(c) S_R = D_+(c) S_L = \lambda,
\quad
\text{at $z=z_L$.}
\end{eqnarray}
In particular, for $c=0$ we have
\begin{eqnarray}
C_L(z;\lambda,0) &=& C_R(z;\lambda,0) = \cos(\lambda(z-z_L)),
\nonumber \\
S_L(z;\lambda,0) &=& - S_R(z;\lambda,0) = \sin(\lambda(z-z_L)).
\label{eq:cond_c=0}
\end{eqnarray}

%%%%%%%%%%%%%%%%%%%%%%%%%%%%%%%%%%%%%%%%%%%%%%%%%%%%%%%%%%%%%%%%%%%%%%%%%%%%%%%%%%%%
\section{Gauge boson wave functions}\label{sec:bosons}

Wave functions for a charged vector boson $V_C=W^{(n)},\,W_R^{(m)}$ ($n=0,1,2,\cdots$, $m=1,2,\cdots$) are given by
\begin{eqnarray}
\begin{pmatrix} h^L_{V_C} \\ h^R_{V_C} \\ \hat{h}_{V_C} \end{pmatrix}
&=& \begin{pmatrix} v^L_{V_C} C(z;\lambda_{V_C}) \\ v^R_{V_C} C(z;\lambda_{V_C}) \\ \hat{v}_{V_C} \hat{S}(z;\lambda_{V_C}) 
\end{pmatrix},
\end{eqnarray}
where
\begin{eqnarray}
\begin{pmatrix} v^L_{V_C} \\ v^R_{V_C} \\ \hat{v}_{V_C} \end{pmatrix}
&=& \frac{1}{\sqrt{r_{V_C}}}
\begin{cases}
\begin{pmatrix} 
\displaystyle \frac{1+c_H}{\sqrt{2}} \\ 
\displaystyle \frac{1-c_H}{\sqrt{2}} \\
\displaystyle -s_H 
\end{pmatrix}  & V_C = W^{(n)}
\\
\frac{\displaystyle 1}{\displaystyle \sqrt{1+c_H^2}}
\begin{pmatrix} 
\displaystyle \frac{+1-c_H}{\sqrt{2}} \\
\displaystyle \frac{-1-c_H}{\sqrt{2}} \\ 0 \end{pmatrix} 
& V_C = W_R^{(m)}
\end{cases}
\end{eqnarray}
with $C = C(z;\lambda_{V_C})$ etc.
$c_H,s_H \equiv \cos\theta_H,\sin\theta_H$.
We have defined
\begin{eqnarray}
\hat{S}(z;\lambda) \equiv \frac{C(1;\lambda)}{S(1;\lambda)} S(z;\lambda).
\end{eqnarray}
The mass spectrum $\{m_{V_C} = k\lambda_{V_C} \}$ is determined by 
\begin{eqnarray}
2 S C'(1,\lambda_{W^{(n)}}) + \lambda_{W^{(n)}}s_H^2 &=& 0,
\nonumber \\
C(1;\lambda_{W_R^{(m)}}) &=& 0,
\end{eqnarray}
and normalization factors are given by
\begin{eqnarray}
r_{W_R^{(m)}} &=& \int_1^{z_L} \frac{dz}{kz} C(z;\lambda_{W_R^{(m)}})^2,
\nonumber \\
r_{W^{(n)}} &=& \int_1^{z_L} \frac{dz}{kz} \bigl\{
(1+c_H^2) C(z;\lambda_{W^{(n)}})^2 + s_H^2 \hat{S}(z;\lambda_{W^{(n)}})^2 \bigr\}.
\end{eqnarray}

Wave functions for the photon $\gamma = \gamma^{(0)}$ is given by
\begin{eqnarray}
\begin{pmatrix} 
h^L_{\gamma^{(0)}} \\ h^R_{\gamma^{(0)}} \\ h^B_{\gamma^{(0)}}\end{pmatrix}
&=& \frac{1}{\sqrt{(1 + s_\phi^2)L}} \begin{pmatrix} s_\phi \\s_\phi \\ c_\phi \end{pmatrix},
\quad \hat{h}_{\gamma^{(0)}} = 0,
\label{eq:photon-wave}
\end{eqnarray}
where $s_\phi \equiv \sin\phi$ and $c_\phi \equiv \cos\phi$.
Wave functions for a massive neutral vector boson $V=Z^{(n)}$, $\gamma^{(m)}$ and $Z_R^{(m)}$ ($n=0,1,2,\cdots$, $m=1,2,\cdots$) are given by
\begin{eqnarray}
\begin{pmatrix} h_V^L \\ h_V^R \\ \hat{h}_V \\ h_V^B \end{pmatrix}
= 
\begin{pmatrix} v_V^L C(z;\lambda_V) \\ v_V^R C(z;\lambda_V) \\ \hat{v}_V \hat{S}(z;\lambda_V) \\ v_V^B C(z;\lambda_V) \end{pmatrix},
\quad
%\begin{matrix}
%V = \gamma^{(m)},\, Z^{(n)},\, Z_R^{(n)},
%\\
%m = 1,2,\cdots, \\
%n = 0,1,2,\cdots,
%\end{matrix}
\end{eqnarray}
where
\begin{eqnarray}
\begin{pmatrix} v_V^L \\ v_V^R \\ \hat{v}_V \\ v_V^B \end{pmatrix}
&=& \frac{1}{\sqrt{r_V}}
\begin{cases}
\displaystyle \frac{1}{\displaystyle \sqrt{1+s_\phi^2}}
\begin{pmatrix} 
\displaystyle \frac{(1+s_\phi^2)(1+c_H) - 2s_\phi^2}{\sqrt{2}} \\ 
\displaystyle \frac{(1+s_\phi^2)(1-c_H) - 2s_\phi^2}{\sqrt{2}} \\
- (1+s_\phi^2) s_H  \\ 
-\sqrt{2} s_\phi c_\phi \end{pmatrix}
& V = Z^{(n)}
\\
\displaystyle \frac{1}{\displaystyle \sqrt{1+s_\phi^2}}
\begin{pmatrix} s_\phi \\ s_\phi \\ 0 \\ c_\phi \end{pmatrix} & V = \gamma^{(m)},
\\
\displaystyle \frac{1}{\displaystyle \sqrt{1 + (1 + 2t_\phi^2) c_H^2}}
\begin{pmatrix} 
\displaystyle\frac{+ 1 - c_H}{\sqrt{2}} \\
\displaystyle \frac{-1 - c_H}{\sqrt{2}} \\ 0 \\ \sqrt{2} t_\phi c_H \end{pmatrix} 
& V = Z_R^{(m)},
\end{cases}
\end{eqnarray}
where $t_\phi = \tan\phi$.
The mass spectrum $\{ m_{V} = k\lambda_V \}$ is determined by
\begin{eqnarray}
C'(1;\lambda_{\gamma^{(m)}})  &=& 0,
\nonumber \\
2 SC'(1;\lambda_{Z^{(n)}}) + (1 + s_\phi^2) \lambda_{Z^{(n)}} \sin^2\theta_H &=& 0,
\nonumber \\
C(1;\lambda_{Z_R^{(m)}}) &=& 0,
\end{eqnarray}
and normalization factors are given by $r_{V}  = \int_1^{z_L} \frac{dz}{kz}{\cal F}_V $ where
\begin{eqnarray}
{\cal F}_V 
&=& 
\begin{cases}  
C(z;\lambda_V)^2 & V = Z_R^{(m)},\, \gamma^{(m)},
\\
\displaystyle \begin{matrix}
c_\phi^2 C(z;\lambda_V)^2
+ (1 +s_\phi^2) [c_H^2 C(z;\lambda_V)^2  + s_H^2 \hat{S}(z;\lambda_V)^2]
\end{matrix}
& V = Z^{(n)}.
\end{cases}
\end{eqnarray}

%%%%%%%%%%%%%%%%%%%%%%%%%%%%%%%%%%%%%%%%%%%%%%%%%%%%%%%%%%%%%%%%%%%%%%%%%%%%%%%%%%%
\section{Masses and wave functions of $SO(5)$-vector fermions}\label{sec:fermions}
\subsection{Quark sector}
\subsubsection{$Q_{\rm em}=+5/3$ quark partners $(t_{T})$}

$(T^{3_L},T^{3_R})=(+\frac{1}{2},+\frac{1}{2})$ of $\Psi_1^{q,\g=3}$ state has 
an expansion
\begin{eqnarray}
T(x,z) &=& \sqrt{k}z^2 \sum_{n=1}^\infty \biggl\{ 
t_{T,L}^{(n)} (x) \frac{1}{\sqrt{r_{t_{T,L}^{(n)}}}} C_L(z,\lambda_{t_T^{(n)}},c_1) +
t_{T,R}^{(n)} (x) \frac{1}{\sqrt{r_{t_{T,R}^{(n)}}}} S_R(z,\lambda_{t_T^{(n)}},c_1) 
\biggr\},
\label{eq:T-expansion}
\nonumber\\
\end{eqnarray}
where $\lambda_{t_T^{(n)}} = m_{t_T^{(n)}}/k$. 
The KK mass $m_{t_T^{(n)}}$ is determined by
\begin{eqnarray}
C_L(1,\lambda_{t_T^{(n)}},c_1) &=& 0.
\end{eqnarray}
Normalization factors $r_{t_{T,L/R}^{(n)}}$ are determined so that they satisfy
\begin{eqnarray}
 \frac{1}{r_{t_{T,L}^{(n)}}}\int_1^{z_L} C_L(z,\lambda_T^{(n)},c_1)^2 dz 
=\frac{1}{r_{t_{T,R}^{(n)}}}\int_1^{z_L} S_R(z,\lambda_T^{(n)},c_1)^2 dz = 1,
\end{eqnarray}
and one finds $r_{t_{T,L}^{(n)}} = r_{t_{T,R}^{(n)}}$.
%%%%%%%%%%%%%%%%%%%%%%%%%%%%%%%%%%%%%%%%%%%%%%%%%%%%%%%%
\subsubsection{$Q_{\rm em}=-4/3$ quark partners $(b_Y)$}

$(T^{3_L},T^{3_R})=(-\frac{1}{2},-\frac{1}{2})$ of $\Psi_2^{q,\g=3}$ state has 
an expansion
\begin{eqnarray}
Y(x,z) &=& \sqrt{k}z^2 \sum_{n=1}^\infty \biggl\{
  b_{Y,L}^{(n)} (x) \frac{1}{\sqrt{r_{b_{Y,L}^{(n)}}}} C_L(z,\lambda_{b_{Y}^{(n)}},c_2)
+ b_{Y,R}^{(n)} (x) \frac{1}{\sqrt{r_{b_{Y,R}^{(n)}}}} S_R(z,\lambda_{b_{Y}^{(n)}},c_2) 
\biggr\},
\label{eq:Y-expansion}
\nonumber\\
\end{eqnarray}
where $\lambda_{b_Y^{(n)}} = m_{b_{Y}^{(n)}}/k$ and $m_{b_Y^{(n)}}$ is the KK mass,
which is determined by
\begin{eqnarray}
C_L(1,\lambda_{b_Y^{(n)}}, c_2) &=& 0.
\end{eqnarray}
Factors $r_{b_{Y,L/R}^{(n)}}$ are normalized so that they satisfy
\begin{eqnarray}
  \frac{1}{r_{b_{Y,L}^{(n)}}}\int_1^{z_L} C_L(z,\lambda_{b_Y^{(n)}},c_2)^2 dz
= \frac{1}{r_{b_{Y,R}^{(n)}}}\int_1^{z_L} S_R(z,\lambda_{b_Y^{(n)}},c_2)^2 dz = 1,
\end{eqnarray}
and one finds that $r_{b_{Y,L}^{(n)}} = r_{b_{Y,R}^{(n)}}$.

%%%%%%%%%%%%%%%%%%%%%%%%%%%%%%%%%%%%%%%%%%%%%%%%%%%%%%%%
\subsubsection{$Q_{\rm em}=+2/3$ quark and its partners $(t$, $t_B$, $t_U)$}

$(T^{3_L},T^{3_R})=(+\frac{1}{2},-\frac{1}{2})$, $(-\frac{1}{2},+\frac{1}{2})$
and $(0,0)$ of $\Psi_1$ states  $B,t,t'$ together with
$(+\frac{1}{2},+\frac{1}{2})$ of $\Psi_2$ state $U$ have $Q_{\rm em}=+2/3$ states.
For the third generation $\Psi_{a=1,2}^{q,\g=3}$
contain $\hat{t}$, $\hat{t}_B$ and $\hat{t}_U$.
We have an expansion as follows.
\begin{eqnarray}
\begin{pmatrix} U \\ t \\ B \\ t' \end{pmatrix}(x,z)
 = 
 \sqrt{k}z^2 \sum_{n=0}^\infty \biggl\{
 t_{L}^{(n)}(x) \frac{1}{\sqrt{r_{t_L^{(n)}}}} 
\begin{pmatrix} 
 a^{(t)}_{U} C_L^{(t^{(n)})}(z) \\
 a^{(t)}_{t} C_L^{(t^{(n)})}(z) \\
 a^{(t)}_{B} C_L^{(t^{(n)})}(z) \\ 
 a^{(t)}_{t'} \hat{S}_L^{(t^{(n)})}(z) \end{pmatrix}
+ 
 t_{R}^{(n)}(x) \frac{1}{\sqrt{r_{t_R^{(n)}}}} 
\begin{pmatrix} 
a^{(t)}_{U} S_R^{(t^{(n)})}(z) \\ 
a^{(t)}_{t} S_R^{(t^{(n)})}(z) \\ 
a^{(t)}_{B} S_R^{(t^{(n)})}(z) \\ 
a^{(t)}_{t'} \hat{C}_R^{(t)}(z) \end{pmatrix}
\biggr\}
\nonumber\\ +
 \sqrt{k}z^2 \sum_{n=1}^\infty \biggl\{
 t_{B,L}^{(n)}(x) \frac{1}{\sqrt{r_{t_{B,L}^{(n)}}}} 
\begin{pmatrix} a^{(t_B)}_{U} C_L^{(t_B^{(n)})}(z) \\ a^{(t_B)}_{t} C_L^{(t_B^{(n)})}(z) \\ a^{(t_B)}_{B} C_L^{(t_B^{(n)})}(z) \\ a^{(t_B)}_{t'} \hat{S}_L^{(t_B^{(n)})}(z) \end{pmatrix} + 
 t_{B,R}^{(n)}(x) \frac{1}{\sqrt{r_{t_{B,R}^{(n)}}}} 
\begin{pmatrix} a^{(t_B)}_{U} S_R^{(t_B^{(n)})}(z) \\ a^{(t_B)}_{t} S_R^{(t_B^{(n)})}(z) \\ a^{(t_B)}_{B} S_R^{(t_B^{(n)})}(z) \\ a^{(t_B)}_{t'} \hat{C}_R^{(t_B^{(n)})}(z) \end{pmatrix}
\biggr\}
\nonumber\\
+\sqrt{k}z^2 \sum_{n=1}^\infty \biggl\{
 t_{U,L}^{(n)}(x) \frac{1}{\sqrt{r_{t_{U,L}^{(n)}}}} 
 \begin{pmatrix} a^{(t_U)}_{U} C_L^{(t_U^{(n)})}(z) \\ a^{(t_U)}_{t} C_L^{(t_U^{(n)})}(z) \\ a^{(t_U)}_{B} C_L^{(t_U^{(n)})}(z) \\ a^{(t_U)}_{t'} \hat{S}_L^{(t_U^{(n)})}(z) \end{pmatrix} 
+  t_{U,R}^{(n)}(x) \frac{1}{\sqrt{r_{t_{U,R}^{(n)}}}} 
 \begin{pmatrix} a^{(t_U)}_{U} S_R^{(t_U^{(n)})}(z) \\ a^{(t_U)}_{t} S_R^{(t_U^{(n)})}(z) \\ a^{(t_U)}_{B} S_R^{(t_U^{(n)})}(z) \\ a^{(t_U)}_{t'} \hat{C}_R^{(t_U^{(n)})}(z) \end{pmatrix} 
\biggr\},
\nonumber\\
\label{eq:up-expansion}
\end{eqnarray}
where $C_L^{(t_B^{(n)})}(z) \equiv C_L(z,\lambda_{t_B^{(n)}},c)$, $\lambda_{t_B^{(n)}} = m_{t_{B}^{(n)}}/k$ etc. We have defined
\begin{eqnarray}
\{ \hat{S}_L(z,\lambda,c),\, \hat{C}_R(z,\lambda,c)\}
 &\equiv& \frac{C_L(1,\lambda,c)}{S_L(1,\lambda,c)} \{ S_L(z,\lambda,c),\,C_R(z,\lambda,c) \}.
\end{eqnarray}
KK masses $m_{t^{(n)}}$, $m_{t_B^{(n)}}$ and $m_{t_{U}^{(n)}}$
are determined by
\begin{eqnarray}
s_H^2 \frac{(\mu_2^q)^2}{(\mu_2^q)^2 + (\tilde{\mu}^q)^2} + 2 S_R S_L (z=1;\lambda_{t^{(n)}},c)
&=& 0, 
\quad c_1 = c_2 \equiv c,
\label{eq:up-kk}
\end{eqnarray}
and
\begin{eqnarray}
C_L(1,\lambda_{t_B^{(n)}}, c_1) &=& 0,
\nonumber
\\
C_L(1,\lambda_{t_U^{(n)}}, c) &=& 0,
\end{eqnarray}
respectively.
%We note that from \eqref{eq:up-kk} and \eqref{eq:down-kk} brane mass terms should satisfy
%\begin{eqnarray}
%\frac{\tilde{\mu}^q}{\mu_2^q} &=& \frac{m_b}{m_t}.
%\end{eqnarray}
%
Common coefficients are given by
\begin{eqnarray}
&&\begin{pmatrix} a^{(t)}_{U} \\ a^{(t)}_{t} \\ a^{(t)}_{B} \\ a^{(t)}_{t'} \end{pmatrix} 
= \begin{pmatrix}
-\sqrt{2}\tilde{\mu}/\mu_2 \\
(1+c_H) / \sqrt{2}  \\
(1-c_H) / \sqrt{2} \\
- s_H  \end{pmatrix},
\nonumber
\\
&&\begin{pmatrix} a^{(t_B)}_{U} \\ a^{(t_B)}_{t} \\ a^{(t_B)}_{B} \\ a^{(t_B)}_{t'} \end{pmatrix} 
= \begin{pmatrix}
0 \\ (c_H-1)/\sqrt{2} \\ (c_H+1)/\sqrt{2} \\ 0 \end{pmatrix},
\quad
\begin{pmatrix} a^{(t_U)}_{U} \\ a^{(t_U)}_{t} \\ a^{(t_U)}_{B} \\ a^{(t_U)}_{t'} \end{pmatrix} 
=
\begin{pmatrix}
 1 + c_H^2 \\
 (\tilde{\mu}/\mu_2) (1+c_H) \\
 (\tilde{\mu}/\mu_2) (1-c_H) \\
  0 \end{pmatrix},
\end{eqnarray}

Normalization factors $r_{f_L^{(n)}}$, $r_{f_R^{(n)}}$ ($f=t,t_B,t_U$) are determined by
\begin{eqnarray}
&& \frac{1}{r_{f_L^{(n)}}} \int_1^{z_L} \{ [ (a^{(f)}_U)^2 + (a^{(f)}_{t})^2 + (a^{(f)}_{B})^2 ] (C_L^{(f^{(n)})})^2 + (a^{(f)}_{t'})^2 (\hat{S}_L^{(f^{(n)})})^2 ] \} dz 
\nonumber
\\
&=& \frac{1}{r_{f_R^{(n)}}} \int_1^{z_L} \{ [ (a^{(f)}_U)^2 + (a^{(f)}_{t})^2 + (a^{(f)}_{B})^2 ] (S_R^{(f^{(n)})})^2 + (a^{(f)}_{t'})^2 (\hat{C}_R^{(f^{(n)})})^2 ] \} dz 
= 1,
\end{eqnarray}
and one finds that $r_{f_L^{(n)}} = r_{f_R^{(n)}}$ are satisfied.

%%%%%%%%%%%%%%%%%%%%%%%%%%%%%%%%%%%%%%%%%%%%%%%%%%%%%%%%%%%%%%%%%%%%
\subsubsection{$Q_{\rm em}=-1/3$ quark and its partners $(b$, $b_D$, $b_X)$}

$(T^{3_L},T^{3_R})=(-\frac{1}{2},-\frac{1}{2})$ of $\Psi_1$ states $b$ together with
$(+\frac{1}{2},-\frac{1}{2})$, $(-\frac{1}{2},+\frac{1}{2})$ and $(0,0)$ of $\Psi_2$ states $X,D,b'$ 
have $Q_{\rm em}=-1/3$ states.
For the third generation the corresponding towers are $\hat{b}$, $\hat{b}_D$ and $\hat{b}_X$.
Hence we have an expansion 
\begin{eqnarray}
\begin{pmatrix} b \\ X \\ D \\ b' \end{pmatrix}(x,z)
= 
 \sqrt{k} z^2 \sum_{n=0}^\infty \biggl\{
 b_{L}^{(n)}(x) \frac{1}{\sqrt{r_{b_{X,L}^{(n)}}}}
\begin{pmatrix} 
a^{(b)}_{b} C_L^{(b^{(n)})}(z) \\ 
a^{(b)}_{X} C_L^{(b^{(n)})}(z) \\ 
a^{(b)}_{D} C_L^{(b^{(n)})}(z) \\ 
a^{(b)}_{b'} \hat{S}_L^{(b^{(n)})}(z) \end{pmatrix} + 
 b_{R}^{(n)}(x) \frac{1}{\sqrt{r_{b_{X,R}^{(n)}}}} 
\begin{pmatrix} 
a^{(b)}_{b} S_R^{(b^{(n)})}(z) \\ 
a^{(b)}_{X} S_R^{(b^{(n)})}(z) \\ 
a^{(b)}_{D} S_R^{(b^{(n)})}(z) \\ 
a^{(b)}_{b'} \hat{C}_R^{(b^{(n)})}(z) \end{pmatrix}
\biggr\}
\nonumber\\
+\sqrt{k} z^2 \sum_{n=1}^\infty \biggl\{
 b_{X,L}^{(n)}(x) \frac{1}{\sqrt{r_{b_{X,L}^{(n)}}}}
\begin{pmatrix} a^{(b_X)}_{b} C_L^{(b_X^{(n)})}(z) \\ a^{(b_X)}_{X} C_L^{(b_X^{(n)})}(z) \\ a^{(b_X)}_{D} C_L^{(b_X^{(n)})}(z) \\ a^{(b_X)}_{b'} \hat{S}_L^{(b_X^{(n)})}(z) \end{pmatrix} + 
 b_{X,R}^{(n)}(x) \frac{1}{\sqrt{r_{b_{X,R}^{(n)}}}} 
\begin{pmatrix} a^{(b_X)}_{b} S_R^{(b_X^{(n)})}(z) \\ a^{(b_X)}_{X} S_R^{(b_X^{(n)})}(z) \\ a^{(b_X)}_{D} S_R^{(b_X^{(n)})}(z) \\ a^{(b_X)}_{b'} \hat{C}_R^{(b_X^{(n)})}(z) \end{pmatrix}
\biggr\}
\nonumber\\
+
\sqrt{k} z^2 \sum_{n=1}^\infty \biggl\{
 b_{D,L}^{(n)}(x) \frac{1}{\sqrt{r_{b_{D,L}^{(n)}}}}
\begin{pmatrix} a^{(b_D)}_{b} C_L^{(b_D^{(n)})}(z) \\ a^{(b_D)}_{X} C_L^{(b_D^{(n)})}(z) \\ a^{(b_D)}_{D} C_L^{(b_D^{(n)})}(z) \\ a^{(b_D)}_{b'} \hat{S}_L^{(b_D^{(n)})}(z) \end{pmatrix} + 
 b_{D,R}^{(n)}(x) \frac{1}{\sqrt{r_{b_{D,R}^{(n)}}}} 
\begin{pmatrix} a^{(b_D)}_{b} S_R^{(b_D^{(n)})}(z) \\ a^{(b_D)}_{X} S_R^{(b_D^{(n)})}(z) \\ a^{(b_D)}_{D} S_R^{(b_D^{(n)})}(z) \\ a^{(b_D)}_{b'} \hat{C}_R^{(b_D^{(n)})}(z) \end{pmatrix}
\biggr\},
\label{eq:down-expansion}
\nonumber\\
\end{eqnarray}
where $\lambda_{b_X^{(n)}} \equiv m_{b_{X}^{(n)}}/k$ etc.
Mass spectra $m_{b^{(n)}}$, $m_{b_{X}^{(n)}}$, $m_{b_{D}^{(n)}}$ are determined by
\begin{eqnarray}
s_H^2 \frac{(\tilde{\mu}^q)^2}{(\mu_2^q)^2 + (\tilde{\mu}^q)^2}
+ 2 S_R S_L (1;\lambda_{b^{(n)}},c)
&=& 0,
\quad c_1 = c_2 \equiv c \label{eq:down-kk}
\end{eqnarray}
and
\begin{eqnarray}
C_L(1;\lambda_{b_X^{(n)}}, c_1) &=& 0,
\nonumber\\
C_L(1;\lambda_{b_D^{(n)}}, c) &=& 0.
\end{eqnarray}
Combining \eqref{eq:up-kk} and \eqref{eq:down-kk}, one finds
\begin{eqnarray}
\left(\frac{\tilde{\mu}^q}{\mu_2^q}\right)^2 
&=&  - \left\{ 1 + \frac{s_H^2}{2S_L S_R(1;\lambda_{t^{(n)}},c)}\right\}
= - \left\{ 1 + \frac{s_H^2}{2S_L S_R(1;\lambda_{b^{(n)}},c)}\right\}^{-1},
\end{eqnarray}
and $c$ and $\tilde{\mu}^q/\mu_2^q$ are determined from the masses of top and bottom quarks.
Common coefficients are given by
\begin{eqnarray}
&&
\begin{pmatrix} a^{(b)}_{b} \\ a^{(b)}_{X} \\ a^{(b)}_{D} \\ a^{(b)}_{b'} \end{pmatrix} 
= -\begin{pmatrix}  
-\sqrt{2} \mu_2/\tilde{\mu} \\
(1-c_H)/\sqrt{2} \\ 
(1+c_H)/\sqrt{2} \\ 
s_H
\end{pmatrix},
\nonumber
\\
&&
\begin{pmatrix} a^{(b_D)}_{b} \\ a^{(b_D)}_{X} \\ a^{(b_D)}_{D} \\ a^{(b_D)}_{b'} \end{pmatrix} 
= \begin{pmatrix}  
(\tilde{\mu}/\mu_2)(1+c_H^2) \\ 
1-c_H \\ 
1+c_H \\ 
0
\end{pmatrix},
\quad
\begin{pmatrix} a^{(b_X)}_{b} \\ a^{(b_X)}_{X} \\ a^{(b_X)}_{D} \\ a^{(b_X)}_{b'} \end{pmatrix} 
= \begin{pmatrix}  0 \\ (1+c_H)/\sqrt{2} \\ (1-c_H)/\sqrt{2} \\ 0
\end{pmatrix},
\end{eqnarray}
Factors $r_{f_{L/R}^{(n)}}$  ($f=b,b_X,b_D$) are normalized so that
\begin{eqnarray}
&& \frac{1}{r_{f_L^{(n)}}} \int_1^{z_L} \{ [ (a^{(f)}_{b})^2 + (a^{(f)}_{X})^2 + (a^{(f)}_{D})^2 ] (C_L^{(f)})^2 + (a^{(f)}_{b'})^2 (\hat{S}_L^{(f)})^2 ] \} dz 
\nonumber
\\
&=& \frac{1}{r_{f_R^{(n)}}} \int_1^{z_L} \{ [ (a^{(f)}_{b})^2 + (a^{(f)}_{X})^2 + (a^{(f)}_{D})^2 ] (S_R^{(f)})^2 + (a^{(f)}_{b'})^2 (\hat{C}_R^{(f)})^2 ] \} dz 
= 1,
\end{eqnarray}
and one finds $r_{f_L^{(n)}} = r_{f_R^{(n)}}$ are satisfied.

%\clearpage
In Table~\ref{tbl:fermion-masses5} and \ref{tbl:fermion-masses4}, masses of KK fermions are tabulated. In tables masses of exotic partners of up- and down-type quarks are
\begin{eqnarray}
 && M_{u_U^{(n)}} = M_{u_B^{(n)}} = M_{u_T^{(n)}} \equiv M_{u_x^{(n)}},
\nonumber \\
 &&M_{d_D^{(n)}} = M_{d_X^{(n)}} = M_{d_Y^{(n)}} \equiv M_{d_x^{(n)}},
\end{eqnarray}
and $M_{u_x^{(n)}} = M_{d_x^{(n)}}$ are satisfied.

Here we note that
KK masses of exotics largely depend on their bulk mass parameters.
In particular, since the bulk mass parameter of top and bottom quarks approaching to zero for smaller $z_L$ ($\theta_H$),  the mass spectrum for exotic partners of top and bottom quarks are approximately given by
$C_L(1;\lambda_{f^{(1)}},c_{t}=0) = \cos((z_L-1) m_{f^{(1)}}/k) = 0$ so that
\begin{eqnarray}
m_{f^{(1)}} &\simeq& \frac{m_{KK}}{2},
\quad
f = t_{T,U,B}, b_{Y,X,D}.
\end{eqnarray}

\begin{table}[htbp]
\caption{
Masses of KK fermions for $N_F=4$, $z_L = 10^5$ ($\theta_H = 0.115$) in the unit of TeV.
$M_{q_x^{(n)}}$ is the mass of the $n$-th KK excited 
state of exotic partners of $q$-quark (see text). }\label{tbl:fermion-masses5}
\begin{tabular}{c|cccc}
     $n$      & 1 & 2 & 3 & 4 \\
\hline
$M_{u^{(n)}}$ & 9.19 & 12.23 & 16.71 & 20.00 \\
$M_{d^{(n)}}$ & 9.19 & 12.23 & 16.71 & 20.00 \\
$M_{t^{(n)}}$ & 6.62 &  8.17 & 13.99 & 15.62 \\
$M_{b^{(n)}}$ & 6.64 &  8.15 & 14.01 & 15.60 \\
$M_{u_x^{(n)}} = M_{d_x^{(n)}}$& 9.19 & 16.71 & 24.15 & 31.58 \\
$M_{t_x^{(n)}} = M_{b_x^{(n)}}$& 4.64 & 11.99 & 19.38 & 26.78 \\
\hline
\end{tabular}
\end{table}
\begin{table}[htbp]
\caption{Same as Table~\ref{tbl:fermion-masses5} but for $N_F=4$, $z_L=10^4$ ($\theta_H = 0.0737$).}\label{tbl:fermion-masses4}
\begin{tabular}{c|cccc}
     $n$      & 1 & 2 & 3 & 4 \\
\hline
$M_{u^{(n)}}$ & 14.02 & 18.24 & 24.61 & 29.16 \\
$M_{d^{(n)}}$ & 14.02 & 18.24 & 24.61 & 29.16 \\
$M_{t^{(n)}}$ & 10.11 & 10.59 & 20.45 & 20.95 \\
$M_{b^{(n)}}$ & 10.18 & 10.52 & 20.52 & 20.88 \\
$M_{u_x^{(n)}} = M_{d_x^{(n)}}$& 14.02 & 24.61 & 35.06 & 45.46 \\
$M_{t_x^{(n)}} = M_{b_x^{(n)}}$&  5.40 & 15.73 & 26.07 & 36.42 \\
\hline
\end{tabular}
\end{table}

%\clearpage
%%%%%%%%%%%%%%%%%%%%%%%%%%%%%%%%%%%%%%%%%%%%%%%%%%%%%%%%%
\subsection{Lepton sector}

For charged lepton, neutrino and their exotic partners, KK states are given as follows.

\subsubsection{$Q_{\rm em}=+1$ and $-2$ lepton partners}
$(T^{3_L},T^{3_R}) = (-\frac{1}{2},-\frac{1}{2})$ of $\Psi_3$, $L_{1Y}$, 
and 
$(\frac{1}{2},\frac{1}{2})$ of $\Psi_4$, $L_{2X}$, have $Q_{\rm em} = -2$ and $+1$, respectively.
For the third generation, they are expanded as
\begin{eqnarray}
%\nu_{\tau 2X}
L_{2X}(x,z) &=& \sqrt{k} z^2 \sum_{n=1}^\infty \biggl\{
 \nu_{\tau 2X,L}^{(n)}(x) 
 \frac{1}{\sqrt{r_{\nu_{\tau 2X,L}^{(n)}}}}C_L(z; \lambda_{\nu_{\tau 2X}^{(n)}}, c_4)
\nonumber\\&&
+\nu_{\tau 2X,R}^{(n)}(x) 
 \frac{1}{\sqrt{r_{\nu_{\tau 2X,R}^{(n)}}}}S_R(z; \lambda_{\nu_{\tau 2X}^{(n)}}, c_4)
\biggr\},
\label{eq:2X-expansion}
\\
%\tau_{1Y}
L_{1Y}(x,z) &=& \sqrt{k} z^2 \sum_{n=1}^\infty \biggl\{
 \tau_{1Y,L}^{(n)}(x) \frac{1}{\sqrt{r_{\tau_{1Y,L}^{(n)}}}} C_L(z; \lambda_{\tau_{1Y}^{(n)}}, c_3)
\nonumber\\&&
+\tau_{1Y,R}^{(n)}(x) \frac{1}{\sqrt{r_{\tau_{1Y,R}^{(n)}}}} S_R(z; \lambda_{\tau_{1Y}^{(n)}}, c_3) 
\biggr\},
\label{eq:1Y-expansion}
\end{eqnarray}
where the KK masses are given by $(m_{\nu_{\tau 2X}^{(n)}}, m_{\tau_{1Y}^{(n)}}) = k(\lambda_{\nu_{\tau 2X}}^{(n)}, \lambda_{\tau_{1Y}^{(n)}})$
and determined by
\begin{eqnarray}
C_L(1; \lambda_{\nu_{\tau 2X}^{(n)}}, c_4) = 
C_L(1; \lambda_{\tau_{1Y}^{(n)}}, c_3) = 0.
\end{eqnarray}

Normalization factors are determined by
\begin{eqnarray}
\frac{1}{r_{f_L^{(n)}}} \int_1^{z_L} C_L(z; \lambda_{f^{(n)}}, c)^2 dz 
= \frac{1}{r_{f_R^{(n)}}} \int_1^{z_L} S_R(z;\lambda_{f^{(n)}},c)^2 dz  = 1,
\end{eqnarray}
for $(f,c) = (\tau_{1Y},c_3)$, $(\nu_{\tau 2X},c_4)$.

%%%%%%%%%%%%%%%%%%%%%%%%%%%%%%%%%%%%%%%%%%%%%%%%%%%%%%%%%%%%
\subsubsection{$Q_{\rm em} = -1$ charged lepton and its partners}
$(T^{3_L},T^{3_R}) = (-\frac{1}{2},\frac{1}{2})$, $(\frac{1}{2},-\frac{1}{2})$ and $(0,0)$ of $\Psi_3$ together with $(-\frac{1}{2},-\frac{1}{2})$ of $\Psi_4$ are $Q_{\rm em} = -1$ states.
They are expanded as
\begin{eqnarray}
\begin{pmatrix} L_{3Y} \\ L_{1X} \\ \tau \\ \tau' \end{pmatrix}(x,z) 
&=& \sqrt{k} z^2 \sum_{f} \sum_{n} \biggl\{
f_L^{(n)} \frac{1}{\sqrt{r_{f_L^{(n)}}}} \begin{pmatrix}
a_{3Y}^{(f)} C_L^{f^{(n)}}(z) \\ 
a_{1X}^{(f)} C_L^{f^{(n)}}(z) \\
a_{\tau}^{(f)} C_L^{f^{(n)}}(z) \\
a_{\tau'}^{(f)} \hat{S}_L^{f{(n)}}(z) \end{pmatrix}
+ f_R^{(n)} \frac{1}{\sqrt{r_{f_R^{(n)}}}} \begin{pmatrix}
a_{3Y}^{(f)} S_R^{f^{(n)}}(z) \\ 
a_{1X}^{(f)} S_R^{f^{(n)}}(z) \\
a_{\tau}^{(f)} S_R^{f^{(n)}}(z) \\
a_{\tau'}^{(f)} \hat{C}_R^{f{(n)}}(z) \end{pmatrix}
\biggr\},
\label{eq:lepton-expansion}
\nonumber\\
\end{eqnarray}
where $f = \tau$, $\tau_{1X}$ and $\tau_{3Y}$ and $C_L^{f^{(n)}}(z) \equiv C_L(z; \lambda_{f^{(n)}},c)$ etc.
Common coefficients are given by
\begin{eqnarray}
\begin{pmatrix}
a_{3Y}^{(f)} \\ a_{1X}^{(f)} \\ a_{\tau}^{(f)}  \\ a_{\tau'}^{(f)} \end{pmatrix}
&=& 
\begin{pmatrix} \frac{ \sqrt{2} \tilde{\mu}}{\mu_3} \\
\frac{1-c_H}{\sqrt{2}} \\
\frac{1+c_H}{\sqrt{2}} \\
s_H
\end{pmatrix},
\,
\begin{pmatrix} 0 \\ \frac{c_H + 1}{\sqrt{2}} \\ \frac{c_H -1 }{\sqrt{2}} \\ 0 \end{pmatrix}
\text{ and }
\begin{pmatrix} 1 + c_H^2 \\ (1-c_H)\frac{\tilde{\mu}^\ell}{\mu_3^\ell} \\ (1+c_H) \frac{\tilde{\mu}^\ell}{\mu_3^\ell}
\\ 0 
\end{pmatrix},
\end{eqnarray}
for $f=\tau$, $\tau_{1X}$ and $\tau_{3Y}$, respectively.
The mass of $\tau^{(n)}$ is given by $m_{\tau^{(n)}} = k\lambda_{\tau^{(n)}}$ where
$\lambda_{\tau^{(n)}}$ are determined by
\begin{eqnarray}
s_H^2 \frac{(\mu_3^\ell)^2}{(\mu_3^\ell)^2 + (\tilde{\mu}^\ell)^2}
+ 2 S_L S_R (z=1;\lambda_{\tau^{(n)}},c_{\ell})
&=& 0,
\label{eq:lepton-mass}
\end{eqnarray}
and $\tau^{(0)}$ corresponds to the tau lepton.
%$\mu_3^\ell$ and $\tilde{\mu}_\ell$ are chosen so that
%\begin{eqnarray}
%\frac{\tilde{\mu}^\ell}{\mu_3^\ell} = \frac{m_{\nu_\tau}}{m_{\tau}}. 
%\label{eq:lepton-ratio}
%\end{eqnarray}
For $\tau_{\cal E}^{(n)}$ (${\cal E} = 3Y$ and $1X$, $n=1,2,\cdots$), the KK masses are
determined by
\begin{eqnarray}
C_L(1; \lambda_{\tau_{\cal E}^{(n)}}, c_\ell) = 0, 
\quad m_{\tau_{\cal E}^{(n)}} \equiv k \lambda_{\tau_{\cal E}^{(n)}}.
\end{eqnarray}
Normalization factors are determined by
\begin{eqnarray}
&& \frac{1}{r_{f_L^{(n)}}} \int_1^{z_L} \left\{ \left[ (a^{(f)}_{3Y})^2 + (a^{(f)}_{1X})^2 + (a^{(f)}_{\tau})^2 ] (C_L^{(f)})^2 + (a^{(f)}_{\tau'})^2 (\hat{S}_L^{(f)})^2 \right] \right\} dz 
\nonumber
\\
&=& \frac{1}{r_{f_R^{(n)}}} \int_1^{z_L} \left\{ \left[ (a^{(f)}_{3Y})^2 + (a^{(f)}_{1X})^2 + (a^{(f)}_{\tau})^2 ] (S_R^{(f)})^2 + (a^{(f)}_{\tau'})^2 (\hat{C}_R^{(f)})^2 \right] \right\} dz 
= 1,
\end{eqnarray}
where $f = \tau$,$\tau_{3Y}$ and $\tau_{\tau_{1X}}$. 
One finds $r_{f_L^{(n)}} = r_{f_R^{(n)}}$ are satisfied.

%%%%%%%%%%%%%%%%%%%%%%%%%%
\subsubsection{$Q_{\rm em} = 0$ neutrino and its partners}
$(T^{3_L},T^{3_R}) = (\frac{1}{2},\frac{1}{2})$ of $\Psi_3$, 
and $(\frac{1}{2},-\frac{1}{2})$ and $(-\frac{1}{2},\frac{1}{2})$ and $(0,0)$ of $\Psi_4$ are $Q_{\rm em} = 0$ states. They are expanded as 
\begin{eqnarray}
\begin{pmatrix} \nu \\ L_{3X} \\ L_{2Y} \\ \nu' \end{pmatrix}(x,z) 
&=& \sqrt{k} z^2 \sum_{f} \sum_{n} \biggl\{
f_L^{(n)} \frac{1}{\sqrt{r_{f_L^{(n)}}}} \begin{pmatrix}
a_{\nu}^{(f)} C_L^{f^{(n)}}(z) \\ 
a_{3X}^{(f)} C_L^{f^{(n)}}(z) \\
a_{2Y}^{(f)} C_L^{f^{(n)}}(z) \\
a_{\nu'}^{(f)} \hat{S}_L^{f{(n)}}(z) \end{pmatrix}
+ f_R^{(n)} \frac{1}{\sqrt{r_{f_R^{(n)}}}} \begin{pmatrix}
a_{\nu}^{(f)} S_R^{f^{(n)}}(z) \\ 
a_{3X}^{(f)} S_R^{f^{(n)}}(z) \\
a_{2Y}^{(f)} S_R^{f^{(n)}}(z) \\
a_{\nu'}^{(f)} \hat{C}_R^{f{(n)}}(z) \end{pmatrix}
\biggr\},
\label{eq:nu-expansion}
\nonumber\\
\end{eqnarray}
where $f = \nu$, $\nu_{\tau{3X}}$ and $\nu_{\tau{2Y}}$.
$C_L^{f^{(n)}} = C_L(z; \lambda_{f^{(n)}},c)$, $c_3 = c_4 \equiv c^\ell$ etc.
Common coefficients are given by
\begin{eqnarray}
\begin{pmatrix}
a_{\nu}^{(f)} \\ a_{3X}^{(f)} \\ a_{2Y}^{(f)}  \\ a_{\nu'}^{(f)} \end{pmatrix}
&=& 
\begin{pmatrix} \frac{ \sqrt{2} \mu_3^\ell}{\tilde{\mu}^\ell} \\
- \frac{1+c_H}{\sqrt{2}} \\
- \frac{1-c_H}{\sqrt{2}} \\
s_H
\end{pmatrix},
\,
\begin{pmatrix} 0 \\ \frac{1 - c_H}{\sqrt{2}} \\ \frac{1 + c_H}{\sqrt{2}} \\ 0 \end{pmatrix},
\text{ and }
\begin{pmatrix} (1 + c_H^2) \frac{\tilde{\mu}^\ell}{\mu_3^\ell} \\ 1+c_H \\ 1-c_H \\ 0 
\end{pmatrix}
\end{eqnarray}
for $f=\tau$, $\nu_{\tau{2Y}}$ and $\nu_{\tau{3X}}$, respectively.
KK masses of $\nu_{\tau}^{(n)}$, $m_{\nu_{\tau}^{(n)}} \equiv k \lambda_{\nu_{\tau}^{(n)}}$,($n=0,1,2,\cdots$) are determined by
\begin{eqnarray}
s_H^2 \frac{(\tilde{\mu}^\ell)^2}{(\mu_3^\ell)^2 + (\tilde{\mu}^\ell)^2}
+ 2 S_L S_R (z=1;\lambda_{\nu^{(n)}},c_{\ell})
&=& 0,
\label{eq:neutrino-mass}
\end{eqnarray}
and $\nu_{\tau}^{(0)}$ corresponds to the tau neutrino.
From \eqref{eq:lepton-mass} and \eqref{eq:neutrino-mass}, one finds
\begin{eqnarray}
\left( \frac{\mu_3^\ell}{\tilde{\mu}^\ell}\right)^2
&=&  - \left\{ 1 + \frac{s_H^2}{2S_L S_R(1;\lambda_{\nu_{\tau}^{(n)}},c)}\right\}
\nonumber\\
&=& -\left\{ \frac{s_H^2}{2S_L S_R(1;\lambda_{\tau^{(n)}},c)}\right\}^{-1},
\end{eqnarray}
and $c$ and $\tilde{\mu}^\ell / \mu_3^\ell$ are determined from the masses of $\tau$ and $\nu_\tau$.
For $\nu_{{\tau{\cal N}}}^{(n)}$ (${\cal N} = 3X$ and $1Y$, $n=1,2,\cdots$), the KK masses are
determined by
\begin{eqnarray}
C_L(1; \lambda_{\nu_{\tau{\cal N}}}^{(n)}, c_\ell) = 0, 
\quad m_{\nu_{\tau{\cal N}}^{(n)}} \equiv k \lambda_{\nu_{\tau{\cal N}}^{(n)}}.
\end{eqnarray}
Normalization factors are determined by 
\begin{eqnarray}
&& \frac{1}{r_{f_L^{(n)}}} \int_1^{z_L} \left\{ \left[ (a^{(f)}_{\nu})^2 + (a^{(f)}_{3X})^2 + (a^{(f)}_{2Y})^2 ] (C_L^{(f)})^2 + (a^{(f)}_{\nu'})^2 (\hat{S}_L^{(f)})^2 \right] \right\} dz 
\nonumber
\\
&=& \frac{1}{r_{f_R^{(n)}}} \int_1^{z_L} \left\{ \left[ (a^{(f)}_{\nu})^2 + (a^{(f)}_{3X})^2 + (a^{(f)}_{2Y})^2 ] (S_R^{(f)})^2 + (a^{(f)}_{\nu'})^2 (\hat{C}_R^{(f)})^2 \right] \right\} dz 
= 1,
\end{eqnarray}
for $f= \nu$, $\nu_{3X}$ and $\nu_{2Y}$.
One finds $r_{f_L^{(n)}} = r_{f_R^{(n)}}$ are satisfied.

%%%%%%%%%%%%%%%%%%%%%%%%%%%%%%%%%%%%%%%%%%%%%%%%%%%%%%%%%%%%%%%%%%%%%%%%%%%%%%%%%%%%%%%%%%%%%%%%
\section{Fermion couplings}\label{sec:fermion-couplings}

The KK expansions \eqref{eq:T-expansion}, \eqref{eq:Y-expansion}, \eqref{eq:up-expansion} \eqref{eq:down-expansion},
\eqref{eq:2X-expansion},
\eqref{eq:1Y-expansion},
\eqref{eq:lepton-expansion} and
\eqref{eq:nu-expansion} are written in the form of
\begin{eqnarray}
T(x,z) &=& \sqrt{k}z^2 \sum_{n=1}^\infty [
t_{T,L}^{(n)}(x) f_{TL}^{t_T^{(n)}}(z) + t_{T,R}^{(n)}(x) f_{TR}^{(t_T^{(n)})}(z)
],
\nonumber
\\
Y(x,z) &=& \sqrt{k}z^2 \sum_{n=1}^\infty [
b_{Y,L}^{(n)}(x) f_{YL}^{b_Y^{(n)}}(z) + b_{Y,R}^{(n)}(x) f_{YR}^{(b_Y^{(n)})}(z)
],
\\
\begin{pmatrix} U \\ t \\ B \\ t' \end{pmatrix}(x,z)
&=& \sqrt{k}z^2 \sum_{t_u = t,t_B,t_U} 
\sum_n \biggl\{
t_{u,L}(x) \begin{pmatrix}
f_{UL}^{t_u^{(n)}}(z) \\
f_{tL}^{t_u^{(n)}}(z) \\
f_{BL}^{t_u^{(n)}}(z) \\
f_{t'L}^{t_u^{(n)}}(z) 
\end{pmatrix}
+
t_{u,R}(x) \begin{pmatrix}
f_{UR}^{t_u^{(n)}}(z) \\
f_{tR}^{t_u^{(n)}}(z) \\
f_{BR}^{t_u^{(n)}}(z) \\
f_{t'R}^{t_u^{(n)}}(z) 
\end{pmatrix}
\biggr\},
\\
\begin{pmatrix} b \\ X \\ D \\ b' \end{pmatrix} (x,z)
&=& \sqrt{k} z^2 \sum_{b_d = b,b_X,b_D} \sum_n \biggl\{
b_{d,L}(x) \begin{pmatrix}
f_{bL}^{b_d^{(n)}}(z) \\
f_{XL}^{b_d^{(n)}}(z) \\
f_{DL}^{b_d^{(n)}}(z) \\
f_{b'L}^{t_u^{(n)}}(z)  \end{pmatrix} +
b_{d,R}(x) \begin{pmatrix}
f_{bR}^{b_d^{(n)}}(z) \\
f_{XR}^{b_d^{(n)}}(z) \\
f_{DR}^{b_d^{(n)}}(z) \\
f_{b'R}^{b_d^{(n)}}(z) 
 \end{pmatrix}
\biggr\}.
\end{eqnarray}
In terms of these wave functions we write gauge-boson couplings and Yukawa couplings as follows.

%%%%%%%%%%%%%%%%%%%%%%%%%%%%%%%%%%%%%%%%%%%%%%%%%%%%%
\subsection{Vector boson couplings}

\subsubsection{$\bar{\psi}^{(-1/3)}_n V^- \psi^{(2/3)}_m$ and $\bar{\psi}^{(-1)}_n V^- \psi^{(0)}_m$ couplings}

For $b_d = b, b_D, b_X$, $t_u = t, t_U, t_B$ and $V^-=W^-,W_R^-$
we have
\begin{eqnarray}
\lefteqn{
\int \frac{dz}{kz^4}
\bar{\Psi}_1 [\gamma^\mu g_A A_\mu + \bar{\Psi}_2 \gamma^\mu g_A A_\mu \Psi_2 ]} \nonumber\\
&\supset&
 \bar{b}_{dL}^{(n)} V^-_\mu t_{uL}^{(m)} \cdot  
g_w \sqrt{L} \int_1^{z_L} dz \frac{1}{\sqrt{2}}\biggl\{
h^L_V \left[ f_{bL}^{b_b^{(n)}} f_{tL}^{t_u^{(m)}} + f_{DL}^{b_d^{(n)}} f_{UL}^{t_u^{(m)}}\right]
+ h^R_V \left[ f_{bL}^{b_d^{(n)}} f_{BL}^{t_u^{(m)}} + f_{XL}^{b_d^{(n)}} f_{UL}^{t_u^{(m)}} \right]
\nonumber\\&&
+ \hat{h}_V \left[ f_{bL}^{b_d^{(n)}} f_{t'L}^{t_u^{(m)}} - f_{b'L}^{b_d^{(n)}} f_{UL}^{t_u^{(m)}} \right]
\biggr\} + \text{H.c.}
\nonumber\\
&\equiv& \bar{b}_{dL}^{(n)} V^-_\mu t_{uL}^{(m)} \cdot g_{Vd_d^{(n)}t_u^{(m)}}^L
+ \text{H.c.}
\label{eq:Wud-coupling}
\end{eqnarray}
and right-handed couplings with replacements $L \to R$ in spinors and their wave functions.

For leptons couplings we obtain $\bar{\ell} V^- \nu$ couplings from the above formula with replacements
\begin{eqnarray}
b_d \to \tau_{{\cal E}},
\quad t_u \to \nu_{\tau{\cal N}},
\quad (b,b_D,b_X,b') \to (\tau, \tau_{3Y}, \tau_{1X},\tau'),
\nonumber\\
\quad (t,t_U,t_B,t') \to (\nu_\tau, \nu_{\tau 2Y}, \nu_{\tau 3X},\nu'_\tau).
\label{eq:quark-lepton_replace}
\end{eqnarray}

%%%%%%%%%%%%%%%%%%%%%%%%%%%
\subsubsection{$\bar{\psi}^{(2/3)}_n V_\mu \psi^{(2/3)}_m$ and $\bar{\psi}^{(-1/3)}_n V_\mu \psi^{(-1/3)}_m$}
For up-type quarks and their exotic partners $t_u,t_{u'} = t, t_B, t_U$, 
down-type quarks and their exotic partners $b_d,b_{d'} = b, b_D, b_X$
and neutral vector boson $V=\gamma^{(l)},Z^{(l)},Z_R^{(l)}$,
we have $\bar{t}_u V t_{u'}$ and $\bar{b}_d V b_{d'}$ couplings as
\begin{eqnarray}
\lefteqn{
g_A \bar{\Psi}_1 \gamma^\mu \left[A_\mu + \left(\frac{2}{3}\right)\frac{g_B}{g_A}B_\mu \right]\Psi_1
+ g_A \bar{\Psi}_2 \gamma^\mu \left[A_\mu + \left(-\frac{1}{3}\right)\frac{g_B}{g_A} B_\mu \right] \Psi_2
}\nonumber\\
&\supset& \bar{t}_{u'L}^{(n)} \gamma^\mu V_\mu t_{uL}^{(m)} \cdot
g_w \sqrt{L} 
\int_1^{z_L} dz \biggl\{
\frac{1}{2} (h^L_V-h^R_V) \left[-f_{BL}^{t_{u'}^{(n)}} f_{BL}^{t_{u}^{(m)}}
+f_{tL}^{t_{u'}^{(n)}} f_{tL}^{t_{u}^{(m)}}\right]
\nonumber\\&&
+\frac{1}{2} (h^L_V + h^R_V) f_{UL}^{t_{u'}^{(n)}} f_{UL}^{t_{u}^{(m)}}
\nonumber\\&&
+ \frac{1}{2}\hat{h}_V \left[ 
f_{t'L}^{t_{u'}^{(n)}} \left(f_{BL}^{t_{u}^{(m)}} + f_{tL}^{t_{u}^{(m)}}\right)
+ \left(f_{BL}^{t_{u'}^{(n)}} + f_{tL}^{t_{u'}^{(n)}}\right) f_{t'L}^{t_{u}^{(m)}}
\right]
\nonumber\\&&
+\frac{g_B}{g_A} h^B_V
\left[
\frac{2}{3} \left(f_{tL}^{t_{u'}^{(n)}} f_{tL}^{t_{u}^{(m)}}
+f_{BL}^{t_{u'}^{(n)}} f_{BL}^{t_{u}^{(m)}}
+f_{t'L}^{t_{u'}^{(n)}} f_{t'L}^{t_{u}^{(m)}}\right) 
- \frac{1}{3} \left(f_{UL}^{t_{u'}^{(n)}} f_{UL}^{t_{u}^{(m)}}\right)
\right]
\biggr\}
+ \text{H.c.}
\nonumber\\&&
+
\bar{b}_{d'L}^{(n)} \gamma^\mu V_\mu b_{dL}^{(m)} \cdot 
g_w \sqrt{L} 
\int_1^{z_L} dz \biggl\{
\frac{1}{2} (h^L_V - h^R_V) \left[-f_{DL}^{b_{d'}^{(n)}} f_{DL}^{b_{d}^{(m)}}
+f_{XL}^{d_{b'}^{(n)}} f_{XL}^{d_{b}^{(m)}}\right]
\nonumber\\&&
-\frac{1}{2} (h^L_V + h^R_V) f_{bL}^{b_{d'}^{(n)}} f_{bL}^{b_{d}^{(m)}}
\nonumber\\&&
+ \frac{1}{2}\hat{h}_V \left[ 
f_{b'L}^{b_{d'}^{(n)}} \left(f_{XL}^{b_{d}^{(m)}} + f_{DL}^{b_{d}^{(m)}}\right)
+ \left(f_{XL}^{b_{d'}^{(n)}} + f_{DL}^{b_{d'}^{(n)}}\right) f_{b'L}^{b_{d}^{(m)}}
\right]
\nonumber\\&&
+\frac{g_B}{g_A} h^B_V
\left[
-\frac{1}{3} \left(
 f_{XL}^{b_{d'}^{(n)}} f_{XL}^{b_{d}^{(m)}}
+f_{DL}^{b_{d'}^{(n)}} f_{DL}^{b_{d}^{(m)}}
+f_{b'L}^{b_{d'}^{(n)}} f_{b'L}^{b_{d}^{(m)}}\right) 
+\frac{2}{3} \left(f_{bL}^{b_{d'}^{(n)}} f_{bL}^{b_{d}^{(m)}}\right)
\right]
\biggr\}
+ \text{H.c.}
\nonumber\\
&\equiv&
\bar{t}_{u'L}^{(n)} \gamma^\mu V_\mu t_{uL}^{(m)} \cdot g_{Vt_{u'}^{(n)}t_{u}^{(m)}}^L + \text{H.c.}
+ \bar{b}_{d'L}^{(n)} \gamma^\mu V_\mu b_{dL}^{(m)} \cdot g_{V b_{d'}^{(n)} b_{d}^{(m)}}^L + \text{H.c.}
\label{eq:Zuu_Zdd_coupling}
\end{eqnarray}
and right-handed couplings.
Lepton couplings $\bar{\psi}^{(-1)} V_\mu \psi^{(-1)}$ 
and $\bar{\psi}^{(0)} V_\mu \psi^{(0)}$ are obtained from the above formula with replacements \eqref{eq:quark-lepton_replace}.

We note that photon wave functions \eqref{eq:photon-wave} which can be rewritten as
\begin{eqnarray}
h^L_{\gamma^{(0)}} &=& h^R_{\gamma^{(0)}} = \frac{g_B}{g_A} h^B_{\gamma^{(0)}}
= \frac{\sin\theta_W}{\sqrt{L}}, \quad \hat{h}_{\gamma^{(0)}}=0
\end{eqnarray}
 yield proper electromagnetic couplings $Q_{\rm em} e \bar{\psi}\gamma^\mu A^\gamma_\mu \psi$.

%%%%%%%%%%%%%%%%%%%%%%%
\subsubsection{$\bar{\psi}^{(2/3)}_n V^-_\mu \psi^{(5/3)}_m$ and $\bar{\psi}_n^{(0)} V_\mu^- \psi_m^{(+1)}$}
For $t_T$, $t_u = t, t_U, t_B$ and $V^-=W^-,W_R^-$
we have
\begin{eqnarray}
\lefteqn{
\int \frac{dz}{kz^4}
\bar{\Psi}_1 \gamma^\mu g_A A_\mu \Psi_1 }\nonumber\\
&\supset&
 \bar{t}_{uL}^{(n)} \gamma^\mu V^-_\mu t_{TL}^{(m)} \cdot  
g_w \sqrt{L} \int_1^{z_L} dz \frac{1}{\sqrt{2}}\biggl\{
h^L_V \left[ f_{BL}^{t_u^{(n)}} f_{TL}^{t_T^{(m)}} \right]
+ h^R_V \left[ f_{tL}^{t_u^{(n)}} f_{TL}^{t_T^{(m)}} \right]
\nonumber\\&&
- \hat{h}_V \left[ f_{t'L}^{t_u^{(n)}} f_{TL}^{t_T^{(m)}} \right]
\biggr\} + \text{H.c.}
\nonumber\\
&\equiv&
\bar{t}_{uL}^{(n)} \gamma^\mu V^-_\mu t_{TL}^{(m)} \cdot g_{Vt_u^{(n)}t_T^{(m)}}^L
+ \text{H.c.}\label{eq:WuT_coupling}
\end{eqnarray}
and corresponding right-handed couplings.

Lepton couplings are obtained from the above formula with replacements \eqref{eq:quark-lepton_replace} and
\begin{eqnarray}
t_T \to \nu_{\tau 2X},
\quad f_T \to f_{\nu_{\tau 2X}}.
\end{eqnarray}
%%%%%%%%%%%%%%%%%%%%%%%%%%
\subsubsection{$\bar{\psi}^{(-1/3)}_n V^+_\mu \psi^{(-4/3)}_m$ and $\bar{\psi}_n^{(-1)} V^+_\mu \psi^{(-2)}_m$ couplings}

For $b_Y$, $b_d = b, b_X, b_D$ and $V^+=W^+,W_R^+$,
we have
\begin{eqnarray}
\lefteqn{
\int \frac{dz}{kz^4}
\bar{\Psi}_2 \gamma^\mu g_A A_\mu \Psi_2 }\nonumber\\
&\supset&
 \bar{b}_{dL}^{(n)} V^+_\mu b_{YL}^{(m)} \cdot  
g_w \sqrt{L} \int_1^{z_L} dz \frac{1}{\sqrt{2}}\biggl\{
h^L_V \left[ f_{XL}^{b_d^{(n)}} f_{YL}^{b_Y^{(m)}} \right]
+ h^R_V \left[ f_{DL}^{b_d^{(n)}} f_{YL}^{b_Y^{(m)}} \right]
\nonumber\\&&
+ \hat{h}_V \left[ f_{b'L}^{b_d^{(n)}} f_{YL}^{b_Y^{(m)}} \right]
\biggr\} + \text{H.c.}
\nonumber\\
&\equiv& \bar{b}_{dL}^{(n)} V^+_\mu b_{YL}^{(m)} \cdot g_{Vb_d^{(n)}b_Y^{(m)}}^L + \text{H.c.}
\label{eq:WbY_coupling}
\end{eqnarray}
and corresponding right-handed couplings.

Lepton couplings are obtained from the above formula with replacements \eqref{eq:quark-lepton_replace} and 
\begin{eqnarray}
b_Y  \to \tau_{1Y},
\quad
f_Y \to f_{1Y}. 
\end{eqnarray}

%%%%%%%%%%%%%%%%%%%%%%%%%%%%%%%%%%%%%%%%%%%
%\clearpage
\subsubsection{Numerical values}

In Tables~\ref{tbl:WuD-couplings} -  \ref{tbl:ZdD-couplings}, KK fermions' couplings to $W$ and $Z$ bosons for $N_F=4$, $z_L=10^5$ ($\theta_H = 0.115$) are tabulated.

\begin{table}[htbp]
\caption{
Couplings of the $W$ boson to the up-quark and KK excited down-type states
in unit of $g_w/\sqrt{2}$ for $N_F=4$, $z_L=10^5$ ($\theta_H = 0.115$).}\label{tbl:WuD-couplings}
\begin{tabular}{ccccc}
   & $n=1$ & 2 & 3 & 4  \\ 
\hline
$g_{W u^{(0)} d^{(n)}}^L/(g_w/\sqrt{2})$ &
$-4.28\times 10^{-9}$ & $-2.98\times 10^{-7}$ & $-5.77\times 10^{-8}$ & $9.24\times 10^{-6}$ \\
$g_{W u^{(0)} d^{(n)}}^R/(g_w/\sqrt{2})$ & 
$-1.30\times 10^{-2}$ & $-3.67\times 10^{-13}$& $-2.16\times 10^{-3}$ & $-1.87\times 10^{-13}$ \\
\hline
$g_{Wu^{(0)}d_{D}^{(n)}}^L/(g_w/\sqrt{2})$ & 
$1.59\times 10^{-8}$ & $-3.54\times 10^{-9}$ & $2.52\times 10^{-9}$ & $-1.58\times 10^{-9}$\\
$g_{Wu^{(0)}d_{D}^{(n)}}^R/(g_w/\sqrt{2})$ & 
$2.95\times 10^{-2}$ &  $4.91\times 10^{-3}$ & $1.62\times 10^{-3}$ &  $7.27\times 10^{-4}$\\ 
\hline
$g_{Wu^{(0)}d_{X}^{(n)}}^L/(g_w/\sqrt{2})$ & 
$8.90\times 10^{-8}$ & $-1.22\times 10^{-7}$ & $1.50\times 10^{-7}$ & $-1.73\times 10^{-7}$\\
$g_{Wu^{(0)}d_{X}^{(n)}}^R/(g_w/\sqrt{2})$ & 
$1.23\times 10^{-14}$ & $-9.31\times 10^{-15}$ & $7.87\times 10^{-15}$ & $-6.97\times 10^{-15}$\\ 
\hline
\end{tabular}
\end{table}
\begin{table}[htbp]
\caption{
Couplings of the $W$ boson to the down-quark and KK excited up-type states
in unit of $g_w/\sqrt{2}$ for $N_F=4$, $z_L=10^5$ ($\theta_H = 0.115$).}\label{tbl:WdU-couplings}
\begin{tabular}{ccccc}
   & $n=1$ & 2 & 3 & 4  \\ 
\hline
$g_{W d^{(0)} u^{(n)}}^L/(g_w/\sqrt{2})$ &
$-3.36\times 10^{-8}$ & $-1.21\times 10^{-7}$ & $6.54\times 10^{-9}$ & $3.46\times 10^{-8}$ \\
$g_{W d^{(0)} u^{(n)}}^R/(g_w/\sqrt{2})$ & 
$-2.95\times 10^{-2}$ & $-3.68\times 10^{-13}$& $-4.92\times 10^{-3}$ & $-1.87\times 10^{-13}$ \\
\hline
$g_{Wd^{(0)}u_{B^{(n)}}}^L/(g_w\sqrt{2})$ & 
$1.44\times 10^{-24}$ & $-2.48\times 10^{-24}$ & $-1.99\times 10^{-24}$ & $-4.23\times 10^{-24}$\\
$g_{Wd^{(0)}u_{B}^{(n)}}^R/(g_w/\sqrt{2})$ & 
$5.21\times 10^{-31}$ & $-4.15\times 10^{-31}$ & $-4.21\times 10^{-32}$ & $-2.99\times 10^{-31}$\\ 
\hline
$g_{Wd^{(0)}u_{U}^{(n)}}^L/(g_w/\sqrt{2})$ & 
$1.59\times 10^{-8}$ & $-3.54\times 10^{-9}$ & $2.52\times 10^{-9}$ & $-1.58\times 10^{-9}$\\
$g_{Wd^{(0)}u_{U}^{(n)}}^R/(g_w/\sqrt{2})$ & 
$1.29\times 10^{-2}$ & $2.15 \times 10^{-3}$ & $7.12\times 10^{-4}$ & $3.19\times 10^{-4}$\\ 
\hline
\end{tabular}
\end{table}
\begin{table}[htbp]
\caption{
Couplings of the $W$ boson to the up- or down-quark and KK excite states with non-SM electric charges in unit of $g_w/\sqrt{2}$ for $N_F=4$, $z_L=10^5$ ($\theta_H = 0.115$).}\label{tbl:WqX-couplings}
\begin{tabular}{ccccc}
   & $n=1$ & 2 & 3 & 4 \\ 
\hline
$g_{Wu^{(0)}u_{T}^{(n)}}^L/(g_w/\sqrt{2})$ & 
$-1.74\times 10^{-7}$ & $3.86\times 10^{-9}$ & $-2.75\times 10^{-9}$ & $1.72\times 10^{-9}$\\
$g_{Wu^{(0)}u_{T}^{(n)}}^R/(g_w/\sqrt{2})$ & 
$-3.22\times 10^{-2}$ & $-5.37\times 10^{-3}$ & $-1.78\times 10^{-3}$ & $-7.94\times 10^{-4}$\\ 
\hline
$g_{Wd^{(0)}d_{Y}^{(n)}}^L/(g_w/\sqrt{2})$ & 
$-3.96\times 10^{-8}$ & $8.82\times 10^{-9}$ & $-6.29\times 10^{-9}$ & $3.93\times 10^{-9}$\\
$g_{Wd^{(0)}d_{Y}^{(n)}}^R/(g_w/\sqrt{2})$ & 
$-3.22\times 10^{-2}$ & $-5.37\times 10^{-3}$ & $-1.78\times 10^{-3}$ & $-7.94\times 10^{-4}$\\ 
\hline
\end{tabular}
\end{table}
\begin{table}[htbp]
\caption{Couplings of the $Z$ boson to the up quark and KK excited states in unit of $g_w/\cos\theta_W$
for $N_F=4$, $z_L=10^5$ ($\theta_H = 0.115$).}\label{tbl:ZuU-couplings}
\begin{tabular}{cccccc}
    & $n=0$ &1 & 2 & 3 & 4  \\ 
\hline
$g_{Zu^{(0)}u^{(n)}}^L/(g_w/\cos\theta_W) $ & $0.345912$ & $-2.26\times 10^{-9}$ & $-5.04\times 10^{-8}$ & $5.61\times 10^{-11}$ & $1.82\times 10^{-8}$\\ 
$g_{Zu^{(0)}u^{(n)}}^R/(g_w/\cos\theta_W) $ &$-0.154263$& 
$-6.47\times 10^{-3}$ & $-8.43\times 10^{-6}$ & $-1.08\times 10^{-3}$ & $-1.20\times 10^{-7}$\\ 
\hline
$g_{Zu^{(0)}u_{B}^{(n)}}^L/(g_w/\cos\theta_W) $ & -- & 
$-8.65\times 10^{-9}$ & $ 1.93\times 10^{-9}$ & $-1.37\times 10^{-9}$ & $8.59\times 10^{-10}$\\
$g_{Zu^{(0)}u_{B}^{(n)}}^R/(g_w/\cos\theta_W) $ & -- & 
$-1.61\times 10^{-2}$ & $-2.68\times 10^{-3}$ & $-8.85\times 10^{-4}$ & $-3.96\times 10^{-4}$\\ 
\hline
$g_{Zu^{(0)}u_{U}^{(n)}}^L/(g_w/\cos\theta_W) $ & -- & 
$-7.98\times 10^{-9}$ & $ 1.78\times 10^{-9}$ & $-1.27\times 10^{-9}$ & $7.92\times 10^{-10}$\\
$g_{Zu^{(0)}u_{U}^{(n)}}^R/(g_w/\cos\theta_W) $ & -- &  
$-1.48\times 10^{-2}$ & $-2.47\times 10^{-3}$ & $-8.17\times 10^{-4}$ & $-3.65\times 10^{-4}$\\ 
\hline
\end{tabular}
\end{table}
\begin{table}[htbp]
\caption{Couplings of the $Z$ boson to the down quark and KK excited states in unit of $g_w/\cos\theta_W$
for $N_F=4$, $z_L=10^5$ ($\theta_H = 0.115$).}\label{tbl:ZdD-couplings}
\begin{tabular}{cccccc}
    & $n=0$ & 1 & 2 & 3 & 4  \\ 
\hline
$g_{Zd^{(0)}d^{(n)}}^L/(g_w/\cos\theta_W) $ & $-0.423018$ & 
$4.46\times 10^{-8}$ & $-3.32\times 10^{-7}$ & $-4.12\times 10^{-8}$ & $5.27\times 10^{-7}$\\ 
$g_{Zd^{(0)}d^{(n)}}^R/(g_w/\cos\theta_W) $ & $0.0771316$ & 
$1.48\times 10^{-2}$ & $4.22\times 10^{-6}$ & $2.47\times 10^{-3}$ & $6.00\times 10^{-8}$\\ 
\hline
$g_{Zd^{(0)}d_{X}^{(n)}}^L/(g_w/\cos\theta_W) $ & -- & 
$-1.77\times 10^{-8}$ & $4.74\times 10^{-8}$ & $-6.02\times 10^{-8}$ & $7.13\times 10^{-8}$\\
$g_{Zd^{(0)}d_{X}^{(n)}}^R/(g_w/\cos\theta_W) $ & -- & 
$ 1.62\times 10^{-2}$ & $2.69\times 10^{-3}$ & $ 8.91\times 10^{-4}$ & $3.99\times 10^{-4}$\\ 
\hline
$g_{Zd^{(0)}d_{D}^{(n)}}^L/(g_w/\cos\theta_W) $ & -- &
$8.00\times 10^{-9}$ & $-1.78\times 10^{-9}$ & $1.27\times 10^{-9}$ & $-7.94\times 10^{-10}$\\
$g_{Zd^{(0)}d_{D}^{(n)}}^R/(g_w/\cos\theta_W) $ & -- &
$6.51\times 10^{-3}$ & $ 1.08\times 10^{-3}$ & $3.58\times 10^{-4}$ & $1.60\times 10^{-4}$\\ 
\hline
\end{tabular}
\end{table}

%\clearpage
%%%%%%%%%%%%%%%%%%%%%%%%%%%%%%%%%%%%%%%%%%%%%%%%%
\subsection{Higgs Yukawa couplings}

Yukawa couplings among Higgs $H^{(k)}$ and quark-sector fermions are read from
\begin{eqnarray}
\lefteqn{
\sum_{i=1,2}
\int_1^{z_L} \sqrt{-g} e_5^z i\Psi_i \Gamma^m (-ig_A A_z) \Psi_i
}\nonumber\\
&\supset&
H^{(k)}\left(\bar{t}_{uL}^{(m)} t_{u'R}^{(n)}  + \bar{t}_{u'R}^{(n)} t_{uL}^{(m)}\right) (x) \nonumber\\&&
\times 
\frac{1}{2} g_w \sqrt{L} \int_1^{z_L}  u_{H^{(k)}}(z) 
\left[ f_{t'L}^{t_u^{(m)}} \left(f_{BR}^{t_{u'}^{(n)}} - f_{tR}^{t_{u'}^{(n)}}\right)
- 
\left(f_{BL}^{t_{u}^{(m)}} - f_{tL}^{t_{u}^{(m)}}\right) f_{t'R}^{t_{u'}^{(n)}}
\right] dz
\nonumber\\&&
-H^{(k)} \left(\bar{t}_{uR}^{(m)} t_{u'L}^{(n)}  + \bar{t}_{u'L}^{(n)} t_{uR}^{(m)}\right) (x)
\nonumber\\&&
\times
\frac{1}{2} g_w \sqrt{L} \int_1^{z_L}  u_{H^{(k)}}(z) 
\left[ 
f_{t'R}^{t_u^{(m)}} \left(f_{BL}^{t_{u'}^{(n)}} - f_{tL}^{t_{u'}^{(n)}}\right)
- 
\left(f_{BR}^{t_{u}^{(m)}} - f_{tR}^{t_{u}^{(m)}}\right)f_{t'L}^{t_{u'}^{(n)}}
\right] dz,
\label{eq:Hfnfn}
\end{eqnarray}
and when $t_u^{(m)} = t_{u'}^{(n)}$, one obtain
\begin{eqnarray}
&& H^{(n)}(x) \left(\bar{t}_{uL}^{(m)} t_{uR}^{(m)} + \bar{t}_{uR}^{(m)} t_{uL}^{(m)}\right)
\nonumber\\&&
\times \frac{1}{2}g_w \sqrt{L} \int_1^{z_L} u_{H^{(n)}}
\left[ 
f_{t'L}^{t_u^{(m)}} \left(f_{BR}^{t_u^{(m)}} - f_{tR}^{t_u^{(m)}}\right)
- \left(f_{BL}^{t_u^{(m)}} - f_{tL}^{t_u^{(m)}}\right) f_{t'R}^{t_u^{(m)}}
\right]dz.
\label{eq:Hfmfn}
\end{eqnarray}
For the Higgs boson $H = H^{(0)}$, the wave function is given by
\begin{eqnarray}
u_H(z) = u_{H^{(0)}}(z) = \sqrt{\frac{2}{k(z_L^2-1)}} z.
\label{eq:Higgs-wave} 
\end{eqnarray}

%%%%%%%%%%%%%%%%%%%%%%%%%%%%%%%
\section{Boson couplings}\label{sec:boson-couplings}
%\subsection{$\gamma WW$, $ZWW$, $Z_R WW$ and $ZWW_R$ couplings}
\subsection{Vector boson trilinear couplings}
The $V^{(l)} W^{(m)} W^{(n)}$ couplings for $V=\gamma, Z, Z_R$ are contained in 
\begin{align}
&\int_1^{z_L} \frac{dz}{kz}\left(-\frac{1}{4}\right)
\text{Tr}\left[F_{\mu\nu}F_{\rho\sigma}\right]
\eta^{\mu \rho}\eta^{\nu\sigma} \cr
&\supset i g_A\int_1^{z_L}\frac{dz}{kz}
\text{Tr} \Big[
 (\partial_\mu \hat{V}_\nu-\partial_\nu \hat{V}_\mu)[\hat{W}^+_\rho,\hat{W}^-_\sigma] \cr & \qquad
+(\partial_\mu \hat{W}^-_\nu-\partial_\nu \hat{W}^-_\mu)[\hat{V}_\rho,\hat{W}^+_\sigma] 
+(\partial_\mu \hat{W}^+_\nu-\partial_\nu \hat{W}^+_\mu)[\hat{V}_\rho,\hat{W}^-_\sigma]\Big] 
\eta^{\mu \rho}\eta^{\nu\sigma}\nonumber\\
&\supset i\sum_{m,n}g_{V^{(l)} W^{(m)} W^{(n)}}\eta^{\mu \rho}\eta^{\nu\sigma}
\Big\{(\partial_\mu Z^{(l)}_\nu -\partial_\nu Z^{(l)}_\mu) W^{+(m)}_\rho W^{-(n)}_\sigma \nonumber\\
&-(\partial_\mu W^{+(m)}_\nu -\partial_\nu W^{+(m)}_\mu) Z^{(l)}_\rho W^{-(n)}_\sigma +(\partial_\mu W^{-(n)}_\nu -\partial_\nu W^{-(n)}_\mu) Z^{(l)}_\rho W^{+(m)}_\sigma \Big\}
\end{align}
so that one finds that
\begin{align}&
g_{V^{(l)} W^{(m)} W^{(n)}} = g_w \sqrt{L} \int_1^{z_L} \frac{dz}{kz}\nonumber\\
&\times\biggl\{ h_{V^{(l)}}^L \bigg( h_{W^{(m)}}^Lh_{W^{(n)}}^L+  \frac{\hat{h}_{W^{(m)}}\hat{h}_{W^{(n)}}}{2}\bigg)
+h_{V^{(l)}}^R \bigg( h_{W^{(m)}}^Rh_{W^{(n)}}^R+\frac{\hat{h}_{W^{(m)}}\hat{h}_{W^{(n)}}}{2}\bigg) \nonumber\\
&\quad+\hat{h}_{V^{(l)}}\bigg( \frac{h_{W^{(m)}}^L\hat{h}_{W^{(n)}}+  h^R_{W^{(m)}}\hat{h}_{W^{(n)}}
+\hat{h}_{W^{(m)}}h_{W^{(n)}}^L+  \hat{h}_{W^{(m)}}h_{W^{(n)}}^R}{2}\bigg)\bigg\} .\label{ZWW1}
\end{align}
Here $C_{W^{(m)}} = C(z; \lambda_{W^{(m)}})$ etc.

$V^{(l)}W^{(m)} W_R^{(n)}$ and $V^{(l)}W_R^{(m)}W_R^{(n)}$ couplings are obtained from above expression with replacements 
$W^{(n)} \to W_R^{(n)}$.

%%%%%%%%%%%%%%%%%%%%%%%%%%%%%%%%%%%%%%%%%%%%%%%%%%%%%%%
\subsection{$HZZ$ and $HZZ_R$ couplings}
The Higgs coupling $HZ^{(m)} Z^{(n)}$ is contained in the $\text{Tr} F_{\mu z} F^{\mu z} $ term 
\begin{align}
& -i g_A k^2 \int_1^{z_L}\frac{dz}{kz} 
\text{Tr}\left[\big(\partial_z \hat{Z}_\mu\big) \big[\hat{H}, \hat{Z}_\nu\big]\right]\eta^{\mu\nu} \nonumber\\
&\supset -\frac{1}{2}\sum_{n}g_{HZ^{(m)} Z^{(n)}} H Z^{(m)}_\mu Z^{(n)}_\nu \eta^{\mu\nu}
-\sum_{m<n}g_{HZ^{(m)} Z^{(n)}} H Z^{(m)}_\mu Z^{(n)}_\nu \eta^{\mu\nu}
\end{align}
so that
\begin{align}
g_{HZ^{(m)} Z^{(n)}}
= & -g_A k^2 \int_1^{z_L}\frac{dz}{kz}\frac{1}{2}u_H(z)\nonumber\\
&\times
\left[-\big(\partial_z \hat{h}_{Z^{(m)}}\big)\big(h^L_{Z^{(n)}}-h^R_{Z^{(n)}}\big)
+\partial_z\big(h^L_{Z^{(n)}}-h^R_{Z^{(n)}}\big)\hat{h}_{Z^{(m)}} + (m \longleftrightarrow n) \right]~,
\label{HZZ}
\end{align}
where the $u_H(z)$ is given in \eqref{eq:Higgs-wave}.

Similarly, 
the Higgs coupling $HZ^{(m)} Z^{(n)}_R$ is contained in the $\text{Tr} F_{\mu z} F^{\mu z} $ term 
\begin{align}
& -i g_A k^2 \int_1^{z_L}\frac{dz}{kz} 
\left\{\text{Tr}\left[\big(\partial_z \hat{Z}_\mu\big) \big[\hat{H}, \hat{Z}_{R\nu}\big]\right]
+\text{Tr}\left[\big(\partial_z \hat{Z}_{R\mu}\big) \big[\hat{H}, \hat{Z}_\nu\big]\right]\right\}\eta^{\mu\nu} \nonumber\\
&\supset 
-\sum_{m, n}g_{HZ^{(m)} Z^{(n)}_R} H Z^{(m)}_\mu Z^{(n)}_{R\nu} \eta^{\mu\nu}
\end{align}
so that
\begin{align}
g_{HZ^{(m)} Z^{(n)}_R}
= & -g_A k^2 \int_1^{z_L}\frac{dz}{kz}\frac{1}{2}u_H(z)\nonumber\\
&\times
\left[-\big(\partial_z \hat{h}_{Z^{(m)}}\big)\big(h^L_{Z^{(n)}_R}-h^R_{Z^{(n)_R}}\big)
-\hat{h}_{Z^{(m)}}\partial_z\big(h^L_{Z^{(n)}_R}-h^R_{Z^{(n)_R}}\big)
\right]~.
\label{HZZR}
\end{align}
$HWW$ and $HWW_R$ couplings are seen in \cite{Funatsu:2015xba}.

%%%%%%%%%%%%%%%%%%%%%%%%%%%%%%%
\section{Decay width}\label{sec:formula-decay}

For a heavy charged vector boson $W'$, the $W' \to WH$
decay width is given by
\begin{eqnarray}
\Gamma(W' \to WH) &=& \frac{M_{W'}}{192\pi}
\left( \frac{g_{W'WH}}{M_W} \right)^2
\nonumber\\&& \times
\left(
1 + \frac{10 M_W^2 - 2 M_H^2}{M_{W'}^2}
+ \frac{M_W^4 + M_H^4 - 2 M_W^2 M_H^2}{M_{W'}^4}
\right),
\end{eqnarray}
and $\Gamma(Z' \to ZH)$ is obtained from the above expression by replacements of $W$ with $Z$.
The decay width for$W' \to WZ$ is given by
\begin{eqnarray}
\Gamma(W' \to ZW) &=& \frac{M_{W'}}{192\pi} g_{W'WZ}^2 \frac{M_{W'}^4}{M_W^2 M_Z^2}
\nonumber\\&& \times
\left( 1 - \frac{(M_Z+M_W)^2}{M_{W'}^2} \right)
\left( 1 - \frac{(M_Z-M_W)^2}{M_{W'}^2} \right)
\nonumber\\&& \times
\left( 1 + \frac{10(M_Z^2 + M_W^2)}{M_{W'}^2}
+ \frac{M_Z^4 + M_W^4 + 10 M_Z^2M_W^2}{M_{W'}^4}\right),
\end{eqnarray}
and $\Gamma(Z' \to W^+W^-)$ is obtained by replacements
 $W'\to Z'$ and $Z\to W$.

For the decay of a heavy fermion $F$ (mass $m_F$) to a light fermion $f$ (mass $m_f$) and a vector boson $V$ (mass $m_V$),
the decay rate is given by
\begin{eqnarray}
\Gamma(F \to fV) &=& \frac{m_F}{32\pi} \sqrt{\lambda(1,m_f/m_F,m_V/m_F)}
\biggl\{
\frac{(g_{VFf}^L)^2 + (g_{VFf}^L)^2}{m_V^2m_F^2}
\bigl[(m_F^2 - m_f^2)^2
\nonumber\\&&
 + m_V^2(m_F^2 + m_f^2) - 2m_V^2 \bigr]
- 12 g_L g_R \frac{m_f}{m_F} \biggr\},
\end{eqnarray}
where
\begin{eqnarray}
\lambda(A,B,C) &\equiv& A^4 + B^4 + C^4 - 2 (A^2B^2 + B^2C^2 + C^2A^2),
\end{eqnarray}
where $g_{VFf}^{L/R}$ are the left- and right- handed coupling of $\bar{f}FV$.

For decay widths of exotic fermions $t_T^{(n)}$ and $b_Y^{(n)}$, we have
\begin{eqnarray}
&&\Gamma(t_T^{(n)} \to tW^+) = 
\frac{1}{32\pi}
M_{t_T^{(n)}}\sqrt{\lambda(1,M_t/M_{t_T^{(n)}},M_W/M_{T^{(n)}})} 
\nonumber\\&&
\times \biggl\{
 \frac{\left(g^L\right)^2+\left(g^R\right)^2}{
M_W^2 M_{t_T^{(n)}}^2}
\left[ (M_{t_T^{(n)}}^2-M_t^2)^2 + M_W^2(M_{t_T^{(n)}}^2+M_t^2) - 2 M_W^4\right]
- 12 g^L g^R \frac{M_t}{M_{t_T^{(n)}}}
\biggr\},
\nonumber\\
\end{eqnarray}
where 
\begin{eqnarray}
g^{L,R} &=& g_{Wt_T^{(n)}t}^{L,R},
\end{eqnarray}
are left- and right-hand couplings of $t_{T}^{(n)}$ to $tW^+$.
Decay width for $b_Y^{(n)}$ to $bW^-$ are obtained by replacements
\begin{eqnarray}
(t_T^{(n)}, t, W^+) &\to& (b_Y^{(n)}, W^-, b).
\end{eqnarray}
%%%%%%%%%%%%%%%%%%%%%%
\section{Cross section}\label{sec:formula-scat}

Cross sections of processes $f\bar{f}' \to W' \to WH$ and $f\bar{f} \to Z' \to ZH$
in the center-of-mass frame are given as follows.
For the process $f(p_1)\bar{f}(p_2)\to V' \to V(k_1)H(k_2)$,
the differential cross section is given by
\begin{eqnarray}
\frac{d\sigma}{d\cos\theta}
&=& \frac{1}{64\pi} \frac{|\bm{k}|}{s\sqrt{s}}
g_{HV'V}^2 (|g_{V'f}^L|^2 + |g_{V'f}^R|^2) 
\nonumber\\&&\times
\frac{2M_V^2 - |\bm{k}|^2(\cos^2\theta-1)}{M_V^2}
\frac{s}{(s-M_{V'^2})^2 + M_{V'}^2 \Gamma_{V'}^2},
\end{eqnarray}
where $\theta$ is the angle between $\bm{p}_1$ and $\bm{k}_1$.
\begin{eqnarray}
 |\bm{k}| \equiv |\bm{k}_1| = |\bm{k}_2|
&=&  \frac{\sqrt{\lambda(M_{V'},M_V,M_H)}}{2 M_{V'}},
\end{eqnarray}
is the momentum of a final state particle.
Integrating with respect to $\theta$, and taking interferences among intermediate bosons into account we obtain
\begin{eqnarray}
\lefteqn{
\sigma(f\bar{f}'\to W,W' \to WH)
}\nonumber\\
 &=& 
\frac{1}{N_c^i}\frac{1}{48\pi}\frac{\bm{|k|}}{\sqrt{s}} \frac{3M_{W}^2 + |\bm{k}|^2}{M_{W}^2}
\biggl\{
\sum_{V=W,W^{(1)}} \frac{g_{HVW}^2 [ |g_{Vff'}^L|^2 + |g_{Vff'}^R|^2]
}{(s - M_V^2)^2 + M_V^2 \Gamma_V^2}
\nonumber\\&&
+ 2 \Re \left[
\frac{g_{H W W} g_{H W^{(1)} W} [ (g_{W ff'}^L) (g_{W^{(1)} ff'}^{L})^*
 + (g_{W ff'}^R)  (g_{W^{(1)} ff'}^{R})^* ]}{
{}[(s-M_{W}^2) + i M_{W}\Gamma_{W}][(s-M_{W^{(1)}}^2) - i M_{W^{(1)}}\Gamma_{W^{(1)}}]}
\right]
\biggl\},
\\
\lefteqn{\sigma(f\bar{f}\to \gamma,Z,Z' \to ZH)}
\nonumber\\
 &=& 
\frac{1}{N_c^i}\frac{1}{48\pi}\frac{\bm{|k|}}{\sqrt{s}} \frac{3M_{Z}^2 + |\bm{k}|^2}{M_{Z}^2}
\biggl\{
\sum_{V=Z,Z^{(1)},\gamma^{(1)},Z_R^{(1)}}
\frac{ g_{HVZ}^2 [ |g_{Vf}^L|^2 + |g_{Vf}^R|^2]}{(s - M_V^2)^2 + M_V^2 \Gamma_V^2}
\nonumber\\&&
+ \sum_{\substack{V_1,V_2 = Z,Z^{(1)},\gamma^{(1)},Z_R^{(1)} \\ V_1 \ne V_2}}
\Re \left[
\frac{
g_{HV_1 Z} g_{HV_2 Z} [ (g_{V_1 f}^L) (g_{V_2 f}^{L})^*
 + (g_{V_1 f}^R) (g_{V_2 f}^{R})^*]
}{[(s-M_{V_1}^2) + i M_{V_1}\Gamma_{V_1}]
{}[(s-M_{V_2}^2) - i M_{V_2}\Gamma_{V_2}]}
\right]
\biggl\},
\end{eqnarray}
where $N_c^i$ is the number of colors of initial-state fermions.

For the process $f(p_1)\bar{f}'(p_2) \to W(k_1) Z(k_2)$, 
we adopt the approximation in which 
the interference term between the SM part and NP part is dropped. 
The cross section formulae in the SM are found in \cite{Brown:1979ux}.
The differential cross section mediated by  heavy charged vector bosons $W'$ in the center-of-mass frame is given by
\begin{eqnarray}
\lefteqn{
\frac{d\sigma}{dt}(f\bar{f} \to W' \to WZ) 
}\nonumber\\
&=& \frac{1}{64\pi s^2} \cdot 4s^2 A(t,u) \biggl\{
\sum_{\substack{
W'\in \{ W^{(n)} \} \\
W' \ne W
}}
\frac{g_{W'WZ}^2(|g_{W'ff'}^L|^2 + |g_{W'ff'}^R|^2)^2}{(s-M_{W'}^2)^2 + M_{W'}^2 \Gamma_{W'}^2}
\nonumber\\&& 
+ \sum_{\substack{W_1,W_2 \in \{ W^{(n)} \}\\ M_W \ll M_{W_1} < M_{W_2}}}
2\Re\left[ 
\frac{g_{W_1WZ} g_{W_2WZ}[
(g_{W_1ff'}^L)(g_{W_2ff'}^L)^* + (g_{W_1ff'}^R)(g_{W_2ff'}^R)^*]
}{ [(s-M_{W_1}^2) + i M_{W_1} \Gamma_{W_1}] [(s-M_{W_2}) - i M_{W_2} \Gamma_{W_2}]}
\right]
\biggr\},
\end{eqnarray}
where $s$, $t$ and $u$ are Mandelstam variables and $A(t,u)$ is given in \cite{Brown:1979ux}
by
\begin{eqnarray}
A(t,u) &=& \left(\frac{ut}{M_Z^2 M_W^2}-1 \right)
\left[\frac{1}{4} - \frac{M_Z^2 + M_W^2}{2s}
+ \frac{(M_W^2 + M_Z^2)^2 + 8 M_W^2 M_Z^2}{4s^2} \right]
\nonumber\\&&
+ \left( \frac{M_W^2 + M_Z^2}{M_W^2 M_Z^2}\right)
\left[ \frac{s}{2} - M_W^2 - M_Z^2 + \frac{(M_W^2 - M_Z^2)^2}{2s}\right].
\end{eqnarray}
Here $t_{\rm min} \le t \le t_{\rm max}$, $t_{\rm min,max} = \frac{1}{2}(M_W^2+M_Z^2-s)\pm\frac{1}{2}s\beta$, $\beta = |\bm{k}|/(\sqrt{s}/2)$
with $|\bm{k}| = |\bm{k}_{1,2}| = \sqrt{\lambda(M_{W'},M_W,M_Z)}/2M_{W'}$.
Integrating $d\sigma/dt$ with respect to $t$ and using
\begin{eqnarray}
\int_{t_{\rm min}}^{t_{\rm max}} A(t,u) dt
&=& \frac{s^3\beta^3}{24 M_W^2 M_Z^2}
\left[ 1 + \frac{10(M_W^2 + M_Z^2)}{s} 
+ \frac{M_W^4 + M_Z^4 + 10M_W^2M_Z^2}{s^2}\right],
\end{eqnarray}
we obtain
\begin{eqnarray}
\lefteqn{\sigma(ff' \to W' \to WZ)}
\nonumber\\
&=& \frac{1}{384\pi} 
%\frac{g_{W'WZ}^2 (|g_{W'ff'}^L|^2 + |g_{W'ff'}^R|^2 )}{
%(s- M_{W'}^2)^2 + M_{W'}^2\Gamma_{W'}^2} 
%\nonumber\\&& \times
\frac{s^3\beta^3}{M_W^2 M_Z^2}
\left[ 1 + \frac{10(M_W^2+M_Z^2)}{s} + \frac{M_W^4 + M_Z^4 + 10 M_W^2 M_Z^2}{s^2}\right]
\nonumber\\&& \times
\biggl\{
\sum_{\substack{W' \in \{W^{(n)} \}
\\
M_{W'} \gg M_W
}}
\frac{g_{W'WZ}^2(|g_{W'ff'}^L|^2 + |g_{W'ff'}^R|^2)^2}{(s-M_{W'}^2)^2 + M_{W'}^2 \Gamma_{W'}^2}
\nonumber\\&& 
\sum_{\substack{
W_1,W_2 \in \{ W^{(n)}\}
\\
M_W \ll M_{W_1} < M_{W_2}
}}
+ 2\Re\left[ 
\frac{g_{W_1 WZ} g_{W_2 WZ}[
(g_{W_1 ff'}^L)(g_{W_2 ff'}^L)^* + (g_{W_1 ff'}^R)(g_{W_2 ff'}^R)^*]
}{ [(s-M_{W_1}^2) + i M_{W_1} \Gamma_{W_1}] [(s-M_{W_2}) - i M_{W_2} \Gamma_{W_2}]}
\right] \biggr\}
.
\end{eqnarray}
Formulae for the processes  $f\bar{f} \to Z' \to W^+W^-$ ($Z' = Z^{(n)},\gamma^{(n)}, Z_R^{(n)}$, $n=1,2,\cdots$) 
can be obtained from the above formulae by replacements $W'\to Z'$  and $Z \to W$.

For the process $f\bar{f}' \to \{ V_i\} \to F\bar{F}$ 
where $f^{(\prime)}$, $F^{(\prime)}$ are massless fermions and $V_i$ are vector bosons,
differential cross section is given by
\begin{eqnarray}
\lefteqn{\frac{d\sigma}{d\cos\theta}(f\bar{f}\to\{ V_i\} \to F\bar{F}')
}\nonumber\\
&=& \frac{N_c^f}{N_c^i} \frac{s}{128\pi}
\biggl\{ \sum_{i} \frac{1}{(s-M_{V_i}^2)^2 + M_{V_i}^2 \Gamma_{V_i}^2}
\nonumber\\&&
\times  \biggl[
 \left(|g_{V_i ff'}^L|^2 + |g_{V_i ff'}^R|^2\right)
\left(|g_{V_iFF'}^L|^2 + |g_{V_iFF'}^R|^2 \right) (1 + \cos^2\theta)
\nonumber\\&&
+ 2 \left(|g_{V_i ff'}^L|^2 - |g_{V_i ff'}^R|^2\right)
\left(|g_{V_iFF'}^L|^2 - |g_{V_iFF'}^R|^2 \right)\cos\theta
\biggr]
\nonumber\\&&
+2\Re\sum_{i>j}
 \frac{1}{(s-M_{V_i}^2)^2 + M_{V_i}^2 \Gamma_{V_i}^2}
 \frac{1}{(s-M_{V_i}^2)^2 + M_{V_i}^2 \Gamma_{V_i}^2}
\nonumber\\&&
\times  \biggl[
\left(g_{V_i ff'}^L g_{V_j ff'}^{L*} + g_{V_i ff'}^R g_{V_j ff'}^{R*} \right)
\left(g_{V_i FF'}^L g_{V_j FF'}^{L*} + g_{V_i FF'}^R g_{V_j FF'}^{R*} \right) (1 + \cos^2\theta)
\nonumber\\&&
+ 2 
\left(g_{V_i ff'}^L g_{V_j ff'}^{L*} - g_{V_i ff'}^R g_{V_j ff'}^{R*} \right)
\left(g_{V_i FF'}^L g_{V_j ff'}^{L*} - g_{V_i FF'}^R g_{V_j FF'}^{R*} \right)
\cos\theta
\biggr]
\biggr\},
\end{eqnarray}
where $\theta$ is the scattering angle. The corresponding distribution in the transverse momentum $p_T \equiv (\sqrt{s}/2) \sin\theta$ is obtained by
\begin{eqnarray}
\frac{d\sigma}{dp_T}(p_T,s) &=& \left.\frac{d\sigma}{d\cos\theta}\right|_{\cos\theta = \sqrt{1 - 4p_T^2/s}} \cdot \frac{4p_T}{s\sqrt{1 - 4p_T^2/s}}.
\end{eqnarray}

%%%%%%%%%%%%%%%%%%%%%%%%%%%%%%%%%%%%%%%%%%%%%%%%%%%%%%%%%%%%%%%%%%%%%%%%%%%

%%%%%%%%%%%%%%%%%%%%%%%%%%%%%%%%%%%%%%%%%%%%%%%
\end{document}